\documentclass{aa}
\usepackage{graphicx}
\usepackage[varg]{txfonts}
\usepackage[colorlinks=True,linkcolor=blue,citecolor=blue, urlcolor=blue]{hyperref}
\usepackage{pdflscape}
\usepackage{natbib,twoopt}
\usepackage{amsmath}
\usepackage{moresize}
\bibpunct{(}{)}{;}{a}{}{,}             

\begin{document} 
        
        \title{
A global view on star formation: The GLOSTAR Galactic plane survey \\ X. Galactic \ion{H}{ii} region catalog using radio recombination lines\thanks{Tables \ref{tab:catalog} and \ref{tab:phy_prop} are only available in electronic form at the CDS via anonymous ftp to \url{cdsarc.u-strasbg.fr} (130.79.128.5) or via \url{http://cdsweb.u-strasbg.fr/cgi-bin/qcat?J/A+A/}. }}
        
        \author{S.~Khan \inst{1,\thanks{Member of the International Max Planck Research School (IMPRS) for Astronomy and Astrophysics at the Universities of Bonn and Cologne.}},
        M.~R.~Rugel \inst{1,2,3,\thanks{M.R.R. is a Jansky Fellow of the National Radio Astronomy Observatory.}},
        A. Brunthaler \inst{1},
        K. M. Menten \inst{1},
        F. Wyrowski \inst{1},
                J. S. Urquhart\inst{4}, 
        Y. Gong \inst{1},
                A. Y. Yang \inst{5,6,1},
        H. Nguyen\inst{1},
        R. Dokara \inst{1},
        S. A. Dzib \inst{1}, 
        S.-N. X. Medina\inst{1,7}, 
        G. N. Ortiz-Le\'{o}n\inst{8,1},
        J. D. Pandian\inst{9},
        H.\,Beuther\inst{10},
        V. S. Veena\inst{1,11},
        S. Neupane\inst{1},
        A. Cheema\inst{1},
        W. Reich\inst{1},
        \and
        N. Roy\inst{12}
        }

        \institute{Max-Planck-Institut f\"{u}r Radioastronomie, Auf dem H\"{u}gel 69, 53121 Bonn, Germany 
                \email{skhan@mpifr-bonn.mpg.de}
                \and
                Center for Astrophysics | Harvard $\&$ Smithsonian, 60 Garden Street, Cambridge, MA 02138, USA 
                \and
                National Radio Astronomy Observatory, P.O. Box O, 1003 Lopezville Road, Socorro, NM 87801, USA
                \and 
                Centre for Astrophysics and Planetary Science, University of Kent, Canterbury, CT2 7NH, UK
                \and
                National Astronomical Observatories, Chinese Academy of Sciences, A20 Datun Road, Chaoyang District, Beijing, 100101, P. R. China
                \and 
                Key Laboratory of Radio Astronomy and Technology, Chinese Academy of Sciences, A20 Datun Road, Chaoyang District, Beijing, 100101, P. R. China
                \and
                German Aerospace Center, Scientific Information, 51147 Cologne, Germany
                \and
                Instituto Nacional de Astrof\'isica, \'Optica y Electr\'onica, Apartado Postal 51 y 216, 72000 Puebla, Mexico
                \and
                Department of Earth $\&$ Space Sciences, Indian Institute of Space Science and Technology, Trivandrum 695547, India
                \and 
                Max Planck Institute for Astronomy, Koenigstuhl 17, 69117 Heidelberg, Germany
                \and
                I. Physikalisches Institut, Universit\"at zu Köln, Z\"ulpicher Str. 77, 50937 K\"oln, Germany
                \and
                Department of Physics, Indian Institute of Science, Bangalore 560012, India}

        \date{Received ; accepted }
        
        \abstract
        {Studies of Galactic \ion{H}{ii} regions are of crucial importance for studying star formation and the evolution of the interstellar medium. Gaining an insight into their physical characteristics contributes to a more comprehensive understanding of these phenomena. The GLOSTAR project aims to provide a GLObal view on STAR formation in the Milky Way by performing an unbiased and sensitive survey. This is achieved by using the extremely wideband (4$-$8~GHz) \textit{C}-band receiver of the \textit{Karl G. Jansky} Very Large Array and the Effelsberg 100~m telescope. Using radio recombination lines observed in the GLOSTAR survey with the VLA in D-configuration with a typical line sensitivity of 1\,$\sigma \sim \rm 3.0~mJy~beam^{-1}$ at $\sim \rm 5~km~s^{-1}$ and an angular resolution of 25\arcsec, we cataloged 244 individual Galactic \ion{H}{ii} regions ($-$2$\degr \leq$ $\ell$ $\leq$ 60$\degr$ \& |\textit{b}| $\leq$ 1$\degr$ and 76$\degr \leq$ $\ell$ $\leq$ 83$\degr$ \& $-$1$\degr \leq$ \textit{b} $\leq$ 2$\degr$) and derived their physical properties. We examined the mid-infrared (MIR) morphology of these \ion{H}{ii} regions and find that a significant portion of them exhibit a bubble-like morphology in the GLIMPSE 8~$\mu$m emission. We also searched for associations with the dust continuum and sources of methanol maser emission, other tracers of young stellar objects, and ﬁnd that 48\% and 14\% of our \ion{H}{ii} regions, respectively, are coextensive with those. We measured the electron temperature for a large sample of \ion{H}{II} regions within Galactocentric distances spanning from 1.6 to 13.1~kpc and derived the Galactic electron temperature gradient  as $\sim$ 372 $\pm$ 28~K~kpc$^{-1}$ with an intercept of 4248 $\pm$ 161~K, which is consistent with previous studies.}

        
        \authorrunning{S.\, Khan et al.}
        \titlerunning{Galactic \ion{H}{ii} regions in the GLOSTAR survey}
        \maketitle
        \section{Introduction} \label{sec:intro}
        \ion{H}{ii} regions are the zones of ionized plasma formed when the ultraviolet (UV) radiation emitted by massive stars of spectral type B0 or earlier ionizes the surrounding interstellar medium (ISM). Since these massive stars have lifespans as short as a few million years, the presence of \ion{H}{ii} regions serves as a strong indication of recent high-mass star formation (HMSF) activity within the Galaxy. Despite their small numbers and short existence, these regions play a crucial role in the dynamics and evolution of the ISM and the Galaxy as a whole. \ion{H}{ii} regions act as classic indicators of Galactic spiral arms and have played a fundamental role in enhancing our knowledge of our Galaxy's structure~\citep[e.g.,][]{1976A&A....49...57G, 2019ApJ...885..131R}. The chemical abundances within these regions represent the current abundances of the Galaxy, illustrating Galactic chemical evolution. The central stars not only emit a substantial amount of UV and far-ultraviolet (FUV) photons into their surroundings, but also provide feedback mechanisms that influence the subsequent generation of star formation in a sequential manner \citep{1977ApJ...214..725E,2003A&A...399.1135D,2011A&A...527A..62B,2012MNRAS.421..408T}.
                
        Young and dense \ion{H}{ii} regions, including hypercompact (HC) and ultracompact (UC) \ion{H}{II} regions, are powered by massive young stars that are embedded within dusty molecular cocoons. These cocoons are optically thick to visible and UV radiation emitted by the massive young star. As \ion{H}{ii} regions evolve, they progressively disperse the surrounding material. \ion{H}{ii} regions harbor a substantial amount of dust that can absorb a significant portion of the UV radiation emitted by their central OB stars~\citep{2021A&A...645A.110Y,2018ApJ...864..136B,1989ApJS...69..831W}. These absorbed photons are re-emitted at infrared wavelengths. As a result, \ion{H}{ii} regions exhibit a characteristic mid-infrared (MIR) signature, with emission from polycyclic aromatic hydrocarbons (PAHs), in the $\sim$4 to 12~$\mu$m wavelength range, typically enclosing the 24~$\mu$m continuum emission from warm dust, which coexists with the ionized gas traced by radio continuum emission~\citep[see][]{2011ApJS..194...32A}. The distinctive MIR emission has been utilized to identify potential \ion{H}{ii} region candidates~\citep[e.g.,][]{2014ApJS..212....1A}, which can subsequently be validated through observations of radio continuum and radio recombination lines (RRLs).
        
        The radio continuum emission from \ion{H}{ii} regions is thermal bremsstrahlung, which makes them bright at radio wavelengths. Emission from recombining electrons and nuclei in the plasma produces recombination lines, observable from optical to radio wavelengths. Emission from RRLs is a valuable tool to confirm the presence of \ion{H}{ii} regions. They facilitate the exploration of the physical conditions and, given their Doppler velocities, the dynamics of ionized gas within these regions \citep{1978ARA&A..16..445B, 2002ASSL..282.....G}. Additionally, measuring the velocities of RRLs as probes of Galactic rotation enables the determination of ``kinematic distances'' to \ion{H}{II} regions. The relationship between the electron temperatures derived from RRL measurements and Galactocentric distances reveals a Galactic temperature and metallicity gradient \citep{1975A&A....38..451C, 1983MNRAS.204...53S, 1983A&A...127..211W, 2006ApJ...653.1226Q}. The metallicity gradient imposes significant constraints on the chemical evolution of the Galaxy \citep{2015ApJ...806..199B, 2019ApJ...887..114W}. Finally, with a large sample, we can statistically investigate the physical characteristics of Galactic \ion{H}{ii} regions. 
        
        In the past, extensive radio continuum and RRLs surveys have unveiled a large number of Galactic \ion{H}{ii} regions
        \citep[e.g.,][]{1970A&A.....4..357R,1979A&AS...35...23A,1989ApJS...71..469L,1997ApJ...488..224K,2007A&A...461...11U,2009A&A...501..539U,2013MNRAS.435..400U,2018A&A...615A.103K,2019A&A...623A.105G,2020ApJS..248....3C}. \citet{2003A&A...397..213P} created a widely used master catalog of 1442 sources by compiling radio data of Galactic \ion{H}{ii} regions from 24 published works. The Green Bank Telescope \ion{H}{ii} Region Discovery Survey~\citep[GBT HRDS;][]{2010ApJ...718L.106B, 2011ApJS..194...32A} doubled the number of known \ion{H}{ii} regions in the Galactic zone 343$\degr \leq$ $\ell$ $\leq$ 67$\degr$, |\textit{b}| $\leq$ 1$\degr$, by detecting RRLs from 448 previously unknown \ion{H}{ii} regions. Similarly, \citet{2012ApJ...759...96B} conducted an RRL survey with the Arecibo telescope and identified 37 new \ion{H}{ii} regions. Despite these efforts, the census of Galactic \ion{H}{ii} regions remains incomplete, as is indicated by GBT HRDS. In response, \citet{2014ApJS..212....1A} compiled a catalog of around 8000 Galactic \ion{H}{ii} region and \ion{H}{ii} region candidates using data from the all-sky {Wide-Field Infrared Survey Explorer} \citep[WISE;][]{2010AJ....140.1868W}.

    The GLObal view on STAR Formation (GLOSTAR\footnote{\url{https://glostar.mpifr-bonn.mpg.de/glostar/}}) in the Milky Way survey \citep{2019A&A...627A.175M,2021A&A...651A..85B} is an unbiased survey observing the Galactic plane with the \textit{Karl G. Jansky} Very Large Array (VLA) in D- and B-configurations as well as the Effelsberg 100~m radio telescope at C-band (4$-$8 GHz). The survey comprises observations of the continuum emission in full polarization and of spectral lines (namely, the 4.8~GHz transition of formaldehyde ($\rm H_{2}CO$), the 6.7~GHz maser line of methanol ($\rm CH_{3}OH$) maser and numerous RRLs) in order to locate and characterize star-forming regions in the Milky Way. The data contains a wealth of information that has already been used to catalog radio sources \citep[][]{2019A&A...627A.175M, 2023A&A...670A...9D, 2023A&A...680A..92Y, medina2024}, identify supernova remnants \citep[SNRs; ][]{2021A&A...651A..86D,2023A&A...671A.145D}, increase the number of class II methanol masers \citep{2021A&A...651A..87O,2022A&A...666A..59N}, study radio emission of young stellar objects (YSOs) in the Galactic center \citep{2021A&A...651A..88N}, and understand the molecular gas structures on different linear scales with the 4.8~GHz formaldehyde ($\rm H_{2}CO$) absorption line in the Cygnus~X region \citep{2023A&A...678A.130G}. This work uses RRL data observed in the GLOSTAR survey to investigate Galactic \ion{H}{ii} regions and their physical properties.

    In this study, we present a catalog of Galactic \ion{H}{ii} regions distributed over a significant part of the Galactic plane toward which RRL emission was detected in the GLOSTAR survey. We structure this paper as follows. In Sect.~\ref{sec:obs}, we describe the GLOSTAR observations, the imaging processes, and also the ancillary data used in the paper. Section~\ref{sec:source_extraction} details our source extraction criteria and methods. Section~\ref{sec:result} describes the sources catalog and the physical properties of the \ion{H}{ii} regions and provides comparisons with previous studies. In Sect.~\ref{sec:discussion}, we discuss our findings. We present a summary in Sect.~\ref{sec:conclusion}.  
    
        \section{Observations and data analysis} \label{sec:obs}
        
        \subsection{VLA D-configuration observations}\label{sec:obsd}
        The GLOSTAR survey~\citep{2021A&A...651A..85B} was carried out using the VLA in B- and D-configuration and the Effelsberg 100~m radio telescope in the 4$-$8~GHz frequency range (\textit{C}-band). The above paper gives a comprehensive, detailed description of the survey and first results, while a summary follows here. The full survey covers the Galactic longitude range of $-$2$\degr \leq$ $\ell$ $\leq$ 60$\degr$ with latitudes |\textit{b}| $\leq$ 1$\degr$ and, in addition, the Cygnus X region (76$\degr \leq$ $\ell$ $\leq$ 83$\degr$, $-$1$\degr \leq$ \textit{b} $\leq$ 2$\degr$), or 145 square degrees in total. The data products from the survey include continuum emission from the 4.2$-$5.2~GHz and 6.4$-$7.4~GHz ranges in full polarization. Higher-frequency-resolution spectral windows were used to cover the most prominent methanol maser emission line at 6.7~GHz ($\rm 5_1-6_0 A^+$), seven hydrogen RRLs, and the 4.829~GHz ($\rm 1_{1,0}-1_{1,1}$) transition of formaldehyde. The lines were observed in the dual polarization mode with channel spacing of 3.9~kHz for $\rm CH_3OH$ and $\rm H_2CO$, and 62.5~kHz for the RRLs. The details of RRL observations using VLA in D-configuration are summarized in Table~\ref{tab:glostar_line}. The central band velocity with respect to the local standard of rest, $\rm V_{LSR}$, was varied for different regions of the survey based on longitude-velocity plots of CO in the Milky Way \citep[e.g.,][]{2001ApJ...547..792D}. The observations providing the data used in this work were carried out using 380~h of VLA observing time during the time period from December 2011 to April 2017. Further details with program IDs are summarized in Table 1 of \citet{2021A&A...651A..87O}, \citet{2022A&A...666A..59N} and \citet{medina2024}. While the GLOSTAR survey provides C-band RRL observations using the VLA in both B and D configurations and the Effelsberg 100~m telescope, our focus is mainly on the VLA D configuration RRL data. Initial tests on the VLA B-configuration RRL data revealed detections only for the most prominent star-forming regions in our survey area. For most of the regions in this work, the data is heavily affected by spatial filtering and not sensitive enough to detect RRL emission. While adding Effelsberg 100~m RRL data could help fill in the missing flux, we find many large and diffuse \ion{H}{ii} regions in the Effelsberg 100~m data, which are significantly fainter and more extended scales than traced by the VLA D-configuration data presented here. Sources detected in the Effelsberg 100~m data will be discussed in a separate publication after completion of processing, and we shall include comparisons to the VLA D-configuration where applicable. In this work, we focus on RRL emission from \ion{H}{ii} regions on scales of $\sim$0.1-10 pc traced by the VLA D-configuration data.

        \begin{table*}[h!]
            \caption{Observed RRLs, bandwidth, number of channels, spectral resolution, and velocity coverage of VLA observations.}
            \label{tab:glostar_line}
            \centering
            \begin{tabular}{cccccc}
                    \hline \hline
                    RRL & Frequency & Bandwidth & Channels & Resolution & Coverage \\
                        & [MHz]     & [MHz]     &          &   [km~s$^{-1}$] & [km~s$^{-1}$]\\
                    \hline \hline
            H114$\alpha$&4380.954&8&128&4.3&547\\
            H113$\alpha$&4497.776&8&128&4.2&533\\
            H112$\alpha$&4618.789&8&128&4.1&529\\
            H110$\alpha$&4874.157&8&128&3.8&492\\
            H99$\alpha$&6676.076&8&128&2.8&359\\
            H98$\alpha$&6881.486&8&128&2.7&348\\
            H96$\alpha$\tablefootmark{*}&7318.296&8&128&2.5&328\\
                \hline
            \end{tabular}
       \tablefoot{For $\rm \ell = 58\degr - 60\degr$, the RRLs H115$\alpha$ (4268.142~MHz), H102$\alpha$ (6106.855~MHz), and H103$\alpha$ (5931.544~MHz) were observed instead of H110$\alpha$ and H96$\alpha$.
       \tablefootmark{*}{The H96$\alpha$ transition was flagged for the entire survey area due to RFI contamination.}}
    \end{table*}
    
        \subsection{Data reduction and imaging}\label{sec:imaging}
        In this work, we focus on RRL data obtained within the GLOSTAR survey conducted with the VLA in D-configuration. The data underwent calibration using a modified version of the VLA scripted pipeline\footnote{\url{https://science.nrao.edu/facilities/vla/data-processing/pipeline/scripted-pipeline}} (version 1.3.8) for CASA\footnote{\url{https://casa.nrao.edu}} (version 4.6.0) designed for spectral line data. To preserve spectral resolution, no Hanning smoothing was applied during the preliminary calibration. The \texttt{rflag} flagging command was selectively used solely for calibration scans to avoid erroneous flagging of spectral lines. Furthermore, \texttt{statwt} was not used to alter the statistical weighting. The complex gain calibrators used for different fields include: J1804+0101, J1820$-$2528, J1811$-$2055, J1825$-$0737, J1907+0127, J1955+1530, J1925+2106, and J1931+2243, and the flux calibrators are 3C286 and 3C48. After the initial execution of the calibration pipeline, quality checks, manual flagging, and a rerun of the pipeline were carried out to ensure data integrity. The RRLs were imaged individually in fields with dimensions of $\Delta \ell$ $\times$ $\Delta$\textit{b} = 1$\degr~\times$~2$\degr$, except for the Cygnus X region in which the image dimensions were adjusted to 1$\degr~\times$  3$\degr$. To ensure consistent sensitivity across the image borders and account for sources with potential sidelobe effects, we incorporated neighboring pointings during the imaging process. We subtracted the continuum in the \textit{uv}-domain from line-free channels. As line-free channels, we chose all channels excluding velocities between -- 40 and 120~$\rm km~s^{-1}$ for $\ell$ $\geq$ 9$\degr$, and -- 40 and 100~$\rm km~s^{-1}$ for $\ell$ $\leq$ 9$\degr$ as the pure continuum emission. For a few tiles toward the Galactic Center, the channels were selected manually. The line-free continuum and line data were separately imaged and deconvolved with the CASA (version 5.7.0) task \texttt{tclean}, with subparameters {\tt robust=0.5}, \texttt{imagemode=``mosaic''}, \texttt{gridder=``mosaic''}, and \texttt{restoringbeam=``common'' } to restore the common beam. We flagged the higher-frequency H96$\alpha$ line data for the entire survey area due to high contamination from radio frequency interference (RFI) before imaging \citep[for more details see][]{2021A&A...651A..85B}. Other lines that were affected in subsets of the survey are: H99$\alpha$ for the fields centered at $\ell$=17.5$\degr$ and $\ell$=36.5$\degr$; H98$\alpha$, which were flagged at $\ell = 37.5\degr$; the H113$\alpha$ transition, which was flagged for $\ell$=54.5$\degr$; and the H112$\alpha$ transition, for $\ell$=56.5$\degr$. The removal was done using a combination of the automatic flagging routines {\tt rflag} and {\tt tfcrop} prior to imaging within the CASA environment. 

\begin{figure*}[h!]
        \centering
        \includegraphics[scale=0.37]{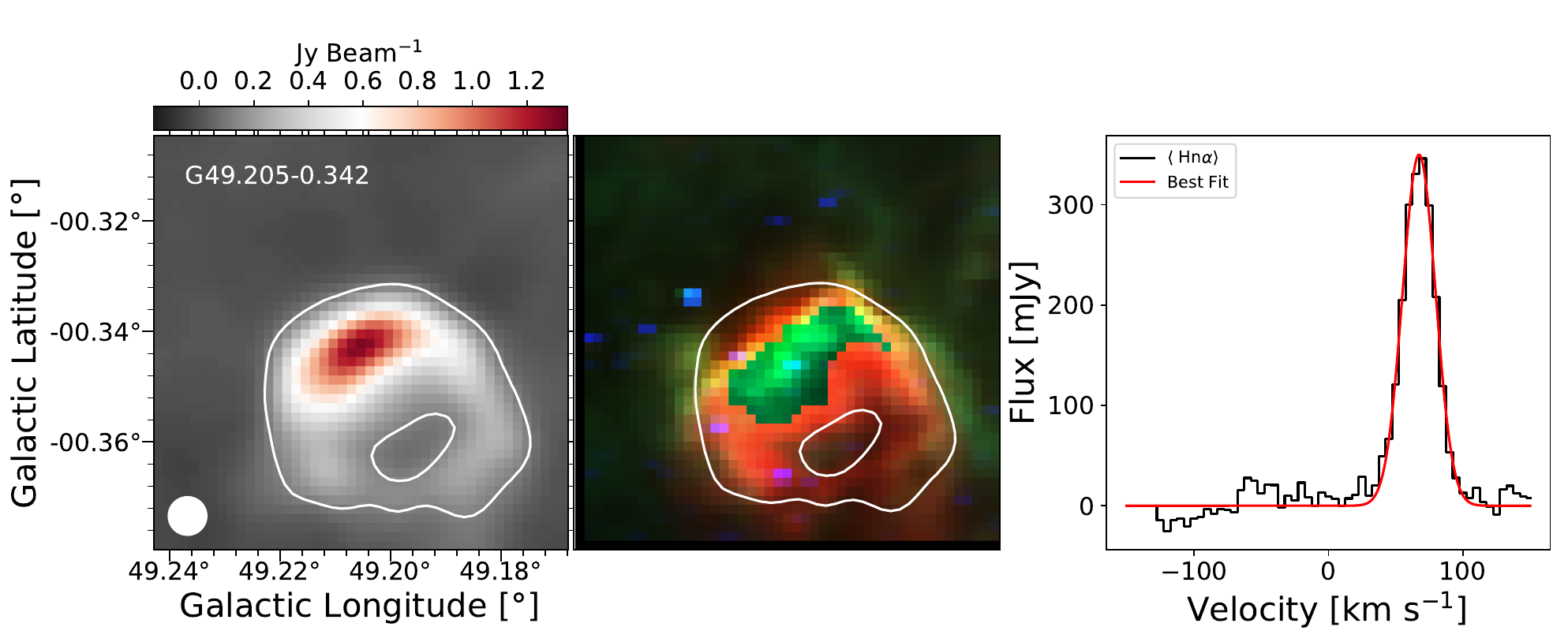}
        \includegraphics[scale=0.37]{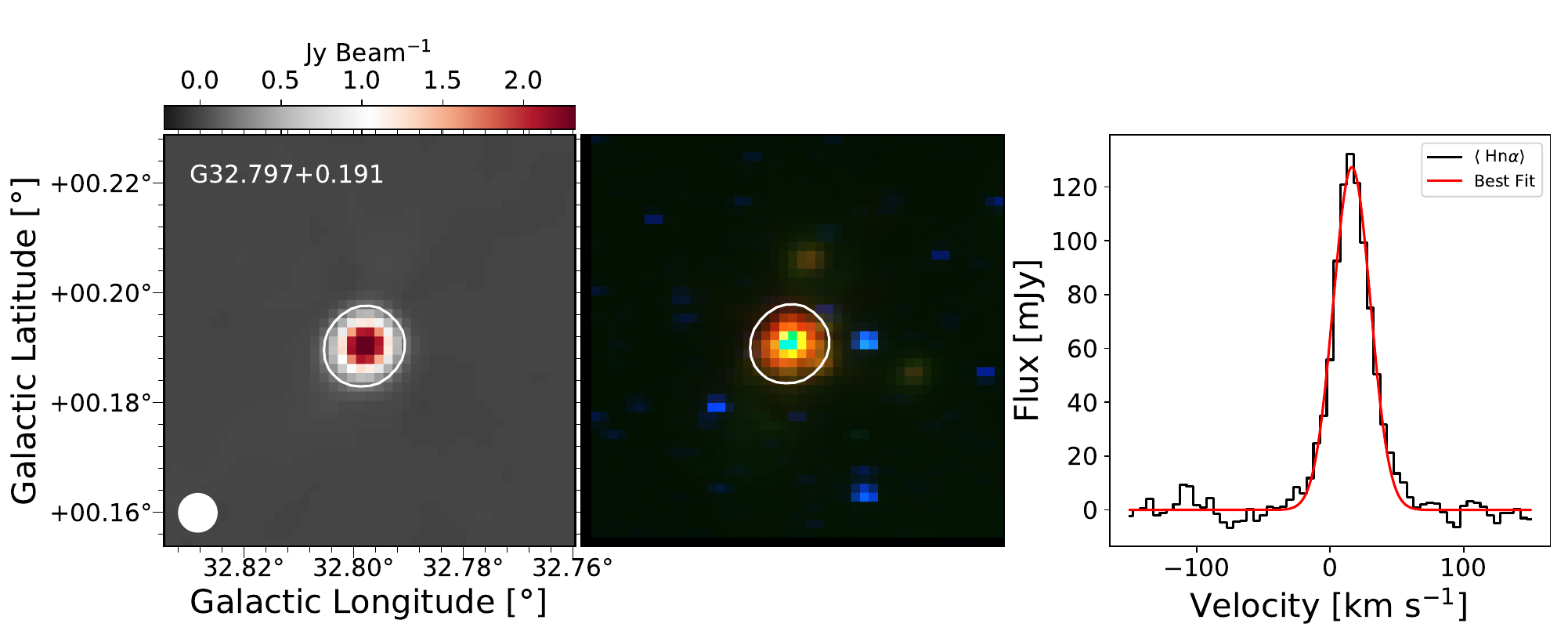}
        \includegraphics[scale=0.37]{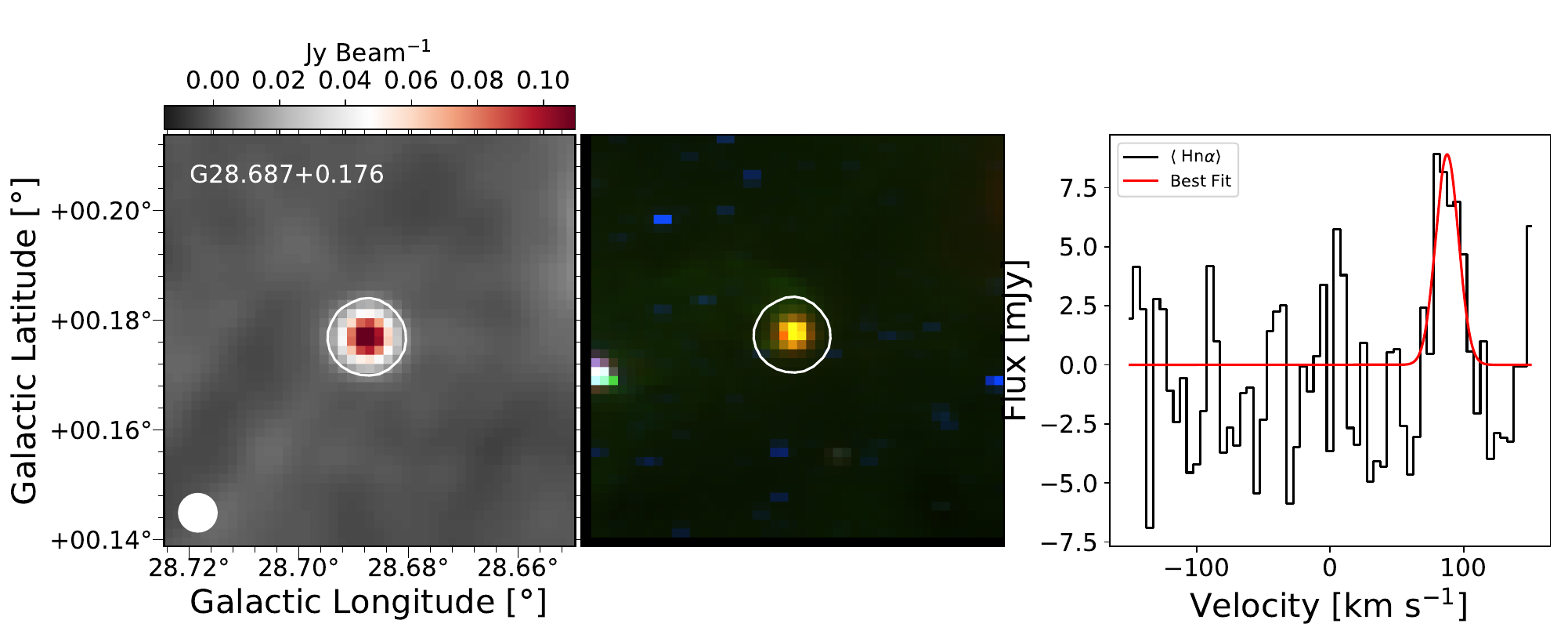}
        \caption{Examples of three GLOSTAR \ion{H}{ii} regions. G049.205-0.342 (\textit{top panel}) and G032.797+0.191 (\textit{middle panel}) exemplify extended and compact radio  \ion{H}{ii} regions found in our catalog, respectively, while G028.687+0.177 (\textit{bottom panel}) represents the weakest source within our catalog. From left to right: GLOSTAR VLA 5.3~GHz radio continuum emission, a three-color image constructed from \textit{Spitzer} infrared data: MIPSGAL 24~$\mu$m (red), GLIMPSE 8.0~$\mu$m (green), and GLIMPSE 3.6~$\mu$m (blue), and stacked $\langle$Hn$\alpha \rangle$ RRL spectra extracted within the region shown as white contour. The filled white circle shows the 25$\arcsec$ GLOSTAR beam. }
        \label{fig:source_example}
    \end{figure*}

        Previous studies \citep[e.g., ][]{2006AJ....132.2326B, 2010MNRAS.405.1654A, 2011ApJS..194...32A,2016A&A...595A..32B,2020ApJS..248....3C} have shown that multiple higher-principle quantum number RRLs (n > 50) can be averaged together to improve the signal-to-noise ratio. Because RRLs with adjacent quantum numbers and the same quantum number difference, $\triangle $n, arise from similar energy levels and have similar oscillator strengths, their line parameters such as intensity, line width, and velocity are nearly identical. \citet{2022A&A...664A.140K} successfully used the GLOSTAR stacked RRLs to study the physical properties of the ionized gas in the W33 Main star-forming region. Thus, we averaged the data of all six observed RRLs to get a more sensitive composite $\langle$Hn$\alpha \rangle$ RRL. However, before stacking RRLs, we re-gridded all the line image cubes to common velocity bins of 5~km~s$^{-1}$ with a velocity range from $-$150~km~s$^{-1}$ to 150~km~s$^{-1}$, and smoothed the data to a common angular resolution of 25$\arcsec$. Fig.~\ref{fig:source_example} provides radio continuum images, corresponding MIR color composite images, and $\langle$Hn$\alpha \rangle$ spectra for three distinct \ion{H}{ii} regions. G049.205-0.342 and G032.797+0.191 exemplify extended and compact \ion{H}{ii} regions found in our catalog, respectively, while G028.687+0.177 represents the weakest source within our catalog.

        In the GLOSTAR VLA survey, \textit{C}-band continuum emission was observed with two 1 GHz basebands in full polarization mode and imaged at an average frequency of 5.8~GHz \citep[see][]{2019A&A...627A.175M,2021A&A...651A..85B}. Images produced from line-free portions of the bands were used to determine the continuum flux density of the sources, which was used to derive physical properties such as the electron temperature, electron density, emission measure (EM), and Lyman continuum photon rate (see Sect.~\ref{sec:phy_prop}). While the sensitivity of the full 2 GHz continuum surpasses that of the line-free continuum, we benefit from an exact frequency coverage match between the line-free continuum and the RRLs. Furthermore, the loss of sensitivity becomes less significant because we are dealing exclusively with relatively bright sources. Similar to RRLs, we also smoothed line-free continuum images to a common angular resolution of  25$\arcsec$ and averaged all line-free continuum images for the six bands that contain the RRLs. Additionally, after excluding the highest frequency RRL, the central frequency of $\langle$Hn$\alpha \rangle$ is 5.3~GHz. For comparisons with the RRLs, one should use the continuum fluxes of \ion{H}{ii} regions mentioned in the current work. For all other purposes, we recommend using fluxes from \citet[][ 2024 in prep]{2019A&A...627A.175M}.

        \subsection{Complementary data}
        In addition to the RRL data, we used data of the $870 \mu$m wavelength sub-millimeter continuum emission collected in the APEX Telescope Large Survey of the Galaxy (ATLASGAL) with a $19.5''$ resolution \citep{2009A&A...504..415S,2013A&A...549A..45C,2014A&A...568A..41U} to search for the dust emission associated with the \ion{H}{ii} regions. To study the MIR morphology of the \ion{H}{ii} regions, we also made use of 3.6 and 8.0~$\mu$m band data from the GLIMPSE survey \citep{2003PASP..115..953B} and $24.0~\mu$m data from the MIPSGAL survey \citep[]{2009PASP..121...76C}, both conducted with the Spitzer Space Observatory.
        
        \section{Source extraction} \label{sec:source_extraction}
        As was mentioned in Sect.~\ref{sec:obs}, we used RRL data to identify Galactic \ion{H}{ii} regions. To accomplish this, a straightforward extraction code can be employed to examine each pixel for the presence of line emission that is brighter than its surroundings. An example of such a code is the source extraction code developed by \citet{2022A&A...666A..59N} for cataloging methanol masers in the GLOSTAR survey. However, we noticed that this method is not effective for detecting weak and broad lines. To address this limitation, we capitalized on the fact that \ion{H}{ii} regions are bright sources in the radio continuum. Therefore, we began by extracting continuum sources from a line-free continuum image. To create the continuum maps, we imaged the line-free channels (Sect.~\ref{sec:imaging}). The source extraction process was performed using the \texttt{BLOBCAT}\footnote{\url{https://blobcat.sourceforge.net/}} software package, which was developed by \citet{2012MNRAS.425..979H}. \texttt{BLOBCAT} is a Python script that uses a flood-fill algorithm to detect and identify clusters of pixels representing sources in two-dimensional radio wavelength images. This package has been employed to create source catalogs for various surveys, including the GLOSTAR survey \citep[][]{2019A&A...627A.175M, 2023A&A...670A...9D, 2023A&A...680A..92Y, medina2024} and the HI/OH/Recombination line survey of the inner Milky Way \citep[THOR; ][]{ 2016A&A...588A..97B, 2018A&A...619A.124W}. We followed the same methodology as \citet{2019A&A...627A.175M}, who created a continuum source catalog within the region of 28$\degr \leq$ $\ell$ $\leq$ 36$\degr$, |\textit{b}|  $\leq$ 1$\degr$. Since the noise in the GLOSTAR survey is position-dependent, we generated independent noise maps using the \texttt{SExtractor} package\footnote{\url{https://www.astromatic.net/software/sextractor/}} \citep{1996A&AS..117..393B} from the line-free continuum data. This algorithm assigns a root mean square (rms) value to each pixel in an image by analyzing the distribution of pixel values within a local mesh until all values converge around a chosen $\sigma$ value.

        To perform automatic source extraction, we used the \texttt{BLOBCAT} software with the noise map as input and used a detection threshold (parameter \texttt{dSNR} in \texttt{BLOBCAT}) of four. This process created a catalog of continuum sources along with their corresponding continuum source masks. To identify sources with RRL brightness higher than the surrounding noise, we developed a simple Python code. This code first applies the continuum source mask generated by the \texttt{BLOBCAT} script to the RRL image and extracts the stacked RRL spectra averaged over the blob. For each blob, the script calculates the signal-to-noise ratio for each velocity channel. If three consecutive channels along the spectral axis  are found to have a signal-to-noise ratios greater than three times the standard deviation, the source was considered for further examination. Implementing these input parameters led to the detection of 265 continuum sources with associated RRL emission within the studied region.
        
\begin{figure}[h!]
        \centering
        \includegraphics[scale=0.7]{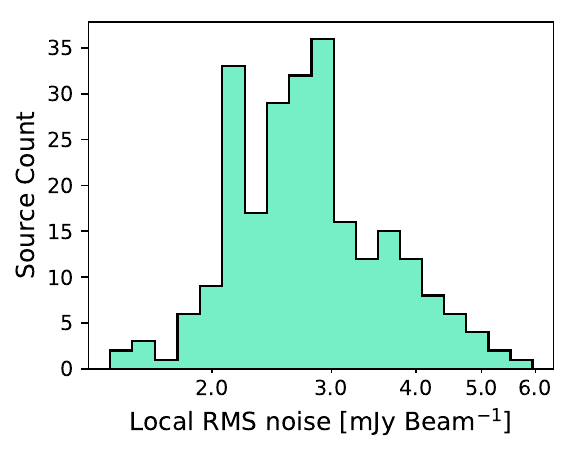}
        \includegraphics[scale=0.7]{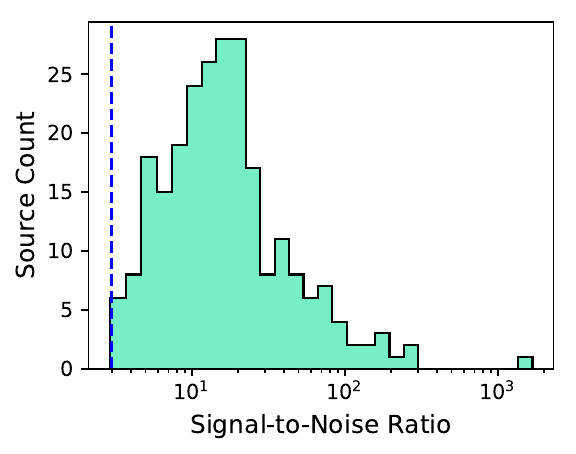}
        \caption{Histograms of the values of the local RMS noise of the velocity channels around the RRL (top) and the signal-to-noise ratio (bottom) for all the RRLs detected in this study. The median value of RMS noise level is $\sim$3.0~mJy~beam~$^{-1}$. The dashed blue line represents the 3$\sigma$ level.}
        \label{fig:local_rms}
    \end{figure}

\begin{figure}[h!]
        \centering
        \includegraphics[scale=0.18]{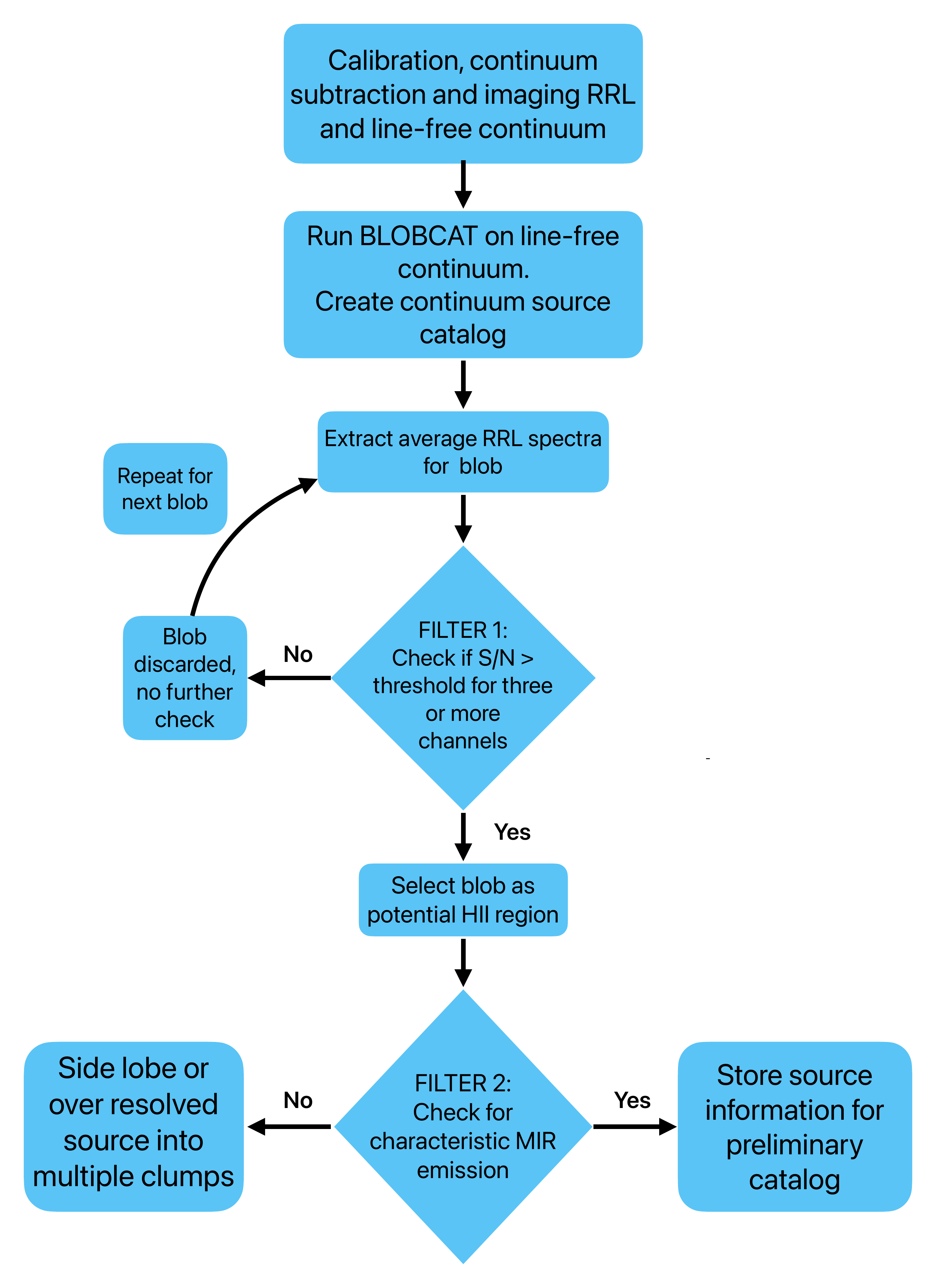}
        \caption{Flow chart illustration of the source extraction process described in Sect.~\ref{sec:source_extraction}.}
        \label{fig:flow_chat}
    \end{figure}

        Due to missing short spacing information, the VLA observations are not sensitive to emission on scales larger than $\sim$120$\arcsec$ in the D (most compact) configuration at $\sim$6.0~GHz. Consequently, some sources that exhibit uniform emission on such large spatial scales may not be detected in the VLA observations. However, the observations are expected to detect clumpy emission within these large \ion{H}{ii} regions. Therefore, no size restriction was imposed when defining the sample. Nevertheless, the sample may contain spurious detections, such as emission from side lobes and large-scale structures that have been over-resolved, causing the emission to split into multiple components. To address these potential false detections, we cross-referenced the sources with characteristic MIR images and also compared them spatially with the WISE \ion{H}{ii} region catalog \citep{2014ApJS..212....1A}. Additionally, we required that the peak amplitude of the $\langle$Hn$\alpha \rangle$ RRL signal must be at least three times the local RMS noise around the source in the RRL image. Fig.~\ref{fig:local_rms} shows the distributions of the local RMS noise values of the velocity channels around the RRLs and the signal-to-noise ratios of the detected RRLs. A flow chart illustrating the method is presented in the Fig.~\ref{fig:flow_chat}. Using this approach, we created an unbiased catalog of 244 Galactic \ion{H}{ii} regions based on RRL observations conducted with VLA in the D-configuration.

        \section{Results} \label{sec:result}
        \subsection{Source catalog}
\begin{figure*}[htbp!]
        \centering
        \includegraphics[width=1\linewidth]{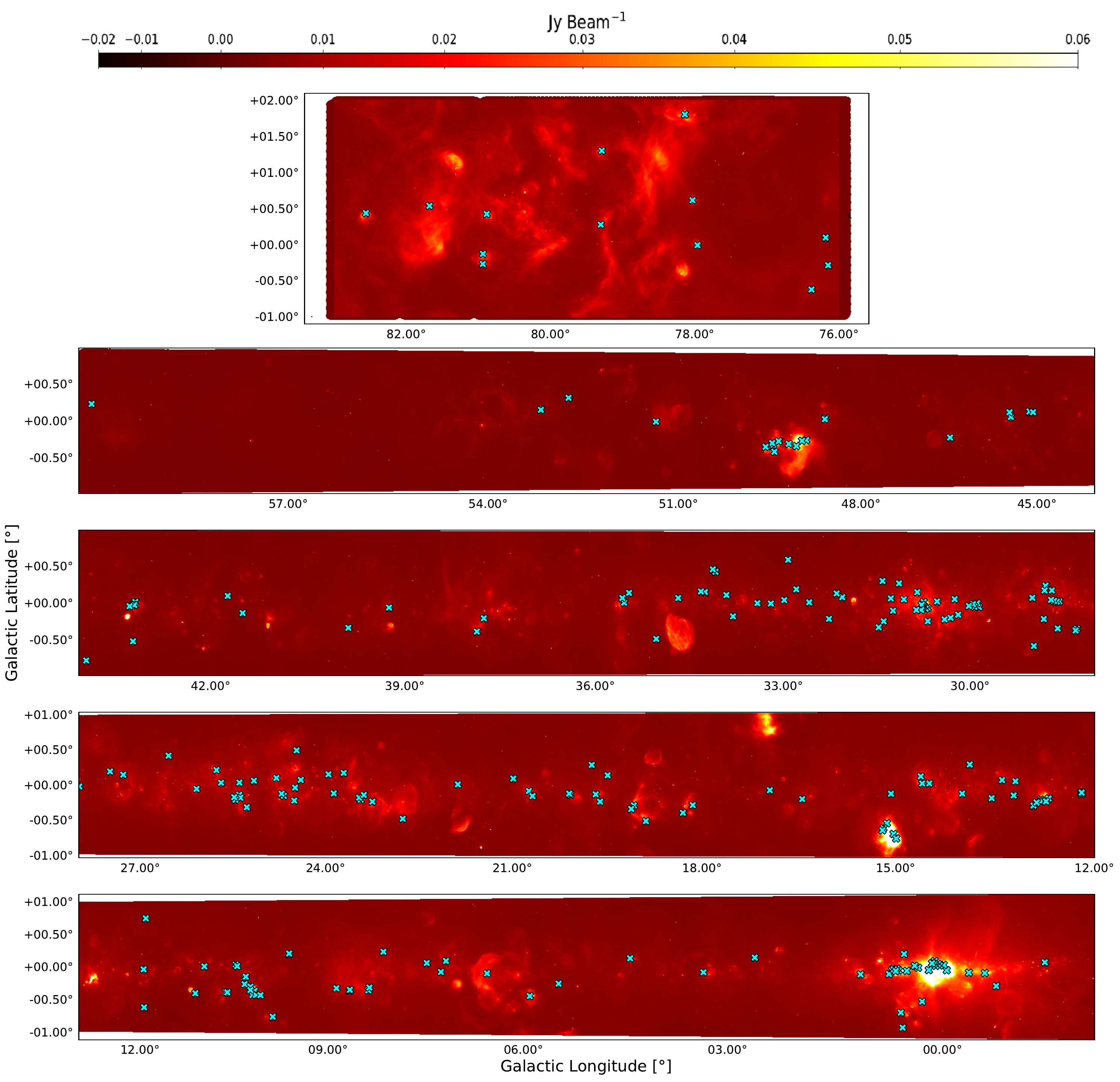}
        \caption{Detected $\langle$Hn$\alpha \rangle$ RRLs plotted as cyan crosses on top of a combination of the VLA D configuration and the Effelsberg~100m single-dish continuum image for the entire GLOSTAR survey region~\citep[$-$2$\degr \leq$ $\ell$ $\leq$ 60$\degr$, |\textit{b}| $\leq$ 1$\degr$ and Cygnus X region; ][]{2019A&A...627A.175M, 2021A&A...651A..85B}. The flux has been limited to between -0.02 and 0.06 Jy beam$\rm ^{-1}$ for visibility. }
        \label{fig:cont_image}
    \end{figure*}

\begin{figure}[h!]
        \centering
        \includegraphics[scale=0.7]{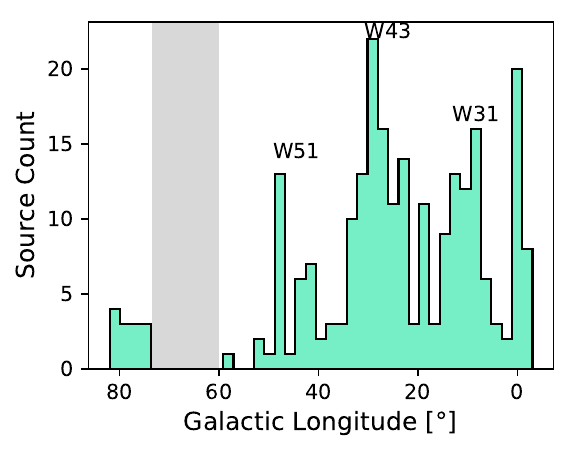}
        \includegraphics[scale=0.7]{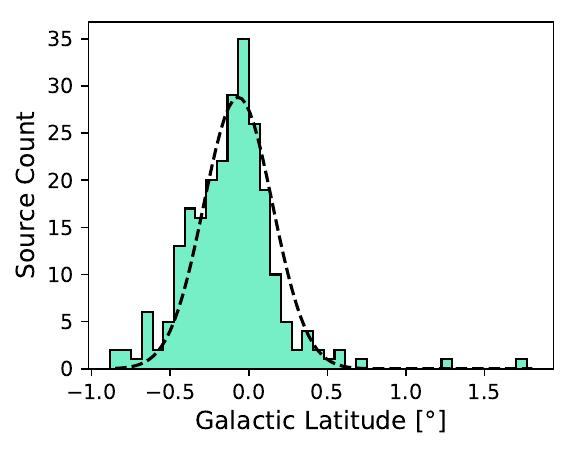}
        \caption{\textit{Top panel}: Histogram showing the distribution of the GLOSTAR \ion{H}{ii} region along the Galactic longitude. The bin size used is 2.0$\degr$ from the $\ell$ = $-$2.0$\degr$ to 83$\degr$. \textit{Bottom panel}: Histogram showing the distribution along the Galactic latitude with a bin width of 0.07$\degr$. The mean of the distribution is \textit{b} = $-$0.07 $\pm$ 0.01$\degr$. The region shaded in gray represents a longitude range of 60$\degr$ to 76$\degr$, which was not included in the GLOSTAR survey.} 
        \label{fig:source_dis}
    \end{figure}

        We identified a total of 244 sources with RRL emission above 3$\sigma$ (where $\sigma$ is local rms noise ranging from 1.6 to 6.0~mJy~beam$^{-1}$, with the median of 3.0~mJy~beam$^{-1}$). Among these \ion{H}{ii} regions, 13 sources are located in the Cygnus X region, and 29 are found toward the Galactic center region within 358$\degr \leq$ $\ell$ $\leq$ 2$\degr$, |\textit{b}| $\leq$ 1$\degr$. Within this subset of 29 \ion{H}{ii} regions toward the Galactic center, five sources are in the close vicinity of the thermally ionized Arched filament~\citep{2001AJ....121.2681L}. In Fig.~\ref{fig:cont_image}, we show the spatial distribution of the detected \ion{H}{ii} regions superimposed on the D-configuration continuum image. Fig.~\ref{fig:source_dis} displays the Galactic longitude and latitude distribution of the detected sources.  Notably, we see peaks in the source distribution at Galactic longitudes corresponding to sites of massive star formation, such as W31, W43, and W51. Along the Galactic latitude, we observe an asymmetric distribution whereby the majority of sources are situated at negative latitudes, as is reported by \citet{2014ApJS..212....1A}. Negative latitudes account for over 61\% of all sources, while positive latitudes make up the remaining 39\%. Furthermore, a few sources at higher Galactic latitudes ($\ell$>1.0$\degr$) are located in the Cygnus X star-forming region. The mean of this distribution is \textit{b} = $-$0.07 $\pm$0.01 $\degr$. The cause of this displacement is commonly attributed to the position of the Sun above the actual Galactic mid-plane \citep{2009A&A...504..415S, 2014ApJS..212....1A, 2019ApJ...871..145A}.
        
\begin{figure*}[h!]
        \centering
        \includegraphics[scale=0.5]{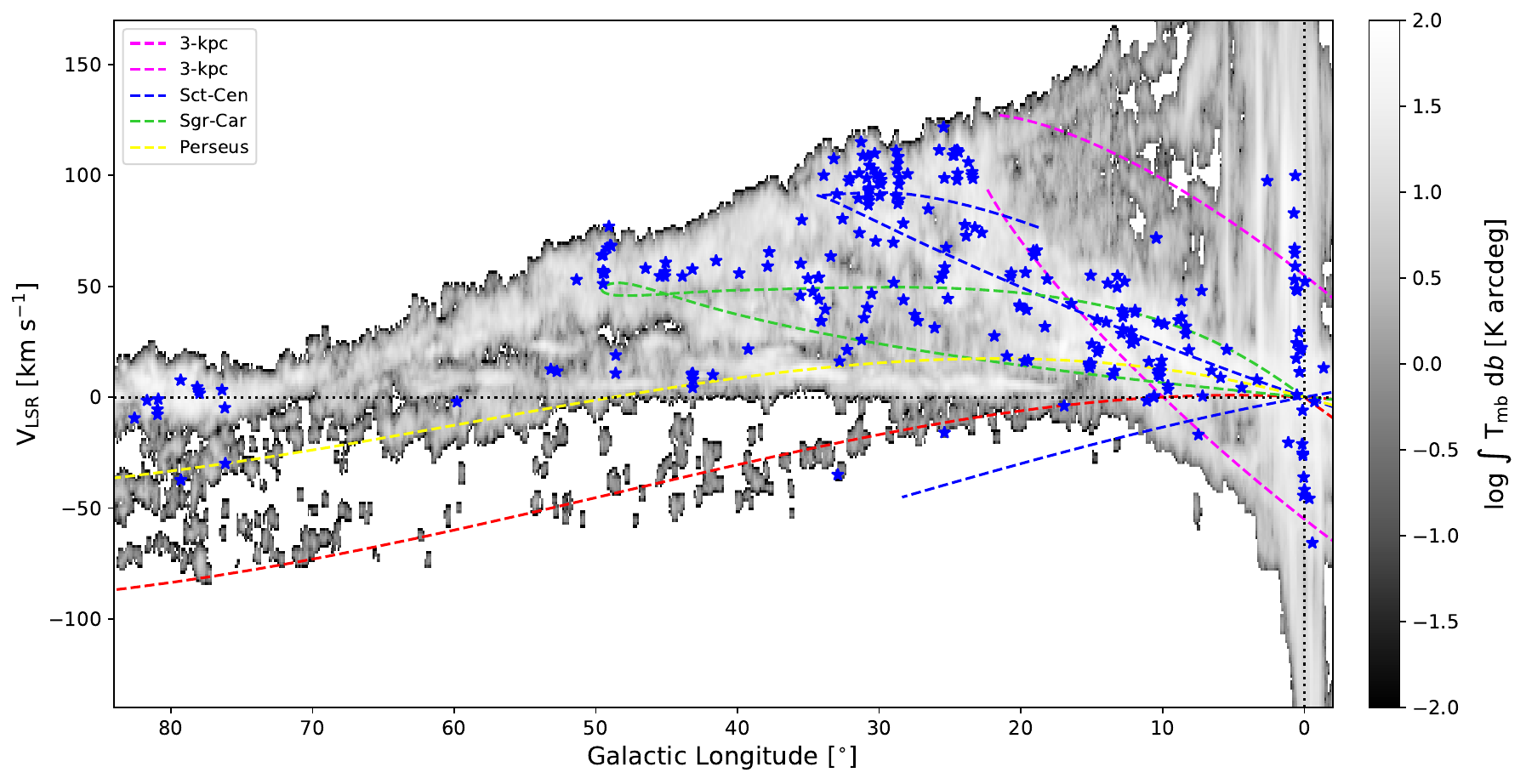}
        \caption{Distribution of detected \ion{H}{ii} regions RRL peak velocity with respect to the Galactic longitude in blue stars. The dashed lines represent the updated spiral arm models of \citet{1993ApJ...411..674T}. The background shows the CO emission from \citet{2001ApJ...547..792D}.}
        \label{fig:pv_image}
    \end{figure*}

        Table~\ref{tab:catalog} presents the continuum flux and $\langle$Hn$\alpha \rangle$ RRL parameters, accompanied by the distances of \ion{H}{ii} regions, along with associations with the dust continuum and 6.7~GHz methanol masers (see Sects. \ref{sec:dust_asso} and \ref{sec:maser_asso}). The Galactic coordinates of the peak continuum emission are listed in Columns 3 and 4, while Columns 5 and 6 display the peak ($\rm S_{peak}$) and integrated continuum flux density ($\rm S_{int}$), respectively. For the RRL parameters, Columns 8, 10, and 12 present the amplitude ($\rm S_{L}$), full-width half maximum (FWHM) line-width ($\Delta V$), and peak LSR velocity ($\rm V_{LSR}$), respectively. These values were derived by fitting a Gaussian to the $\langle$Hn$\alpha \rangle$ RRL. Additionally, Columns 13 and 15 indicate the heliocentric ($\rm D_{sun}$) and Galactocentric distances ($\rm R_{Gal}$), respectively (see Sect.~\ref{sec:distance}). In Fig.~\ref{fig:pv_image}, we show the distribution of detected RRL velocity with the Galactic longitude. As is seen in this ﬁgure, most sources align with the spiral arm models of \citet{1993ApJ...411..674T} and \citet{2019ApJ...885..131R}. Almost no \ion{H}{ii} regions are detected in the more distant outer arms, which is likely due to the limited sensitivity.  


    In addition to \ion{H}{ii} regions, planetary nebulae (PNe) also emit thermal radiation, making them the primary source of contamination in the \ion{H}{ii} region sample. Planetary nebulae are the expanding circumstellar envelopes of the asymptotic giant branch (AGB) stars that are ionized by their hot central white dwarf remnant. Consequently, RRLs from PNe generally exhibit larger line widths ($\sim$ 30 - 50~km~s$^{-1}$) compared to those observed in \ion{H}{ii} regions. Physically, PNe are smaller than \ion{H}{ii} regions, rendering them unresolved and devoid of nebulosity in MIR emission. To ensure the accuracy of our \ion{H}{ii} region sample, we conducted a matching analysis with the \texttt{SIMBAD}\footnote{\url{http://simbad.cds.unistra.fr/simbad/}} database to exclude any previously known PNe within a radius of 1$\arcmin$.  Recently, \citet{2023A&A...670A...9D} and \citet{2023A&A...680A..92Y} identified numerous new PNe candidates in the GLOSTAR regions using high-resolution VLA B-configuration continuum data. However, upon conducting a crossmatch, we did not identify any PNe located within a 25$\arcsec$ radius. Additionally, we employed the characteristic MIR emission to identify any potential contamination from PNe. Our investigation did not reveal any indications of PNe contamination in our \ion{H}{ii} region sample.
    
    \subsubsection{Comparison with other surveys}
    The GLOSTAR survey offers one of the first interferometry blind surveys of RRLs. However, challenges arise in directly comparing radio properties with previous RRL studies due to differences in \textit{uv} coverage, spatial filtering, and angular resolution. Here, our primary objective is to identify common sources with existing \ion{H}{ii} region surveys. The Green Bank Telescope Galactic \ion{H}{ii} region discovery survey (HRDS;~\citealt{2010ApJ...718L.106B, 2011ApJS..194...32A}) is a hydrogen RRL emission survey toward previously unknown Galactic \ion{H}{ii} regions within $-$16$\degr \leq$ $\ell$ $\leq$ 67$\degr$ and |\textit{b}|  $\leq$ 1$\degr$. In its course, 603 discrete hydrogen RRL components were detected at 9~GHz from 448 targets. The HRDS survey has an angular resolution of 82$\arcsec$, which is much larger than our survey resolution of 25$\arcsec$. Crossmatching by eye, we found only 34 common sources. \cite{2018A&A...615A.103K} present a catalog of 239 UC \ion{H}{ii} regions found in the Co-Ordinated Radio `N' Infrared Survey for High-mass star formation survey~\citep[CORNISH; ][]{2012PASP..124..939H} at 5~GHz and 1.5$\arcsec$ angular resolution in the Galactic region of 10$\degr \leq$ $\ell$ $\leq$ 65$\degr$ and |\textit{b}|  $\leq$ 1$\degr$. We crossmatched with CORNISH UC \ion{H}{ii} regions using a separation threshold of 12$\arcsec$, which is half the GLOSTAR beam size. We find 68 GLOSTAR \ion{H}{ii} regions associated with 84 CORNISH UC \ion{H}{ii} regions within 12$\arcsec$ ($\sim$28\%), with an average separation of 4$\arcsec$  between the GLOSTAR and CORNISH peaks. Using data from WISE, \cite{2014ApJS..212....1A} compiled a catalog of over 8000 ``known,'' ``candidate,'' ``group,'' and ``radio-quiet'' Galactic \ion{H}{ii} regions. The WISE \ion{H}{ii} region catalog also includes the HRDS and many other \ion{H}{ii} region catalogs (e.g., follow-up observations of the HRDS and southern sky survey, SHRDS;~\citealp{2021ApJS..254...36W}). Within the region covered by GLOSTAR, the most recent version of the WISE catalog reported $\sim$1100 known, $\sim$650 candidates, $\sim$325 groups, and $\sim$1540 radio-quiet. Out of our detected 244 \ion{H}{ii} regions, 209 are known, 11 are candidates, 20 are groups, and two are radio-quiet \ion{H}{ii} regions in the WISE \ion{H}{ii} region catalog. For the sources spatially associated with multiple sources in the WISE \ion{H}{ii} region catalog, known \ion{H}{ii} regions were designated preferentially. For the sources that have more than one source within the radius of a cataloged WISE \ion{H}{II} region, we assigned them to the same group, whereas for two sources (G028.569+0.020 and G034.322+0.160) we did not find any association with the WISE catalog.  Since the RRLs fluxes are generally about 10\% of the continuum fluxes and are associated with relatively bright \ion{H}{ii} regions, we expect most of the detected sources to be previously known.  

    \subsection{Distances}\label{sec:distance}
    In order to derive the physical properties of the detected \ion{H}{ii} regions, their distances must be estimated first. This has been accomplished through a multi-step process, involving the adoption of reliable distances from the literature if available, determination of the kinematic distances to all remaining sources with available RRL velocities, and resolution of any resulting distance ambiguities using the archival \ion{H}{I} data. These steps are elaborately described in Appendix~\ref{sec:app_dist}.
    
    Deriving kinematic radial distances toward the Galactic center and Cygnus X regions sparks debate due to the degeneracy of Galactic velocities in the direction of Cygnus X (Galactic longitude close to 90$\degr$) and the Galactic center. We have excluded \ion{H}{II} regions within a few degrees of the Galactic center (i.e., 358$\degr \leq$ $\ell$ $\leq$ 3$\degr$) because kinematic distances for sources in this region are highly unreliable. In this study, we include the Galactic center \ion{H}{II} regions in the source catalog (presented in Table~\ref{tab:catalog}) and derive their electron temperatures. We refrain from deriving the remaining physical properties of the Galactic center \ion{H}{II} regions due to uncertainty in the kinematic distances toward the Galactic center.
    
    Determining the distances to individual clouds and \ion{H}{II} regions in Cygnus X has been a persistent challenge~\citep{2006A&A...458..855S}. Previous parallax measurements and line-of-sight extinctions suggest that the bulk of molecular gas should be located at 1.3--1.5~kpc in this complex \citep{2012A&A...539A..79R, 2013ApJ...763..139D, 2013ApJ...769...15X, 2020A&A...633A..51Z, 2020MNRAS.493..351C, 2022A&A...658A.166D}. Given that the molecular clouds in Cygnus X are associated with each other~\citep{2006A&A...458..855S, 2023A&A...678A.130G}, we adopted a distance of 1.4 kpc for the Cygnus X \ion{H}{II} regions.

\begin{figure}[h!]
        \centering
        \includegraphics[width=0.5\textwidth]{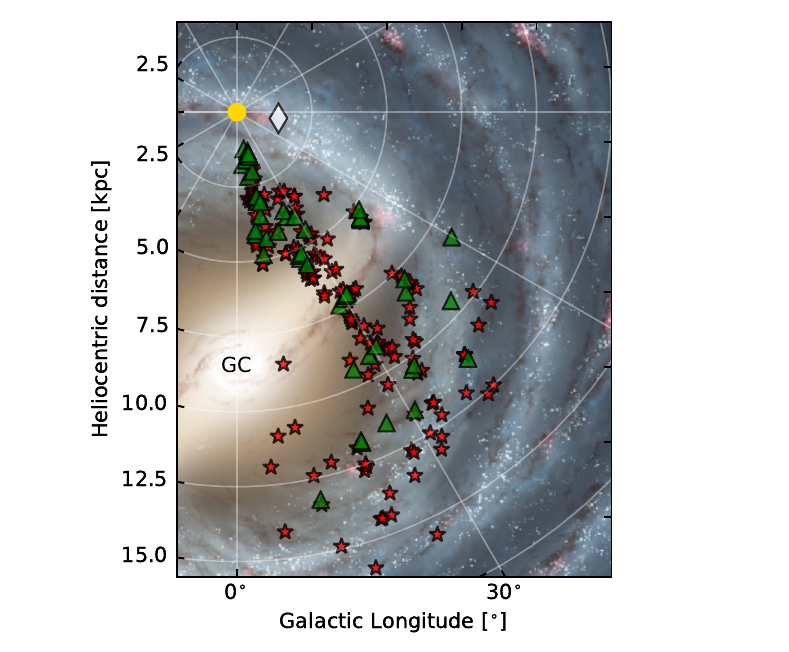}
        \caption{Distances of \ion{H}{ii} region from GLOSTAR D-configuration VLA observations plotted on top of an artist’s rendition of the Milky Way. The red stars represent the positions of \ion{H}{ii} regions with distances adopted from the literature, while the green triangles represent \ion{H}{ii} regions for which we derived the kinematic distances (see text for details). We adopt a distance of 1.4~kpc for Cygnus X \ion{H}{II} regions represented by white diamond.}
        \label{fig:distance}
    \end{figure}

    Using these steps, we obtained the distances from the literature for 149 sources between 3$\degr \leq$ $\ell$ $\leq$ 60$\degr$. For the rest of the 53 sources, we solved the kinematic distance ambiguity using the \ion{H}{i} emission/absorption (\ion{H}{i} E/A) method described by \citet{2003ApJ...582..756K}, \citet{2009ApJ...690..706A}, \citet{2012ApJ...754...62A}, and \citet{2012MNRAS.420.1656U}, resulting in $\sim$92~\% reliable and $\sim$64~\% highly reliable distance estimates (see Appendix~\ref{sec:app_dist} and Table~\ref{tab:dist} for detailed breakdown). In Fig.~\ref{fig:distance}, we show the locations of our \ion{H}{ii} region sample overlaid on an artistic impression of the Milky Way.\footnote{\url{https://photojournal.jpl.nasa.gov/catalog/PIA19341}}
    
    \subsection{Physical properties of the ionized regions}\label{sec:phy_prop}
    In this section, we present the physical properties of the cataloged \ion{H}{ii} regions. Table~\ref{tab:phy_prop} lists the various physical properties derived from measurements of their continuum and RRL emission. The first two columns list the effective angular ($\rm \theta_{source}$) and physical ($\rm d_{eff}$) diameters. Subsequent columns provide the derived electron temperature ($\rm T_e$), EM, electron density ($\rm n_e$), Lyman continuum photon rate ($\rm N_{lyc}$), ionized gas mass ($\rm Mass_{\ion{H}{ii}}$), and their corresponding uncertainties. The methodology used to derive these parameters are described in the following sections. Interferometers are insensitive to large scale diffuse emission, such as the nonthermal radio-continuum emission that permeates the Galactic plane. However, they are well suited for more precise measurements of the total continuum flux density of nebulae, especially when the source's angular size is smaller than the largest angular scale detectable by the telescope. Notably, interferometers like the VLA can simultaneously measure both radio-continuum and RRL emission. This allows precise determinations of the RRL-to-continuum flux ratio that avoid systematic calibration or weather-related issues, bolstering our confidence in the derived physical properties of the \ion{H}{ii} regions. In the following sections, the  detailed processes used to derive these physical properties are discussed. 
      
        \begin{figure*}[h!]
        \centering
        \includegraphics[scale=0.6]{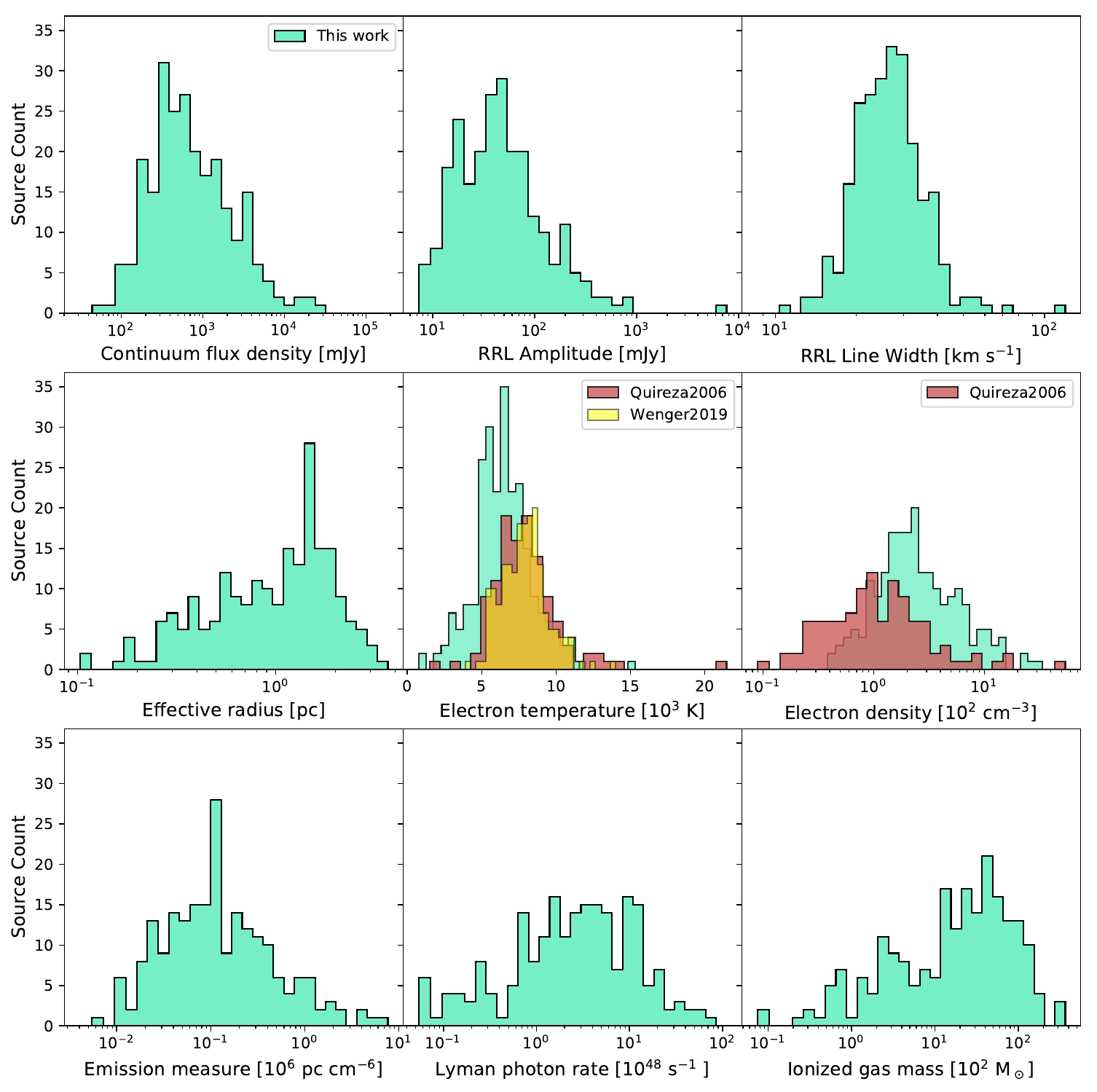}
        \caption{Distributions of various \ion{H}{ii} region physical properties from top left to bottom right, continuum flux density, RRL amplitude, RRL FWHM, effective radius, electron temperature, electron density, EM, Lyman photon rate, and ionized gas mass. The electron temperature reported by \citet{2006ApJ...653.1226Q} is indicated in red, while the one reported by \citet{2019ApJ...887..114W} is shown in yellow, and these are compared to the GLOSTAR \ion{H}{ii} region sample (aquamarine). On average, the electron temperature of GLOSTAR \ion{H}{ii} regions (6707~K) is lower than that of \citet{2006ApJ...653.1226Q} (8214~K) and \citet{2019ApJ...887..114W} (8055~K). Electron density histograms are presented for both the GLOSTAR (aquamarine) and \citet{2006ApJ...653.1226Q} (red) \ion{H}{ii} region samples. On average, the GLOSTAR nebulae have higher electron density than the \citet{2006ApJ...653.1226Q} sources.}
        \label{fig:phy_prop_dist}
    \end{figure*}
        
        \subsubsection{$\langle$Hn$\alpha \rangle$ amplitude and continuum flux density}
        To obtain a precise line-to-continuum ratio, it is necessary to extract both line and continuum data over the same area of a source. Additionally, the measured line-to-continuum ratio may be influenced by diffuse background emission around the sources. This especially applies to extended objects and \ion{H}{ii} regions in crowded star-forming regions in the Galactic plane. Given the sensitivity and \textit{uv} coverage of our RRL observations, we did not anticipate any detections toward the weak diffuse thermal emission and certainly not the nonthermal Galactic background. For these reasons, we first performed background subtraction and then manually determined the continuum flux and generated the $\langle$Hn$\alpha \rangle$ line profile over the same area of the sources. The mean background intensity ($\rm \overline{B}$) in the surrounding regions has been estimated using an iterative sigma-clipping algorithm~\citep{2021A&A...651A..86D, 2023A&A...671A.145D}. However, except for a few sources toward the Galactic center, we found that the relative intensity of the background is lower than 1\% of peak emission. To determine the continuum flux and generate the $\langle$Hn$\alpha \rangle$ line profile over a common area, we began by extracting the line within the range of 10\% to 90\% of the continuum peak emission. We skipped the peak emission to improve the signal-to-noise ratio by spatially averaging the emission and to calculate the angular area of the \ion{H}{ii} region. We selected the region with the highest signal-to-noise ratio for the line. This allowed us to generate the average $\langle$Hn$\alpha \rangle$ line spectra and determine the continuum flux density of the \ion{H}{II} regions. Fig.~\ref{fig:phy_prop_dist} (top middle panel) displays the distribution of the line amplitudes. Among the \ion{H}{ii} regions in our catalog, the brightest one is G015.035$-$0.677 (M17) with a $\langle$Hn$\alpha \rangle$ amplitude of 7.6~Jy, while the weakest is G028.687+0.176 with a $\langle$Hn$\alpha \rangle$ amplitude of 8 mJy. The mean and median values of the amplitude distribution are 0.11~Jy and 0.04~Jy, respectively. We utilized the line-free channels to determine the continuum flux density (S$_\text{C}$). This was done by integrating over the same area as that used to generate the line profile. Hence, the continuum flux density derived here will be lower than that obtained by integrating over the entire source. We estimated S$_\text{C}$ as
        \begin{equation}
          \rm  S_C = \left(\sum\limits_{i=1}^{N_{src}}A_i  - N_{src} \overline{B}\right)\times a_{bm}^{-1},
        \end{equation}
        where $\rm \sum\limits_{i=1}^{N_{src}}A_i$ is the total intensity summed over the $\rm N_{src}$ pixels, $\rm \overline{B}$ is the mean background intensity, and $\rm a_{bm} = \pi\theta_{bm}^2/(4ln(2)\Delta_{pixel}^2)$ is the number of pixels per beam, where $\rm \Delta_{pixel}^2$ is the area of each pixel. To accommodate the uncertainty in flux estimation, we incorporated a GLOSTAR flux density calibration error of 5\%. The top left panel of Fig.~\ref{fig:phy_prop_dist} illustrates the distribution of continuum flux density (S$\rm _C$). Within our catalog, the \ion{H}{ii} region G015.035$-$0.677 (M17) is the brightest, with a continuum flux density of 185.6~Jy, while G025.479$-$0.174 stands out as the weakest, with a continuum flux density of 0.037~Jy. The mean and median values of the S$_C$ distribution are 2.36~Jy and 0.62~Jy, respectively. The angular diameters of the \ion{H}{ii} regions are determined from the number of pixels within the same area of the source used to generate the line profile, $\rm \theta_{source} = 2\sqrt{\frac{N_\mathrm{src}\Delta_\mathrm{pixel}^2}{\pi}}$.

        \subsubsection{Line widths}\label{sec:line_width}
        Fig.~\ref{fig:phy_prop_dist} (top right panel) shows the distribution of the FWHM line widths of the detected $\langle$Hn$\alpha \rangle$ RRLs. The mean $\pm$ standard deviation of this distribution is 28.2 $\pm$ 10.2~km~s$^{-1}$, respectively. Notably, our sample exhibits a comparable, if slightly broader FWHM distribution compared to that of the HRDS~\citep{2011ApJS..194...32A}, \citet{1989ApJS...71..469L} and \citet{2020ApJS..248....3C}, which are 22.3 $\pm$ 5.3, 26.4 $\pm$ 8.1, and 23.6 $\pm$ 2.0~km~s$^{-1}$, respectively. These differences in FWHM may be attributed to the difference in spectral resolution used in the respective studies, which are 5.0, 1.86, 4.0, and 0.4-2.4~km~s$^{-1}$ for the GLOSTAR survey, the HRDS~\citep{2011ApJS..194...32A}, \citet{1989ApJS...71..469L}, and \citet{2020ApJS..248....3C}. 
        
        \begin{figure}[h!]
        \centering
        \includegraphics[scale=0.6]{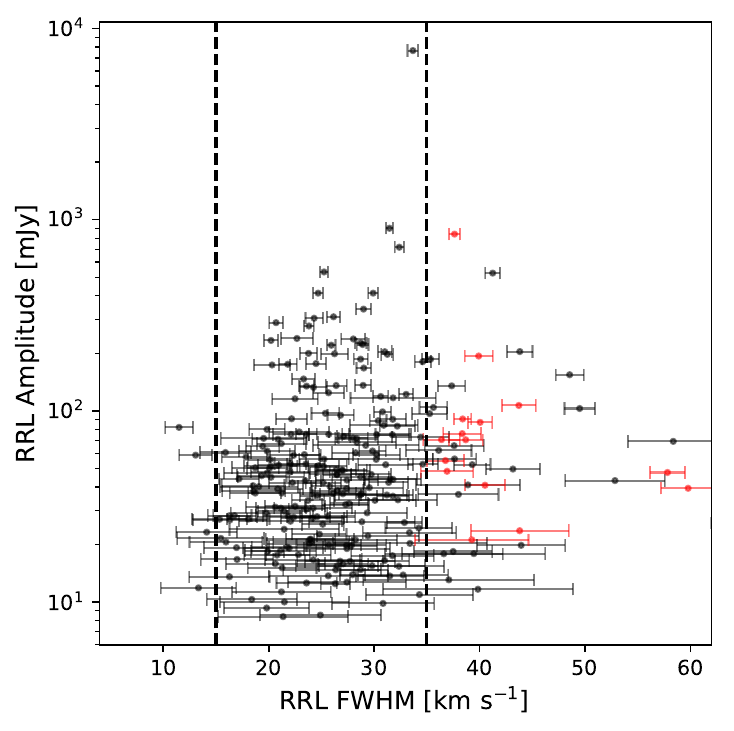}
        \caption{\ion{H}{ii} region's RRL intensity as a function of FWHM line width. Vertical dashed lines represent the line width of 35 and 15~km~s$^{-1}$. The distribution of the low intensity sources is similar to the entire source. The candidate HC \ion{H}{II} regions are represented by the filled red circle.}
        \label{fig:FWHM_amp}
    \end{figure}
    We detected 40 \ion{H}{ii} regions with RRL width $\rm >35$~km~s$^{-1}$. Of these sources, 13 sources are found in the Galactic center region ($-$2$\degr \leq$ $\ell$ $\leq$ 2$\degr$ and |\textit{b}|  $\leq$ 1$\degr$) and two toward the Cygnus X region (76$\degr \leq$ $\ell$ $\leq$ 83$\degr$, $-$1$\degr \leq$ \textit{b} $\leq$ 2$\degr$). \citet{2011ApJS..194...32A} propose that line widths greater than 35~km~s$^{-1}$ should be viewed with caution as they could be due to low signal-to-noise detection, the presence of blended velocity components, or an indication that the source is not an \ion{H}{ii} region.  As the spectral signal-to-noise ratio decreases, the process of Gaussian fitting for deriving the line width becomes increasingly uncertain. In situations with low signal-to-noise, it may become challenging to separate double-velocity components, potentially resulting in an erroneously large line width estimation. In Fig.~\ref{fig:FWHM_amp}, we plot the line width as a function of the line intensity. The plot demonstrates that the RRL amplitude distribution for the broad RRL \ion{H}{ii} regions resembles the overall distribution. Within our sample of 40 broad RRL \ion{H}{ii} regions, 31 sources have signal-to-noise ratios greater than 10$\sigma$. This suggests that the broad line widths we determine for RRL \ion{H}{ii} regions in our sample are not due to a low signal-to-noise ratio.
    
    The line widths of the RRLs can be influenced by a blend of thermal, turbulent, and organized motions, such as those arising from processes like accretion dynamics, infall of matter, and outflow. For typical \ion{H}{ii} regions with electron temperatures of approximately $\sim 10^4$~K, the thermally broadened line width is about 25~km~s$^{-1}$. However, the RRLs line widths of Galactic \ion{H}{ii} regions can exceed this, indicating that they are significantly influenced by turbulent and ordered motions. In addition, pressure broadening is also signiﬁcant for RRLs at centimeter wavelengths~\citep{2008ApJ...672..423K}. Young and compact \ion{H}{ii} regions can show broadened (>40~km~s$^{-1}$) RRLs \citep{2004ApJ...605..285S, 2008ApJ...672..423K} due to the presence of large-scale motion of ionized gas around a young central star. However, it is not yet confirmed that the broad RRL is an intrinsic property of HC \ion{H}{ii} regions. \citet{2019MNRAS.482.2681Y} compiled a catalog of 120 candidate HC \ion{H}{ii} regions located in the Galactic plane region within 10$\degr \leq$ $\ell$ $\leq$ 65$\degr$ and |\textit{b}|  $\leq$ 1$\degr$. In the region overlapping with the GLOSTAR survey coverage (10$\degr \leq$ $\ell$ $\leq$ 60$\degr$ and |\textit{b}|  $\leq$ 1$\degr$), we detected 23 \ion{H}{ii} regions with RRL widths exceeding 35~km~s$^{-1}$. To establish associations with the candidate HC\ion{H}{ii} regions from \citet{2019MNRAS.482.2681Y}, we performed a crossmatch using a distance threshold of 12$\arcsec$, half of the GLOSTAR beam size. As a result, we identified 15 (out of 23) broad RRL \ion{H}{ii} regions that are associated with the candidate HC \ion{H}{ii} regions, with a mean $\pm$ standard deviation separation of 3.3 $\pm$ 1.3$\arcsec$. Furthermore, the mean $\pm$ standard deviation of the spectral indices between 1.4-5 GHz for these 15 broad RRL \ion{H}{ii} regions as reported by \citet{2019MNRAS.482.2681Y} is 0.6 $\pm$ 0.3. Hence, these 15 broad RRL \ion{H}{ii} regions are very likely to be HC \ion{H}{ii} regions.

    Our catalog has only 4 (2\%) \ion{H}{ii} regions with extremely narrow line widths (<15~km~s$^{-1}$). Narrow RRL widths are interpreted to be purely thermal, originating from ``cool''  nebulae with electron temperatures $ \lesssim$~5000~K. \citet{2011ApJS..194...32A} showed that cold nebulae are very rare in our Galaxy; only 6\% of HRDS sources have line widths smaller than 15~km~s$^{-1}$. However, this fraction is three times higher than that found in our survey. This can be attributed to the poorer spectral resolution of our RRL data along with our selection criteria, which rejected sources emitting in only two channels, making the detection of narrower lines more challenging. 

    \begin{table*}[]
        \centering
        \caption{Candidate HC \ion{H}{ii} regions.}
        \label{tab:hyper_HII}
        \begin{tabular}{ccccccc}
        \hline \hline
        S.No. & GLOSTAR Name & CORNISH Name & Separation\tablefootmark{*} & $\Delta$V & $\alpha$\tablefootmark{**} & $\rm d_{eff}$\tablefootmark{***}\\
        & & & [$\arcsec$] &[$\rm km~s^{-1}$] & & [pc]\\
        \hline \hline
        1& G010.624$-$0.384& G010.6234$-$00.3837& 2.43 & 38.69$\pm$1.6 &0.97 &1.24\\
        2& G010.302$-$0.147& G010.3009$-$00.1477& 4.75& 38.36$\pm$1.79 &0.31&0.78\\
        3& G011.937$-$0.615& G011.9368$-$00.6158& 2.89& 40.52$\pm$1.89 &0.36&0.71\\
        4& G012.805$-$0.200& G012.8050$-$00.2007& 2.41& 37.63$\pm$0.5 &0.84&0.88\\
        5& G013.210$-$0.144& G013.2099$-$00.1428& 4.3&  36.39$\pm$1.72 &0.61&1.15\\
        6& G016.944$-$0.074& G016.9445$-$00.0738& 1.92& 43.81$\pm$4.64 &0.55&3.65\\
        7& G026.544+0.415& G026.5444+00.4169& 6.96& 36.92$\pm$2.5 &0.25&3.74\\
        8& G028.287$-$0.364& G028.2879$-$00.3641& 3.22& 39.27$\pm$5.38 &0.23&0.7\\
        9& G029.957$-$0.017& G029.9559$-$00.0168& 4.22& 39.92$\pm$1.33 &0.52&1.31\\
        10& G030.535+0.021& G030.5353+00.0204& 2.42& 36.77$\pm$1.76 &0.2&3.05\\
        11& G034.257+0.153& G034.2572+00.1535& 2.06& 59.79$\pm$2.6 &1.22&0.77\\
        12& G037.874$-$0.399& G037.8731$-$00.3996& 3.97& 43.73$\pm$1.59 &0.55&2.59\\
        13& G043.165$-$0.029& G043.1651$-$00.0283& 2.62& 40.06$\pm$1.16 &1.23&1.93\\
        14& G045.122+0.131& G045.1223+00.1321& 4.0 & 57.83$\pm$1.64 &0.63&1.47\\
        15& G049.490$-$0.369& G049.4905$-$00.3688& 2.03& 38.4$\pm$0.83 &0.93&0.61 \\
        \hline \hline
        \end{tabular}
        \tablefoot{\tablefoottext{*}{Angular separation between GLOSTAR and CORNISH peak emission.} 
        \tablefoottext{**}{Spectral index between 1.4$-$5~GHz as reported by \citet{2019MNRAS.482.2681Y}}
        \tablefoottext{***}{$\rm d_{eff}$ represents the effective diameter of the source (refer to Sect.~\ref{sec:line_width}), and these values might exceed the true size of the sources.}}
    \end{table*}

        \subsubsection{Local thermal equilibrium electron temperatures}\label{sec:elec_temp}
        The detection of RRLs allows us to calculate the electron temperature of the ionized gas. The electron temperature of a nebula is obtained from the RRL-to-continuum ratio, $\rm S_L/S_C$. Since we wanted to derive the average electron temperature, we assumed all sources to be homogeneous, isothermal, and their continuum emission to be optically thin at the GLOSTAR frequency (5.3~GHz). However, this last assumption is not true for candidate HC \ion{H}{ii} regions (see Table~\ref{tab:hyper_HII}), which have a positive spectral index (up to 2) between 1.4 and 5~GHz. We lack the information of the turn-over frequency for these regions, when the emission becomes optically thin. Therefore, accurately estimating the actual optical depth at 5.3~GHz becomes challenging. Consequently, the true electron temperature is likely to be lower than what we derive from our assumption of optically thin continuum emission. In Table~\ref{tab:phy_prop}, we indicate these sources with (*) for the reader's reference. When the radio continuum emission is optically thin, one can estimate the electron temperature using the following equation (see, e.g., \citealt{2019ApJ...887..114W}):

        \begin{equation}
     \label{eq:op_thin_intial}
     \begin{split}
     \rm \left(\frac{T_e}{K}\right)=\Biggl\{3.661\times 10^4 \frac{\Delta n}{n}f_{nm}&\rm \left(\frac{\nu}{GHz}\right)^{1.1}
      \left(\frac{S_C}{S_L}\right)\left(\frac{\Delta v}{km s^{-1}}\right)^{-1}\\ &\rm \left[1+\frac{n(^4He^{+})}{n(H^{+})}\right]^{-1}\Biggr\}^{0.87},
     \end{split}
    \end{equation}    
    where $\rm \Delta v$ is the FWHM line width and $\rm n(^4He^{+})/n(H^{+})$ is the ionic abundance ratio and the expression $\rm f_{nm}$ is the absorption oscillator strength, for which an approximation is given by~\citep{1968Natur.218..756M}:
    \begin{equation}
        \label{eq:oscillator}
        \rm f_{nm} = nM_{\Delta n}\left(1+1.5\frac{\Delta n}{n} \right) 
    \end{equation}          
    The expression $\rm \frac{\Delta n}{n}f_{nm}$ is not a strong function of n for $\Delta {\rm n} = 1$ hydrogen RRLs. 
    For example, ($\rm \frac{\Delta n}{n}f_{nm}$) = 0.1937 and 0.1933 for $\rm \Delta n = 1$ and n = 98 and 114, respectively, obtained using the oscillator strength from~\cite{1968Natur.218..756M}. Since this variation amounts to only about 0.2\% across these H$n\alpha$ transitions, we adopted ($\rm \frac{\Delta n}{n}f_{nm}$)= 0.19345 for $\rm \Delta n = 1$ and n = 107, which is the average of observed principal quantum numbers. After substitution, the final expression for the electron temperature is given by                
    \begin{equation*}
        \label{eq:op_thin}
        \rm \left(\frac{T_e}{K}\right)=\left\{7082.2\left(\frac{\nu}{GHz}\right)^{1.1}\left(\frac{S_C}{S_L}\right)\left(\frac{\Delta v}{km s^{-1}}\right)^{-1}\left[1+\frac{n(^4He^{+})}{n(H^{+})}\right]^{-1}\right\}^{0.87}.
    \end{equation*}
    We detected helium RRLs only in a few bright sources, so we assume the ionic abundance ratio to be a constant value of 0.07 $\pm$ 0.02~\citep{2006ApJ...653.1226Q}.

    The errors in the electron temperature ($\sigma_{T_e}$) were computed by propagating the errors from Gaussian fitting for the line and continuum estimation. To calculate the uncertainty in the derived physical parameters, we utilized the \texttt{SOERP}\footnote{\url{https://github.com/tisimst/soerp}} Python package. SOERP is a Python version of the original Fortran code SOERP, which applies a second-order analysis for error propagation or uncertainty analysis. The electron temperature uncertainty ($\sigma_{T_e}$) ranges from 5\% to 45\%, with a mean of 13\% and a standard deviation of 8\%. For six sources, G000.121+0.043, G020.098$-$0.122, G025.459$-$0.210, G028.687+0.176, G028.789+0.243, and G030.789$-$0.100, $\sigma_{T_e}$ exceeds 30\%. For G020.098$-$0.122, G025.459$-$0.210, G028.687+0.176, G028.789+0.243, and G030.789$-$0.100, the high uncertainty is attributed to the low line intensity, resulting in increased uncertainty in the obtained amplitude and line-width. The high uncertainty for G0.121+0.043 is a consequence of a weak line and crowded continuum emission around this Galactic center region sources. Fig.~\ref{fig:phy_prop_dist} (middle center panel) displays the distribution of the derived electron temperature, which has a mean of 6707~K and a standard deviation of 1974~K. G025.479$-$0.174 and G359.945$-$0.014 exhibit the minimum (1010$\pm$229~K) and maximum (15562$\pm$1119~K) electron temperatures, respectively. The elevated electron temperature in G359.945$-$0.014 can be attributed to contamination by nonthermal continuum emission toward the Galactic center.

    In the Galactic center region, the continuum emission is comprised of a significant amount of nonthermal emission in addition to thermal free-free emission~\citep{2008ApJS..177..515L,1997ApJ...474..275L,2001AJ....121.2681L}. \citet{1996A&ARv...7..289M} determined that thermal or nonthermal emission contributed almost evenly within an area of approximately $\sim$ 400~pc $\times$ 350~pc at 5~GHz toward the Galactic center. Among this region, the free-free flux density was measured to be 580~Jy, with 40\% originating from radio \ion{H}{ii} regions and the remainder from extended low-density \ion{H}{ii} regions. Consequently, we expect depressed values of the line-to-continuum ratio because there is no RRL emission from nonthermal sources. This could lead to a potential overestimation of the reported electron temperature for \ion{H}{II} regions in the Galactic center region. A comprehensive analysis of the proportion between thermal and nonthermal emissions is crucial to accurately determine the electron temperature of \ion{H}{ii} regions toward the Galactic center.

    Accurate calculation of electron temperature requires the incorporation of non-LTE (local thermal equilibrium) effects such as stimulated emission, and pressure broadening from electron impact. These non-LTE effects depend on the local density and \ion{H}{ii} region geometry. However, previous studies~\citep{1980A&A....90...34S,2006ApJ...653.1226Q} showed that under many conditions LTE is a good approximation and that the non-LTE electron temperature, T$^*_e$, is similar to the LTE electron temperature, T$_e$. \citet{1980A&A....90...34S} defines an RRL observing frequency, $\rm \nu_{LTE}$, for which T$^*_e$ $=$ T$_e$. This optimal frequency is a function of EM: $\rm \nu_{LTE}$ $\sim$ 0.081 EM$\rm ^{0.36}$. This is independent of the geometry, density, and temperature structure of the \ion{H}{ii} region. Using our highly reliable values of the EM, we obtain an average frequency of $\rm \nu_{LTE}$ = 8.8~GHz. This value is quite close to our observation frequency of 5.3~GHz. Additionally, \citet{2023MNRAS.526..423W} demonstrate that the assumption of LTE to calculate $\rm T_{e}$ is not valid for the transitions they observed in L band (1--2 GHz), and hence, non-LTE corrections are needed. However, the necessity for non-LTE corrections is insignificant for X and C bands (including the range covered by the GLOSTAR survey). We therefore conclude that our estimated LTE electron temperatures are close to the real electron temperatures of the \ion{H}{ii} regions we detect.     
    
    Despite the significant difference in beam size, we attempted to identify common sources by performing a crossmatch within a threshold distance of 25$\arcsec$ (GLOSTAR beam size) with \citet{2006ApJ...653.1226Q} and \citet{2019ApJ...887..114W}. The average percentage difference between our electron temperature and those of \citet{2006ApJ...653.1226Q} and \citet{2019ApJ...887..114W}, for the common sources are 1.0\% and 4\%, respectively. In Fig.~\ref{fig:phy_prop_dist} (middle center panel), we compare the distributions of electron temperature. A systematic variation in the T$\rm _e$ values is evident. There are several possible explanations for the origin of this difference. As pointed out by \citet{2006ApJ...653.1226Q}, RRL studies were conducted at various frequencies, utilizing different telescopes that have distinct beam sizes. Consequently, each survey investigates different regions within the nebulae. Furthermore, certain \ion{H}{ii} regions exhibit intricate internal structure characterized by significant density and temperature fluctuations. When deriving electron temperatures from different RRL transitions observed with diverse telescopes, differences are likely to arise, particularly when assuming LTE. Calibration discrepancies between line and continuum measurements, as well as variations between telescopes, may also introduce additional challenges. For example, the FWHM of 3$\farcm$20 at 8.6~GHz in \citet{2006ApJ...653.1226Q} was much larger than ours leading to them observing more extended gas compared to our observations, which is reflected in the distribution of electron density in Fig.~\ref{fig:phy_prop_dist} (middle right panel). All things being equal, we expect higher electron density to produce higher electron temperature because the electron density affects the rate of collisional de-excitation. Hence, a high value of n$\rm _e$ inhibits cooling, and thus increases the $\rm T_e$ \citep{1985ApJS...57..349R}. However, in Fig.~\ref{fig:phy_prop_dist} (middle center and right panel), we observe the opposite: the GLOSTAR \ion{H}{ii} region distribution shows higher n$\rm _e$ but lower $\rm T_e$ compared to \citet{2006ApJ...653.1226Q}. \citet{2006ApJ...653.1226Q} assumed \ion{H}{ii} regions to be homogeneous and spherical, deriving the rms electron density using the peak continuum brightness temperature. The rms densities are probably lower than the true densities of the \ion{H}{ii} regions since generally the gas is not homogeneously distributed, showing a hierarchical structure, as was discussed in \citet{2019MNRAS.482.2681Y}. Additionally, the metallicity, which is a function of $\rm R_{Gal}$, primarily regulates the electron temperature. Therefore, a detailed study is required to understand these differences.

    \subsubsection{Emission measure, electron density, Lyman photon rate, and ionized mass}
    Knowing the angular size and distance of the \ion{H}{ii} regions we can estimate further physical properties of the \ion{H}{ii} regions. Considering \ion{H}{ii} regions as fully ionized Str\"{o}mgren spheres without dust and assuming optically thin emission, the relation between EM and flux density is given by \citet{2016A&A...588A.143S}:
    \begin{equation}
        \rm \left(\frac{S_\nu}{Jy}\right) = 2.525 \times 10^{-2} \left(\frac{T_e}{K}\right)^{-0.35}\left(\frac{\nu}{GHz}\right)^{-0.1}\left(\frac{EM}{pc~cm^{-6}}\right)\Omega_{source}
    \end{equation}
    where $\Omega_{\mathrm{source}}$ is the solid angle of the source. The  EM for a circular aperture is given by \citet{2016A&A...588A.143S}:
    \begin{equation}
       \rm \left(\frac{EM}{pc~cm^{-6}}\right) = 3.217\times 10^7\left(\frac{S_\nu}{Jy}\right)\left(\frac{T_e}{K}\right)^{0.35}\left(\frac{\nu}{GHz}\right)^{0.1}\left(\frac{\theta_{source}}{arcsec}\right)^{-2}
    \end{equation}
     From this, we can get this expression for the electron density ($\rm n_e$),

    \begin{equation}
      \begin{split}    
        \rm \left(\frac{n_{e}}{cm^{-3}}\right) = 2.576\times 10^6\left(\frac{S_\nu}{Jy}\right)^{0.5}
        & \rm \left(\frac{T_e}{K}\right)^{0.175} \left(\frac{\nu}{GHz}\right)^{0.05} \\ &\rm \left(\frac{\theta_{source}}{arcsec}\right)^{-1.5}\left(\frac{D}{pc}\right)^{-0.5}   
      \end{split}
    \end{equation}    
    where F$_\nu$ is the flux density at frequency $\nu$, $D$ is the distance to the source, $\rm T_e$ is the electron temperature, and $\rm \theta_{\text{source}}$ is the effective angular diameter of the ionized region. Similarly, the expression for the Lyman photon rate of an \ion{H}{ii} region is taken from \citet{2016A&A...588A.143S},
    \begin{equation}
        \label{eq:n_ly}
        \rm \left(\frac{N_{Lyc}}{s^{-1}}\right) = 4.771\times 10^{42}\left(\frac{S_\nu}{Jy}\right)\left(\frac{T_e}{K}\right)^{-0.45}\left(\frac{\nu}{GHz}\right)^{0.1}\left(\frac{D}{pc}\right)^{2}
    \end{equation}
    Finally, following \citet{Tielens},  we can estimate the ionized mass of the \ion{H}{ii} regions using the following equation, 
    \begin{equation}
        \label{eq:mass}
        \rm \frac{M_{ \ion{H}{ii}}}{M_\sun}= 80 \times \left(\frac{10^3~cm^{-3}}{n_e}\right)\left(\frac{N_{Lyc}}{5\times 10^{49}~s^{-1}}\right)
    \end{equation}
    
    Fig.~\ref{fig:phy_prop_dist}, shows the distribution of the EM, electron density ($\rm n_e$), Lyman photon rate ($\rm N_{Lyc}$) and ionized gas mass ($\rm M_{\ion{H}{ii}}$). The median of EM, $\rm n_e$, $\rm N_{Lyc}$, and $\rm M_{\ion{H}{ii}}$ of our sample are $\rm 1.1~\times~10^5~pc~cm^{-6}$, $\rm 2.5~\times~10^2~cm^{-3}$, $\rm 2.3~\times~10^{48}~s^{-1}$ and $\rm 14.5~M_\sun$, respectively. The distribution of N$_{\rm Lyc}$ ranges from $\rm 3.8~\times~10^{46}~s^{-1}$ to $\rm 1.1~\times~10^{50}~s^{-1}$, corresponding to spectral types of zero-age main sequence (ZAMS) stars spanning from B0 to O4, assuming that all the ionizing radiation comes from a single star \citep{2005A&A...436.1049M}. Additionally, the ranges of the EM and $\rm n_e$ distributions are $\rm 4.7~\times~10^3~to~7.6~\times~10^6~pc~cm^{-6}$ and $\rm 0.47~\times~10^2~to~32.0~\times~10^2~cm^{-3}$, respectively. 
    
\begin{figure}
            \centering
            \includegraphics[scale=0.6]{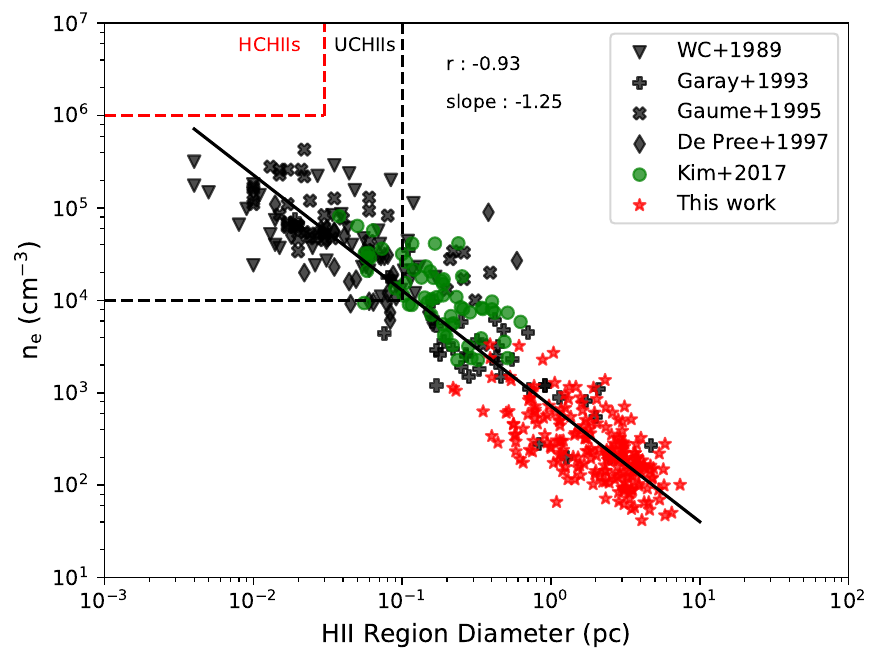}
            \caption{Electron density (n$\rm _e$) as a function of \ion{H}{ii} region diameter (continuation of Fig. 16 of \citealt{2017A&A...602A..37K}). Black symbols indicate values determined in previous radio continuum surveys and green circles shows the distribution found for a sample of mm-RRLs by \citet{2017A&A...602A..37K}, for which the electron densities are derived from mm-RRL emission. The data from the current work are represented by red stars. The solid black lines represent the best-fit line determined by least-square fit to all data with a Spearman’s rank coefficient of r = $-$0.92 and p-value $\ll$ 0.0013. The dashed red line represents HC \ion{H}{ii} regions parameter space with diameter $\leq$ 0.03~pc and n$\rm _e$ $\geq$ 10$\rm ^5$ cm$\rm ^{-3}$ and the dashed black line represents UC \ion{H}{ii} regions parameter space with diameter $\leq$ 0.1~pc and n$\rm _e$ $\geq$ 10$\rm ^4$ cm$\rm ^{-3}$~\citep{2000prpl.conf..299K}.} 
            \label{fig:ne_vs_dia}
        \end{figure}

    Fig.~\ref{fig:ne_vs_dia} demonstrates a strong correlation between $\rm n_e$ and the \ion{H}{ii} region diameter (r = $-$0.92 and p-value $\ll$ 0.0013). A least-squares fit to the data results in a power-law exponent of -1.27$\pm$0.02, which is consistent with the findings of \citet{2017A&A...602A..37K}. Upon comparing the distribution of $\rm n_e$ and the diameter of GLOSTAR \ion{H}{ii} regions to the typical values for HC \ion{H}{ii} regions (dashed red lines) and UC \ion{H}{ii} regions (dashed black lines), we find that our regions are more evolved than UC \ion{H}{ii} regions. The spatial resolution of the GLOSTAR D-configuration RRL image is 25$\arcsec$, which corresponds to 0.1~pc at a distance of 1~kpc and 0.6~pc at a distance of 5~kpc. Hence, this resolution is insufficient to resolve UC/HC \ion{H}{ii} regions. However, it is worth noting that some compact \ion{H}{ii} regions in our survey remain unresolved and show a crossmatch with candidate HC \ion{H}{ii} regions cataloged by \citet{2019MNRAS.482.2681Y} (see Sect.~\ref{sec:line_width}). This indicates that while our resolution limits our ability to study the smallest \ion{H}{ii} regions in detail, we still capture some candidate HC \ion{H}{ii} regions based on spatial correlation and RRL width. To overcome these limitations, the GLOSTAR survey also includes high-resolution observations with the VLA in B-configuration, which will help us detect and study \ion{H}{ii} regions in their early stages. This approach has been validated by \citet{2023A&A...670A...9D} and \citet{2023A&A...680A..92Y}, who discovered many new candidate HC \ion{H}{ii} regions using the VLA B-configuration continuum data. Additionally, we plan to present a catalog of diffuse \ion{H}{ii} regions using Effelsberg~100~m observations in the future. With these capabilities, the GLOSTAR survey can study a wide range of \ion{H}{ii} regions based on their size and evolutionary stages.

    In the following sections, we discuss the association of GLOSTAR \ion{H}{II} regions with GLIMPSE 8~$\mu$m morphology and study the statistical distribution of the physical properties of \ion{H}{II} regions with different morphologies. Later sections will explore the association of GLOSTAR \ion{H}{II} regions with dust emission and methanol masers. Since our detected \ion{H}{II} regions are evolved (Fig.~\ref{fig:ne_vs_dia}), we do not expect them to be deeply embedded. However, investigating these associations will help us understand the statistical differences in the physical properties of \ion{H}{II} regions. Finally, we discuss the Galactic electron temperature gradient.

        \section{Discussion} \label{sec:discussion}
    
        \subsection{Infrared bubbles}\label{sec:mir_morphology}
\begin{figure*}
            \centering
            \includegraphics[scale=0.26]{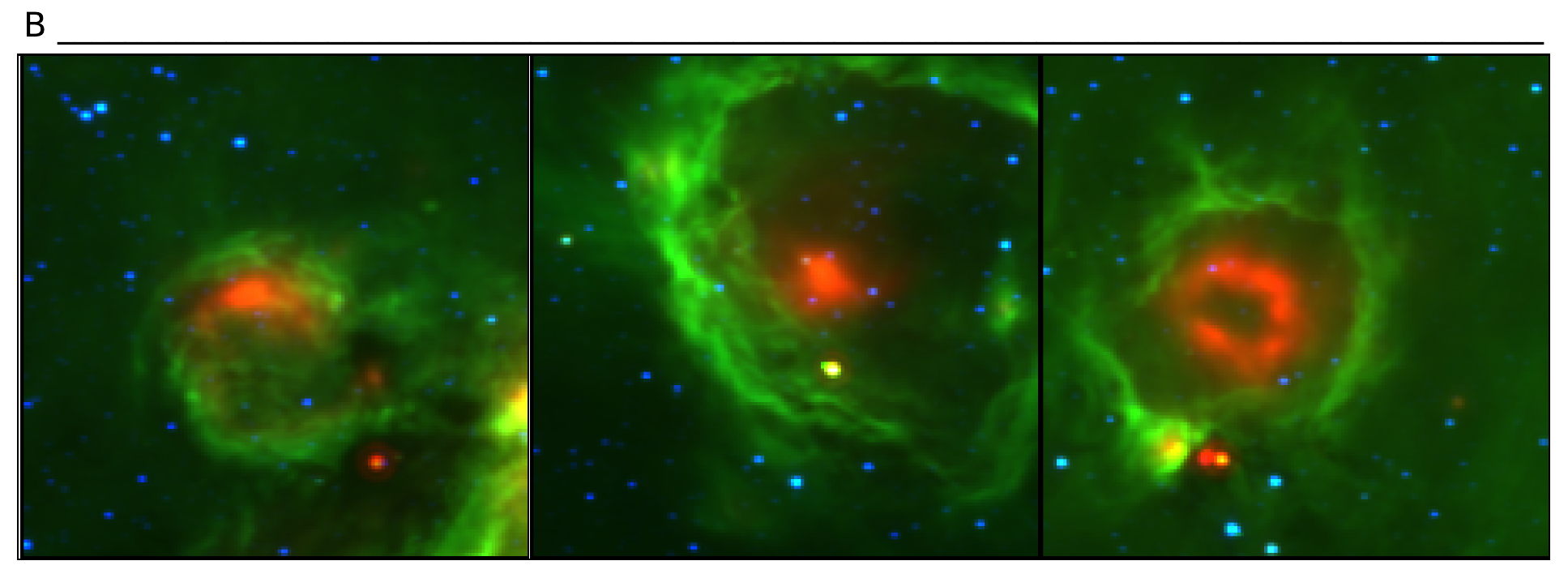}
            \includegraphics[scale=0.26]{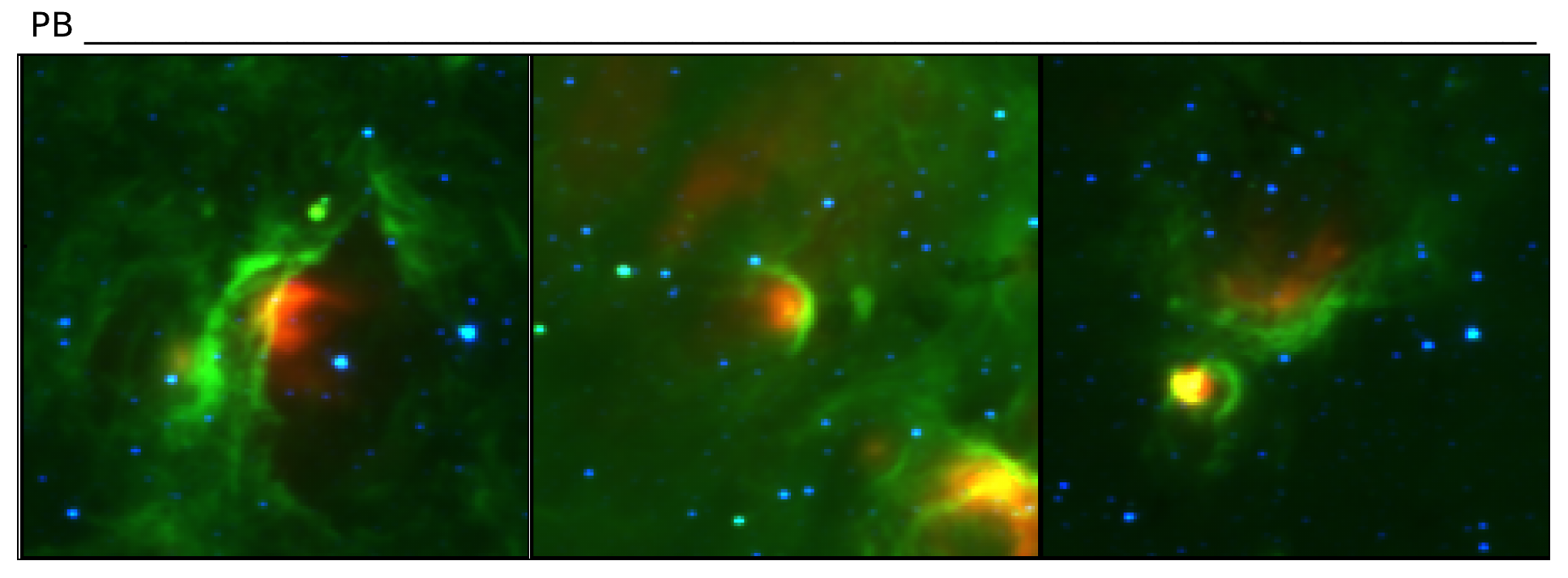}
            \includegraphics[scale=0.26]{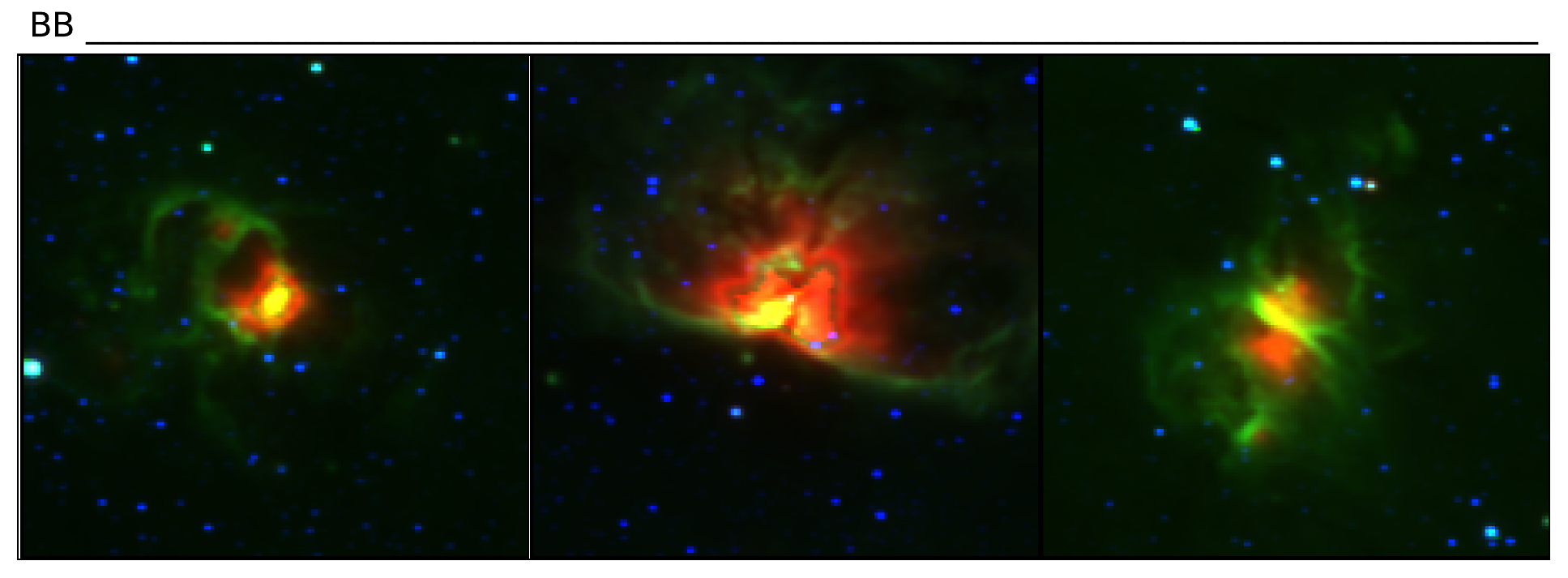}
            \includegraphics[scale=0.26]{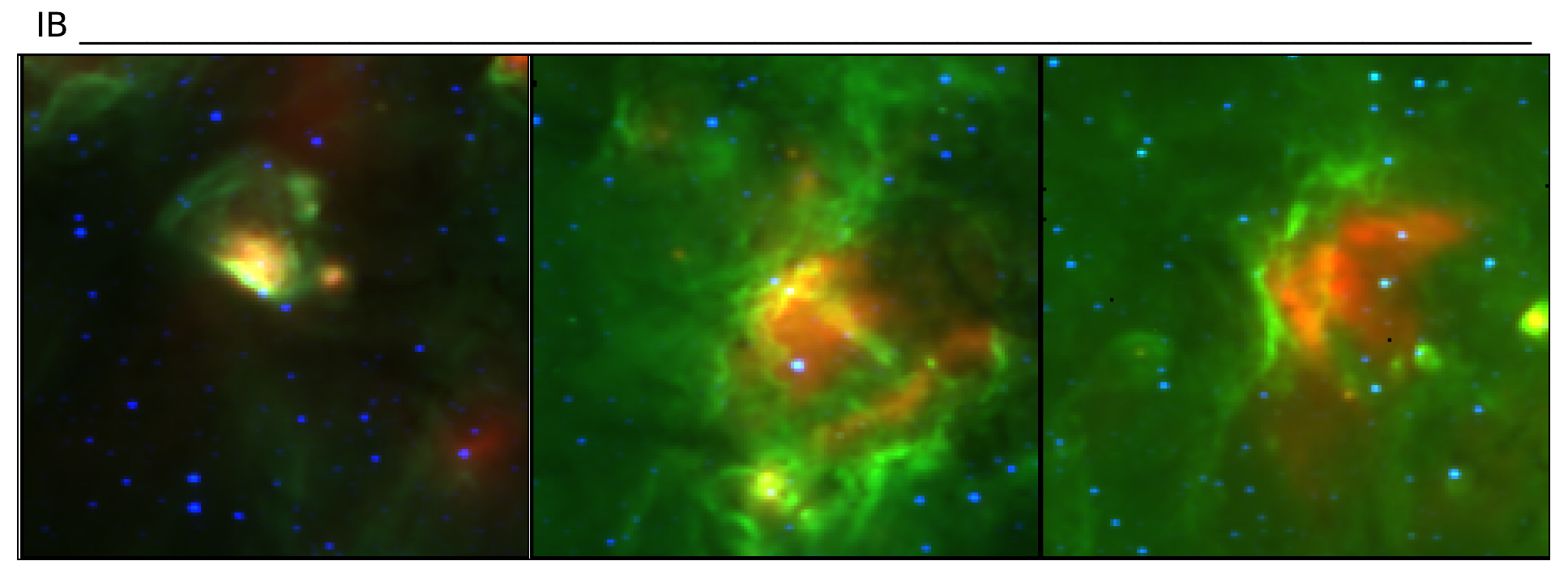}
            \includegraphics[scale=0.26]{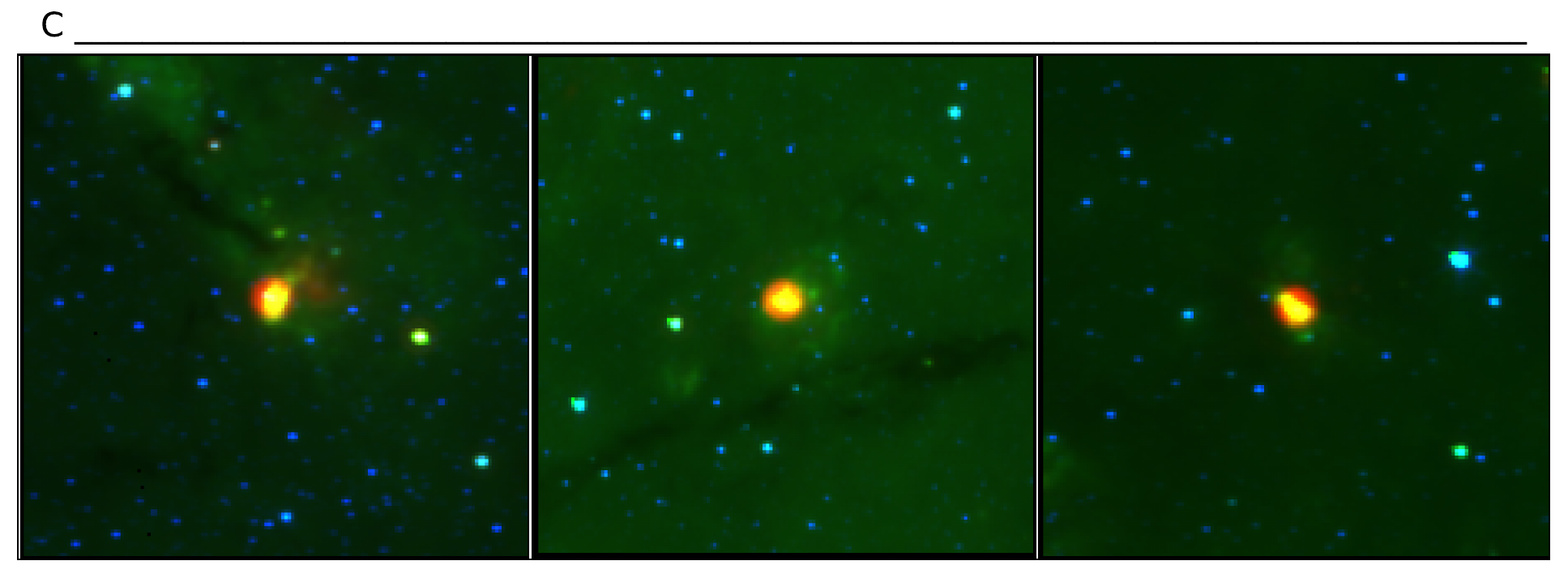}   
            \includegraphics[scale=0.26]{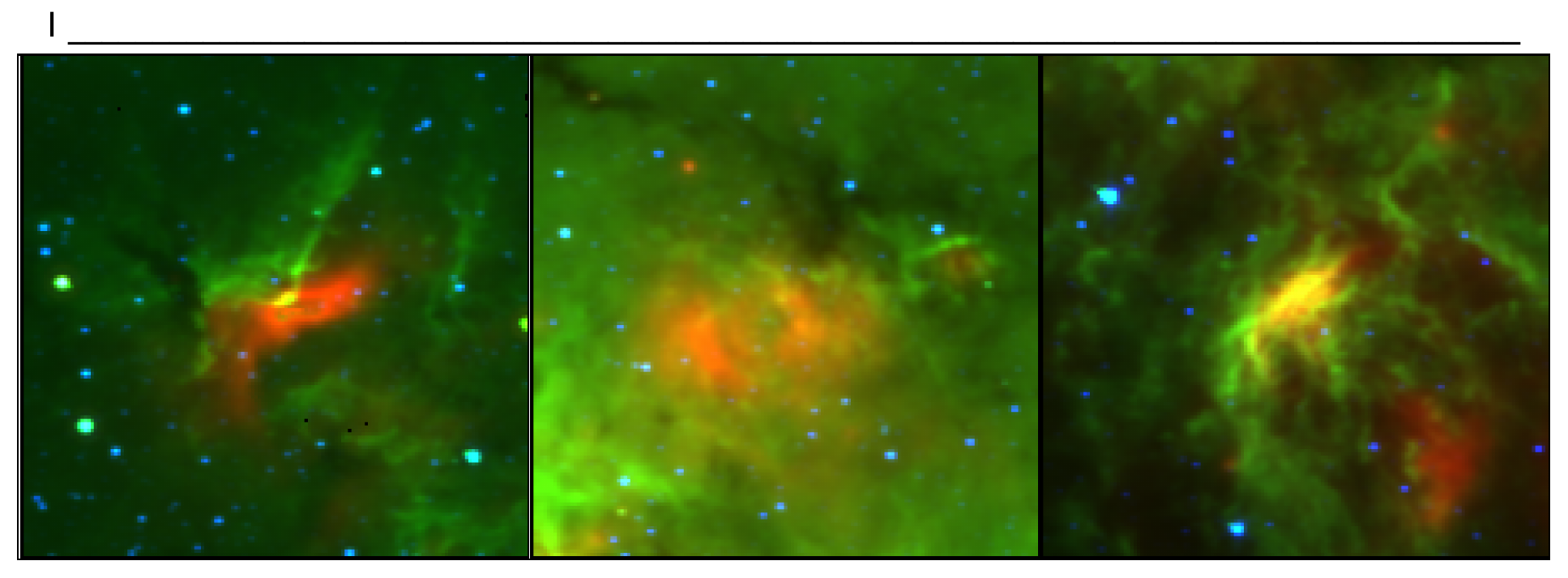}       
            \caption{Example of GLOSTAR \ion{H}{II} region morphological classifications. In each panel, a three-color image is generated using \textit{Spitzer} infrared data: MIPSGAL 24~$\mu$m data represented in red, GLIMPSE 8.0~$\mu$m data in green, and GLIMPSE 3.6~$\mu$m data in blue.}
            \label{fig:mir_mor}
        \end{figure*}

        Many sources identified as ``bubbles'' in the \textit{Spitzer} GLIMPSE data appear to represent  \ion{H}{ii} regions \citep[e.g.,][]{2010ApJ...718L.106B, 2010A&A...523A...6D, 2011ApJS..194...32A}. To investigate their infrared morphologies, we visually examined the Spitzer/GLIMPSE images associated with our RRL sources. Following~\citet{2011ApJS..194...32A}, we categorized the 8~$\mu$m emissions from GLIMPSE as follows: B - Bubble, featuring 8~$\mu$m emission surrounding 24~$\mu$m; BB - Bipolar Bubble, consisting of two connected bubbles with a region of strong infrared emission; PB - Partial Bubble, similar to a bubble but incomplete; IB - Irregular Bubble, lacking a well-defined structure; C - Compact, exhibiting compact 8~$\mu$m emission; I - Irregular Structure, displaying complex morphology that cannot be easily classified; and ND - Not Detected, indicating either a lack of 8~$\mu$m emission or extremely weak emission. Upon analyzing the infrared morphologies, we discovered that approximately 48\% (117) of our nebulae exhibit bubble morphology, while 17\% (42) appeared compact, and 31\% displayed irregular shapes. The remaining 9 sources were classified as `ND' due to their weak or absent infrared emission. These numbers are summarized in Table~\ref{tab:mir_bubble}. Fig.~\ref{fig:mir_mor} displays MIR images depicting different morphological classifications of sampled GLOSTAR \ion{H}{ii} regions. Since we utilized an interferometer, our RRL images were not sensitive large scale \ion{H}{ii} region emission. If RRLs were detected near the periphery of a bubble, we classified its infrared morphology as a bubble. Among the categorized bubble sources, 29\% (34 out of 117) had RRLs detected at the edge, while 70\% (82 out of 117) had RRLs detected at the core. Upon examining the infrared morphology, \cite{2011ApJS..194...32A} found that more than half of the HRDS \ion{H}{ii} regions exhibited bubble morphology, which aligns closely with our findings. 
        \begin{table}[]
            \centering
            \caption{Classification of the MIR morphology of \ion{H}{ii} regions.}
            \label{tab:mir_bubble}
            \begin{tabular}{cccccccc}
            \hline \hline
            &B&BB&PB&IB&C&I&ND\\
            \hline \hline
        Galactic Center&2&0&2&2&9&11&2\\
        Cygnus X&1&0&2&2&1&6&1\\
        Galactic disk\tablefootmark{*}&30&13&31&32&32&59&6\\
        \hline
        Total&33&13&35&36&42&76&9\\
        \hline \hline
            \end{tabular}
            \tablefoot{We categorized the 8~$\mu$m emissions from Spitzer GLIMPSE as follows: B - Bubble, featuring 8~$\mu$m emission surrounding 24~$\mu$m; BB - Bipolar Bubble, consisting of two connected bubbles with a region of strong infrared; PB - Partial Bubble, similar to a bubble but incomplete; IB - Irregular Bubble, lacking a well-defined structure; C - Compact, exhibiting compact 8~$\mu$m emission; I - Irregular Structure, displaying complex morphology that cannot be easily classified; and ND - Not Detected, indicating either a lack of 8~$\mu$m emission or very weak emission.
        \tablefoottext{*}{Here, by Galactic disk, we mean the region of the Galaxy within 2$\degr \leq$ $\ell$ $\leq$ 60$\degr$, |\textit{b}| $\leq$ 1$\degr$}.}
        \end{table}
        
        Through visual examination of images from the GLIMPSE survey, \cite{2006ApJ...649..759C,2007ApJ...670..428C} compiled a catalog of 591 Galactic bubbles. However, we only identified RRLs originating from ionized gas in 15 of the 250 GLIMPSE bubble sources in the GLOSTAR survey region. This limited correlation could be attributed to the limited sensitivity of our interferometry observations for large structures. 
        
\begin{figure*}[h!]
        \centering
        \includegraphics[scale=0.6]{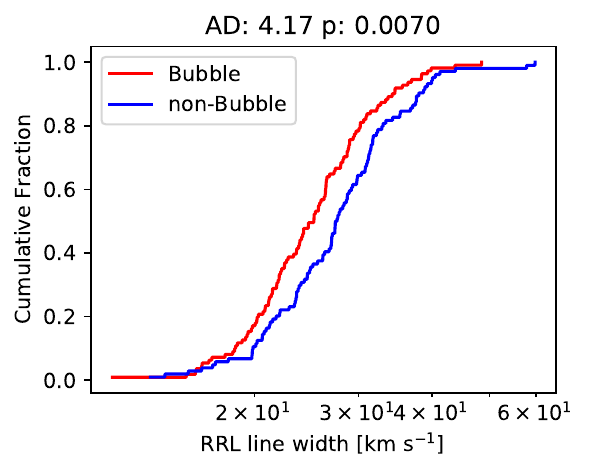}        
        \includegraphics[scale=0.6]{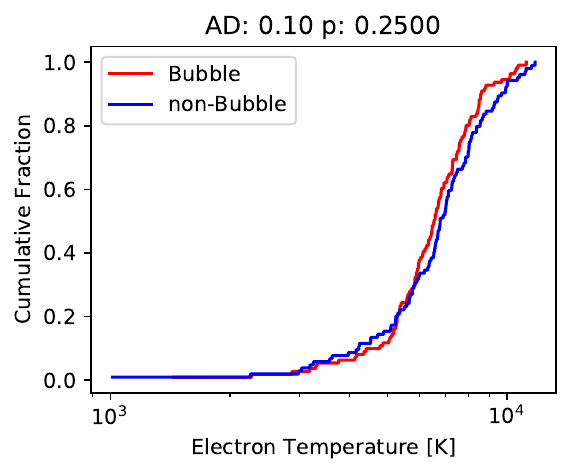}
        \includegraphics[scale=0.6]{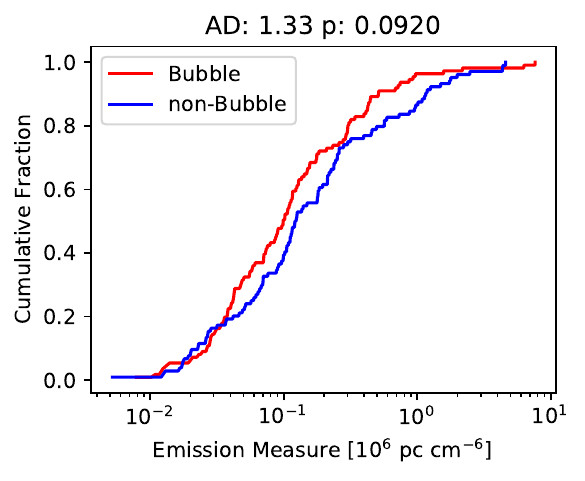}
        \includegraphics[scale=0.6]{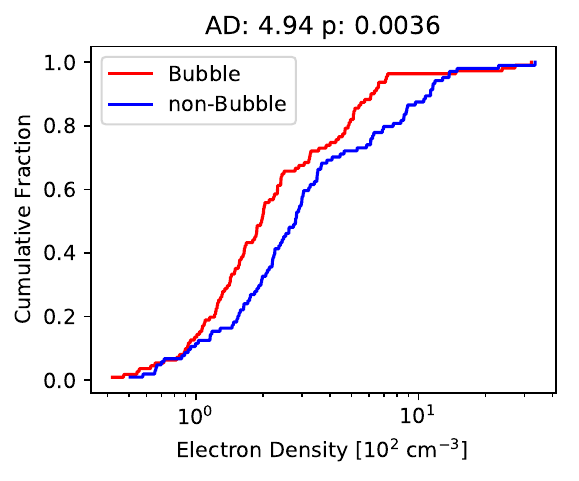}
        \includegraphics[scale=0.6]{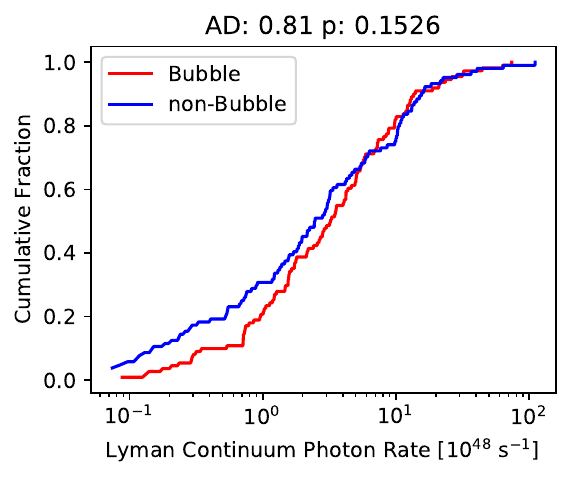}
        \includegraphics[scale=0.6]{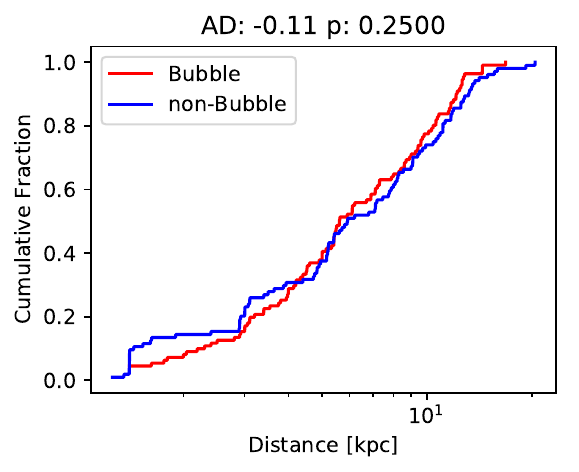}
        \caption{CDFs for various properties of GLOSTAR \ion{H}{ii} regions. We compare the sample of the GLOSTAR \ion{H}{ii} regions with Bubble morphology (in red) to the sample of GLOSTAR \ion{H}{ii} regions non-bubble morphology (in blue). The results of the Anderson-Darling (AD) tests are presented above each plot.}
        \label{fig:bubble_ass}
    \end{figure*}

          The cumulative distribution function (CDF) is a statistical tool that illustrates how data points are distributed across a range of values and facilitates straightforward comparisons between diverse datasets or groups of data. Fig.~\ref{fig:bubble_ass} presents the CDFs of GLOSTAR survey \ion{H}{ii} regions, divided into bubble and non-bubble categories. The non-bubble category includes compact and irregular \ion{H}{ii} regions. To ensure the validity of our analysis, we excluded \ion{H}{ii} regions located in the Galactic center region, as the complicated emission structures that are prevalent there might affect the morphology classification compared to the rest of the Galactic plane. Various \ion{H}{ii} region properties, such as RRL width, electron temperature, electron density, EM, Lyman continuum photon rate, and ionized gas mass, were examined. A comparison was conducted between a total of 111 bubble \ion{H}{ii} region and 104 non-bubble \ion{H}{ii} regions. To assess the statistical significance of differences between the two groups, Anderson-Darling (AD) tests were conducted on all CDFs. The findings are depicted in Fig.~\ref{fig:bubble_ass}. Interestingly, we observed no statistically significant differences in most of the \ion{H}{II} region properties between the two groups, except for RRL width and electron density, where the \textit{p}-value was < 0.05 (significance level of 2$\sigma$). The mean RRL width of non-bubble \ion{H}{ii} regions is 28.4~km~s$\rm ^{-1}$, which is slightly wider than the average line-width of bubble \ion{H}{ii} regions, which is 25.7~km~s$\rm ^{-1}$. This contrasts with \citet{2011ApJS..194...32A}, who found that the average RRL width of bubble sources was identical to the entire HRDS sample. However, the significant level of difference in RRL width is only up to 2$\sigma$, which is not conclusive. According to the ring scenario proposed by \citet{2010ApJ...709..791B}, which suggests that ``bubble'' sources are physically two-dimensional rings, we would expect RRL widths for bubbles to be larger than those of the rest of the \ion{H}{ii} region sample. This is because both red-shifted and blue-shifted ionized gas from the bipolar flow contribute to the bubble line width \citep{2011ApJS..194...32A}. However, our findings show the opposite, suggesting that bubbles are three-dimensional structures.


    \citet{2006ApJ...649..759C,2007ApJ...670..428C} and \citet{2012MNRAS.424.2442S} have demonstrated that bubbles are widespread in IR images of the Galactic plane and compiled a large population of Galactic IR bubbles. These bubble structures in the ISM can arise from various astronomical phenomena, not limited to young, massive stars. Evolved objects such as supernova remnants, PNe, Wolf-Rayet stars, and (post)-AGB stars can also form bubbles. \citet{2010A&A...523A...6D} proposed that at least 86\% of the Galactic bubbles are associated with \ion{H}{ii} regions. Additionally, \citet{2011ApJS..194...32A} and this work showed that approximately 50\% of the \ion{H}{ii} regions exhibit bubble morphology at 8.0~$\mu$m. These findings suggest that almost all Galactic bubbles are \ion{H}{ii} regions, but not all \ion{H}{ii} regions show a bubble-like structure. Studies conducted with a lower angular resolution than ours, indeed confirm a considerably larger population of \ion{H}{ii} regions in the Galactic plane, as is discussed in \citet{2014ApJS..212....1A}.

        \subsection{Association with dust emission}\label{sec:dust_asso}
        Ultracompact and compact \ion{H}{ii} regions usually have associated dust emission from molecular clouds in which their ionizing massive stars have been born. Even later-type O-type stars and early B-type stars are often associated molecular clouds whose dust is heated by the radiation from them~\citep{2002ARA&A..40...27C}. \cite{2006A&A...453.1003T} surveyed the environments of 105 IRAS point sources at 850 and 450~$\mu$m for a comprehensive study on the association of sub-millimeter dust emission with UC \ion{H}{ii} regions. They reported three distinct types of objects: UC cm-wave sources that are not associated with any submillimeter emission (``submillimeter quiet objects''), submillimeter clumps that are associated with UC cm-wave sources (``radio-loud clumps''); and submillimeter clumps that are not associated with any known UC cm-wave sources (``radio-quiet clumps''). Inspired by this, we studied the dust association of the detected 202 \ion{H}{ii} regions within 3$\degr \leq$ $\ell$ $\leq$ 60$\degr$, and |\textit{b}|  $\leq$ 1$\degr$  and categorized these objects into  ``submillimeter quiet \ion{H}{ii} regions'' or  ``submillimeter loud \ion{H}{ii} region.''
        
        We examined dust clumps within the above range from the ATLASGAL compact source catalog~\citep[CSC; ][]{2014A&A...568A..41U}. For ATLASGAL CSC objects, \citet[][]{2018MNRAS.473.1059U, 2022MNRAS.510.3389U} determined clump mass, dust temperature, bolometric luminosity and V$\rm _{LSR}$ velocity. We performed a crossmatch with ATLASGAL sources using a angular separation threshold of 12$\arcsec$, which is three times the uncertainty in ATLASGAL position and half of GLOSTAR RRL image beam size. We found 97 associations within 12$\arcsec$ (48\%). Out of these 97 associated dust clumps, 68 are classified as \ion{H}{ii} regions, whereas 22, 4, and 3 sources are classified as photo-dissociation regions (PDRs), Ambiguous, or Complicated by~\citet{2022MNRAS.510.3389U}. We also observed that the average angular offset for the \ion{H}{ii} regions association is 6.2$\arcsec$ which is  slightly smaller than the angular offset for the other source types which is 8.0$\arcsec$. In the following, we focus our attention on these 68 \ion{H}{ii} regions labeled as submillimeter loud \ion{H}{ii} regions and further explore their association with the corresponding dust clumps.

\begin{figure}[h!]
        \centering
        \includegraphics[width=0.5\textwidth]{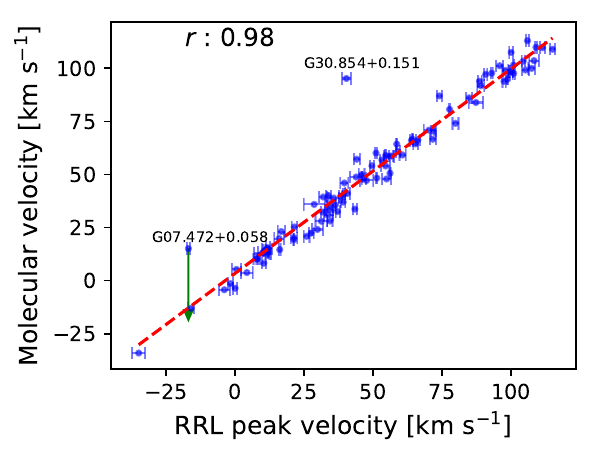}
        \includegraphics[width=0.5\textwidth]{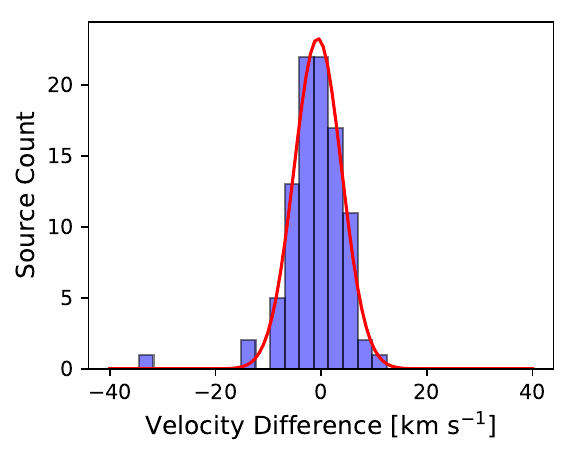}
        \caption{Comparison between the RRL peak velocity and ATLASGAL dust clump velocity. \textit{Top}: The RRL peak velocity is plotted against the ATLASGAL clump velocity. The red line shows a linear fit, with a Spearman’s rank coefficient of r = 0.98 and p-value $\ll$ 0.0013. The linear fit produces a slope of 0.96 $\pm$ 0.02 and a y-intercept of 3.5 $\pm$ 1.3~km~s$^{-1}$. The green arrow shows the direction of the corrected velocity of G007.472+0.058.  \textit{Bottom}: the distribution of velocity offsets between the RRL peak velocities and the molecular line velocities from ATLASGAL is shown. Fitting the distribution with a Gaussian function yields a mean offset of $-$ 0.62 $\pm$ 0.12~km~s$^{-1}$ and a standard deviation of 10.5 $\pm$ 0.3~km~s$^{-1}$.}
        \label{fig:vel_comp_molecule}
    \end{figure}

        In Fig.~\ref{fig:vel_comp_molecule}, the LSR velocities of ATLASGAL dust clumps are compared with the RRL peak velocities measured in GLOSTAR. The velocities of the ATLASGAL sources were determined by observations of spectral lines from multiple molecular line surveys toward them \citep[see Sect. 2.1 of ][ for more details]{2018MNRAS.473.1059U}. A linear fit was performed, resulting in a slope of 0.96 $\pm$ 0.02 and a y-intercept of 3.5 $\pm$ 1.3~km~s$^{-1}$, respectively. The Spearman's correlation coefficient of 0.98 indicates a positive correlation. A Gaussian function was used to fit the distribution of velocity offsets, which yielded a mean offset of $-$0.62 $\pm$ 0.12~km~s$^{-1}$ and a standard deviation of 10.5 $\pm$ 0.3~km~s$^{-1}$. This indicates a small velocity offset between the molecular gas and the ionized gas associated with \ion{H}{II} regions. The results are in excellent agreement with the findings of~\cite{2009ApJS..181..255A}, for their study of molecular properties of the Galactic \ion{H}{ii} regions. This study revealed a standard deviation of 8.5~km~s$^{-1}$ and a mean of 0.2~km~s$^{-1}$ for the distribution of the LSR velocity difference between the RRL and \element[][13]{CO} velocities. These values are also similar to those found in the WISE catalog of Galactic \ion{H}{ii} regions by~\cite{2014ApJS..212....1A}. This shows that the velocities of ionized gas in \ion{H}{ii} regions are in good agreement with those of the molecular gas from which the massive stars formed.
        
        There are two sources with velocity offsets greater than 3$\sigma$: G007.472+0.058 (AGAL007.471+00.059) and G030.854+0.151 (AGAL030.854+00.149). For the latter,~\citet{2018MNRAS.473.1059U} assigned the molecular LSR velocity based on NH$_3$ line observations by~\citet{2012A&A...544A.146W}. \citet{2012A&A...544A.146W} detected strong emission in the NH$\rm _3$ (1,1), (2,2) and (3,3) transitions, toward this source. These transitions revealed the source velocity to be 95.2~km~s$^{-1}$, resulting in a velocity offset of -54.86~km~s$^{-1}$. \citet{2011ApJS..194...32A} reported multiple RRL velocity components for G030.854+0.151 (HRDS name: G030.852+0.149), with the brightest component at 39.6~km~s$^{-1}$, while the other two weaker components are at 88.7 and 114.5~km~s$^{-1}$.  Additionally, this source is in the direction of W43, which is located close to the near end of the Galactic bar and the inner Scutum arm, a region of intricate gas dynamics~\citep{2011A&A...529A..41N,2020ApJ...889...96L}. \citet{2015ApJ...810...42A} suggested that the additional velocity components are associated with the diffuse Warm Ionized Medium (WIM). However, it is unlikely that GLOSTAR D-configuration observation would be able to detect the diffuse gas of the WIM. Hence, in this case the large velocity offset could be attributed to the fact that the velocities of our RRLs and the NH$\rm _3$ transitions observed by \citet{2012A&A...544A.146W} probe different regions along the line of sight.
    In the case of G007.472+0.058 (AGAL007.471+00.059), the source exhibits two peaks of $^{13}$CO emission at 15.07~km~s$^{-1}$ and $-$14.44~km~s$^{-1}$, and a single peak of C$^{18}$O emission at 15.10~km~s$^{-1}$~\citep{2018MNRAS.473.1059U}. However,~\citet{2018MNRAS.473.1059U} assigned the velocity to this source based on the single peak of C\element[][18]{O}.~\citet{2016PASA...33...30R} showed that the molecular velocity of AGAL007.471+00.059 is $-$13.91~km~s$^{-1}$, based on the intensity weighted velocities of HCO$^+$, HCN, and N$_2$H$^+$ lines. Taking the latter, the LSR velocity difference between the molecular and ionized gas emission of AGAL007.471+00.059 is 3.03~km~s$^{-1}$ (< 3$\sigma$), which is consistent with the velocity offset distribution of the sample. 
        
\begin{figure*}[h!]
        \centering
        \includegraphics[scale=0.6]{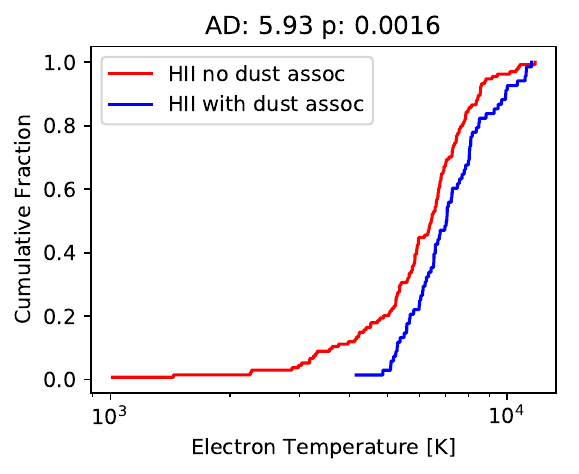}
        \includegraphics[scale=0.6]{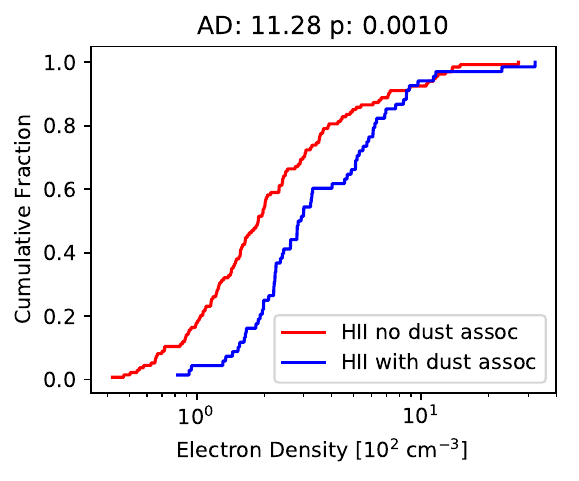}
        \includegraphics[scale=0.6]{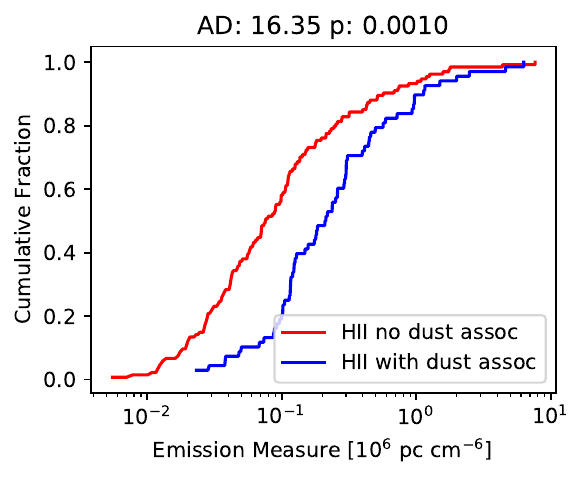}
        \includegraphics[scale=0.6]{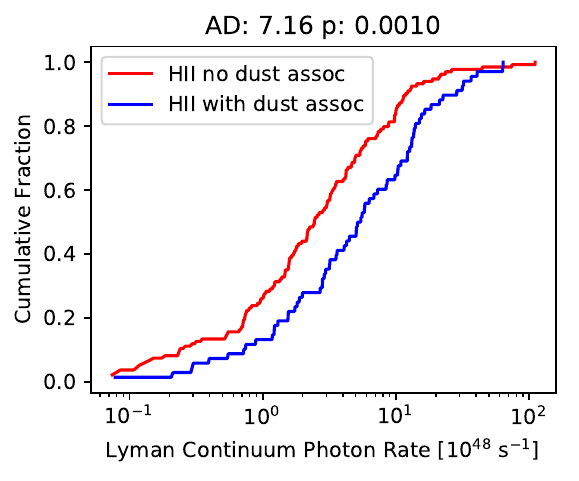}
        \includegraphics[scale=0.6]{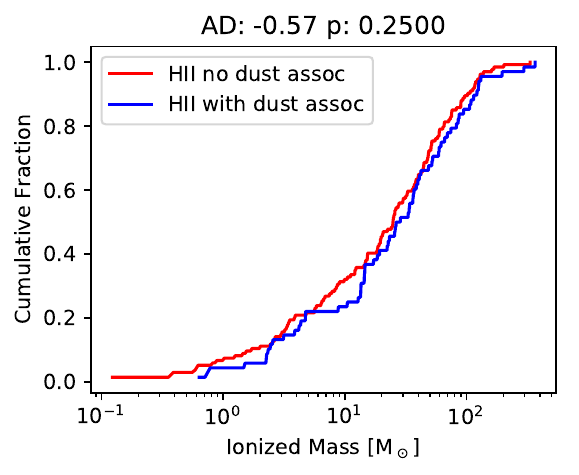}
        \includegraphics[scale=0.6]{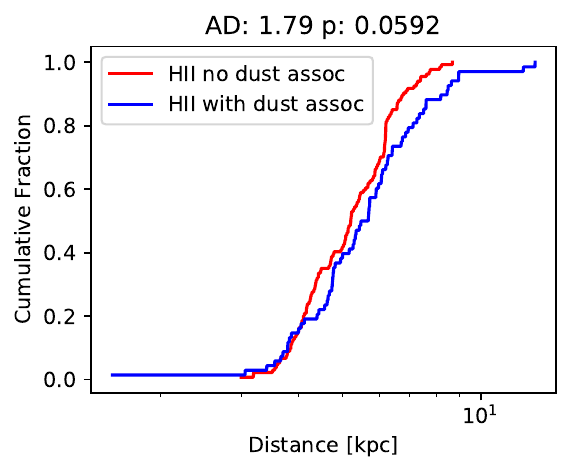}
        \caption{Comparison between the properties of GLOSTAR \ion{H}{ii} regions that are associated with ATLASGAL dust clumps (in red) and those that are not linked to ATLASGAL clumps (in blue). CDFs are presented for various properties of GLOSTAR \ion{H}{ii} regions. The results of the Anderson-Darling (AD) tests, indicated above each plot, provide insights into the statistical significance of the observed differences.}
        \label{fig:dust_without_dust_hist}
    \end{figure*}

    In Fig.~\ref{fig:dust_without_dust_hist}, we show the CDFs for GLOSTAR \ion{H}{ii} region properties, including electron temperature, electron density, EM, Lyman continuum photon rate, ionized gas mass and distance for submillimeter loud and submillimeter quiet \ion{H}{ii} regions. A total of 68 submillimeter loud \ion{H}{ii} regions with associated dust clumps are compared to 134 submillimeter quiet \ion{H}{ii} regions without any dust association. Inspection of the electron density and EM (top middle and right panel of Fig.~\ref{fig:dust_without_dust_hist}), revealed a significant difference between the submillimeter loud and submillimeter quiet \ion{H}{ii} regions (AD test return a \textit{p}-value $\ll$ 0.0013). Fig.~\ref{fig:dust_without_dust_hist} shows that the $\rm n_e$ and EM are higher for submillimeter loud \ion{H}{ii} region as they are likely to be smaller and denser. Turning our attention to the Lyman photon rate (bottom left of Fig.~\ref{fig:dust_without_dust_hist}), we see a significant difference at the 3$\sigma$ level (\textit{p}-value < 0.0013) between submillimeter loud and submillimeter quiet \ion{H}{ii} regions. Additionally, we also notice that the Lyman continuum flux for the submillimeter loud \ion{H}{ii} region is slightly higher compared to the submillimeter quiet \ion{H}{ii} region. It is not very clear why they are hotter and have more Lyman continuum flux. It is possible that as the \ion{H}{ii} regions evolve they become more extended, resulting in underestimation of Lyman continuum flux due to interferometric filtering of diffuse emission. Examining the electron temperature, we notice that the distribution becomes indistinguishable at higher values, which is not the case for lower electron temperatures. The AD test reveals a significant difference at the 3$\sigma$ level (\textit{p}-value $\sim$ 0.0013) between submillimeter loud and submillimeter quiet \ion{H}{ii} regions. In addition to electron density and EM, we also observe that the electron temperature of the submillimeter loud is higher than that of submillimeter quiet \ion{H}{ii} regions. This is not unexpected as the higher value of $\rm n_e$ is expected to produce higher value of $\rm T_e$ (as mentioned in Sect.~\ref{sec:elec_temp}). Examining ionized gas mass and distance, we observe that the distributions of submillimeter loud and submillimeter quiet \ion{H}{ii} regions are indistinguishable from each other.

        \subsection{Association with 6.7 GHz methanol masers}\label{sec:maser_asso}
        Class II methanol maser emission in the 6.7~GHz line was first discovered by \citet{1991ApJ...380L..75M}, and since then, these masers have proven to be valuable tools for investigating high-mass star-forming regions. The 6.7 GHz methanol masers are exclusive tracers of the early evolutionary stages of high-mass stars \citep[e.g.,][]{2015MNRAS.446.3461U, 2020MNRAS.492.1335P}. \citet{2021A&A...645A.110Y} suggests that younger \ion{H}{ii} regions are more likely to be associated with masers as the detection rate of maser emission  decreases as \ion{H}{ii} regions evolve from HC \ion{H}{ii} regions to UC \ion{H}{ii} regions. This is supported by the detection of methanol masers toward Galactic \ion{H}{ii} regions, which seem to be closely correlated with compact \ion{H}{ii} regions~\citep{1998MNRAS.301..640W,2019ApJS..245...12O,2020ApJS..248....3C}. However, the studies conducted by \citet{2016ApJ...833...18H} and \citet{2022A&A...666A..59N} did not reveal any apparent correlation between the flux density from methanol masers and radio continuum sources. This indicates that methanol maser and radio continuum sources exhibit independent luminosity behaviors.    
        
    The GLOSTAR survey~\citep{2021A&A...651A..85B} observed RRLs and Class II 6.7~GHz methanol ($\rm CH_3OH$) masers simultaneously, allowing accurate results of cross matching of their positions. \citet{2021A&A...651A..87O} and \citet{2022A&A...666A..59N} compiled a catalog of 6.7 GHz methanol masers within the region covered by the GLOSTAR survey in the Galactic plane, using the VLA D-configuration data. We conducted an investigation to identify potential correlations between RRLs and 6.7~GHz methanol masers detected in the GLOSTAR VLA D-configuration data~\citep{2022A&A...666A..59N} within the Galactic zone of 2$\degr \leq$ $\ell$ $\leq$ 60$\degr$, |\textit{b}| $\leq$ 1$\degr$. We find a total of 30 ($\sim$ 15\%) cross matches within a radius of 12$\arcsec$, which is half of the size of the GLOSTAR RRL beam. The mean $\pm$ standard deviation of the separation between methanol maser position and continuum peak is 6.0 $\pm$ 2.6$\arcsec$. This number goes down to 17 ($~\sim$ 8\%) sources, when we use an angular distance threshold of 6$\arcsec$, which was used by~\cite{2022A&A...666A..59N} to study the association of methanol masers with GLOSTAR D-configuration continuum sources. \citet{2016ApJ...833...18H} propose two criteria for determining the association between maser and continuum sources based on their projected spatial distance. Firstly, the projected spatial distance between peak positions of the maser and continuum sources should be smaller than 1~pc, and secondly, it should be less than the major axis of the continuum source. In all 30 cases, the projected spatial offset is less than 1~pc, with a mean separation of 0.2 pc. Also, for all cases the projected spatial offsets are smaller than the effective diameters of the host \ion{H}{ii} regions. Using the catalog of methanol masers compiled by \citet{2022A&A...666A..59N} and \citet{2021A&A...651A..87O}, we find only 13\% (30/202) of the \ion{H}{ii} regions have associated methanol maser emission within a radius of 12$\arcsec$. For comparison, \citet{2011ApJS..194...32A} found 10\% of the HRDS \ion{H}{ii} regions to have associated methanol masers. \citet{1998MNRAS.301..640W} observed that 38\% of the sources in their UC \ion{H}{ii} region sample have methanol maser emission. In the study by \citet{2019ApJS..245...12O}, out of 98 radio continuum sources, 23 sources were detected with both RRL and methanol maser emission. \citet{2020ApJS..248....3C} reported that 20\% of the sources with RRL detection were found to be associated with methanol masers. These comparisons suggest that most of the \ion{H}{ii} regions in our sample might not be as young as UC/HC \ion{H}{ii} regions. 
    
\begin{figure}[h!]
        \centering
        \includegraphics[width=0.5\textwidth]{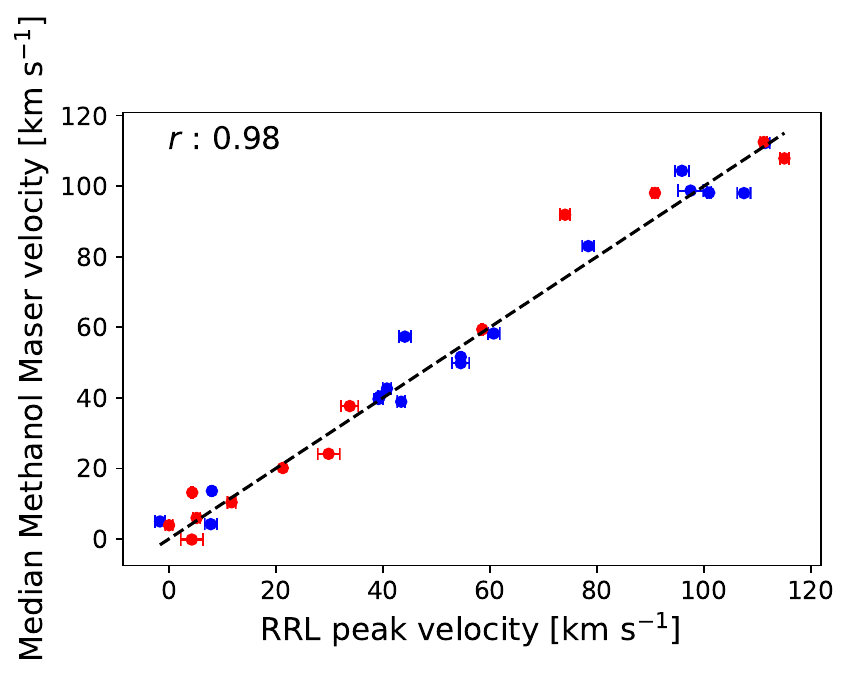}
        \includegraphics[width=0.5\textwidth]{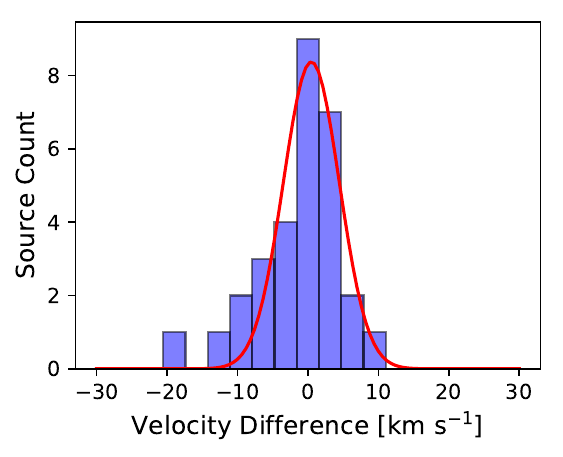}
        \caption{Comparison between the RRL peak velocity and median methanol maser velocity. \textit{Top}: The RRL peak velocity is plotted against the median methanol maser velocity. Sources with a positional offset between the \ion{H}{ii} region and methanol maser of less than 6$\arcsec$ are depicted using blue circles, while sources with a positional offset greater than 6$\arcsec$ but less than 12$\arcsec$ are represented by red circles. The dashed black line represents the line of equality, x=y, with a Spearman’s rank coefficient of r = 0.98 and p-value $\ll$ 0.0013. \textit{Bottom}: The distribution of velocity offsets between the RRL peak velocities and median methanol maser velocity is shown. Fitting the distribution with Gaussian yield a mean of 0.5 $\pm$ 0.3~km~s$^{-1}$ and standard deviation of 4.0 $\pm$ 0.34~km~s$^{-1}$.}
        \label{fig:vel_comp_meth}
    \end{figure}

    Fig.~\ref{fig:vel_comp_meth} shows the correlation of RRL peak velocity and median methanol maser velocity. All the associated RRL and methanol masers source exhibit similar velocities with a Spearman’s rank coefficient of r = 0.98 and \textit{p}-value $\ll$ 0.0013, which intrinsically represents the systemic motion of the sources. Fig.~\ref{fig:vel_comp_meth} presents the distribution of the velocity difference. Fitting the distribution with a Gaussian function yields a mean of 0.5 $\pm$ 0.3~km~s$^{-1}$ and a standard deviation of 4.0 $\pm$ 0.34~km~s$^{-1}$. These results show a strong agreement between RRL and methanol maser velocities and indicating that they are associated with the same molecular cloud. 
    
\begin{figure*}[h!]
        \centering
        \includegraphics[scale=0.4]{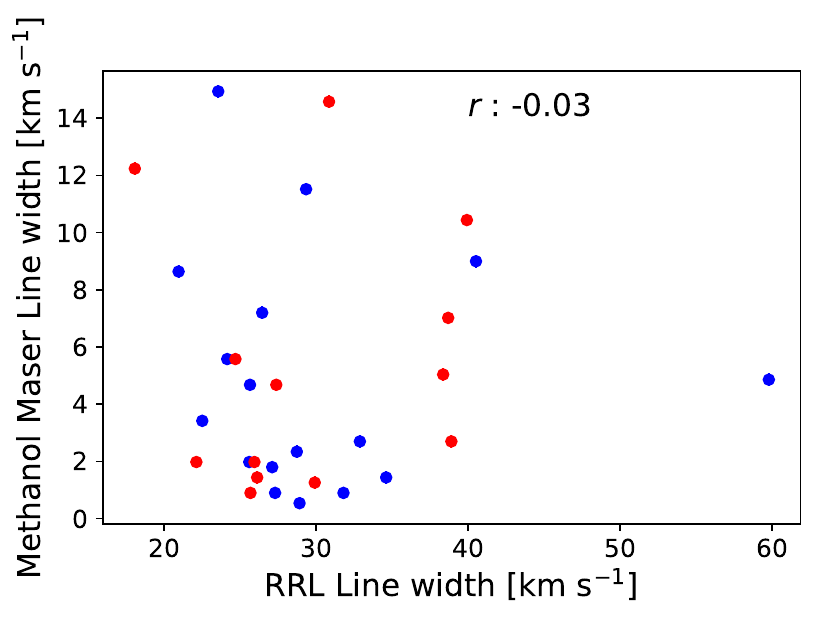}
        \includegraphics[scale=0.4]{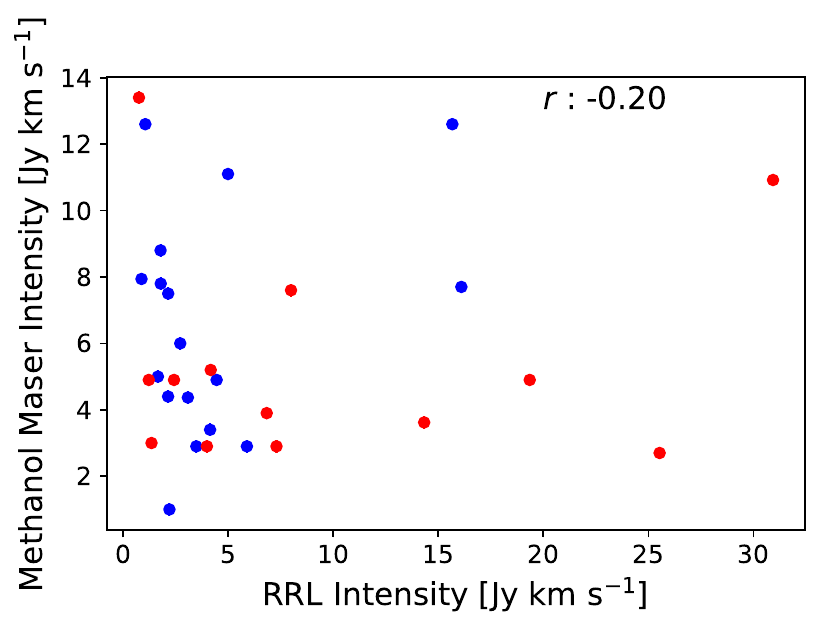}
        \includegraphics[scale=0.4]{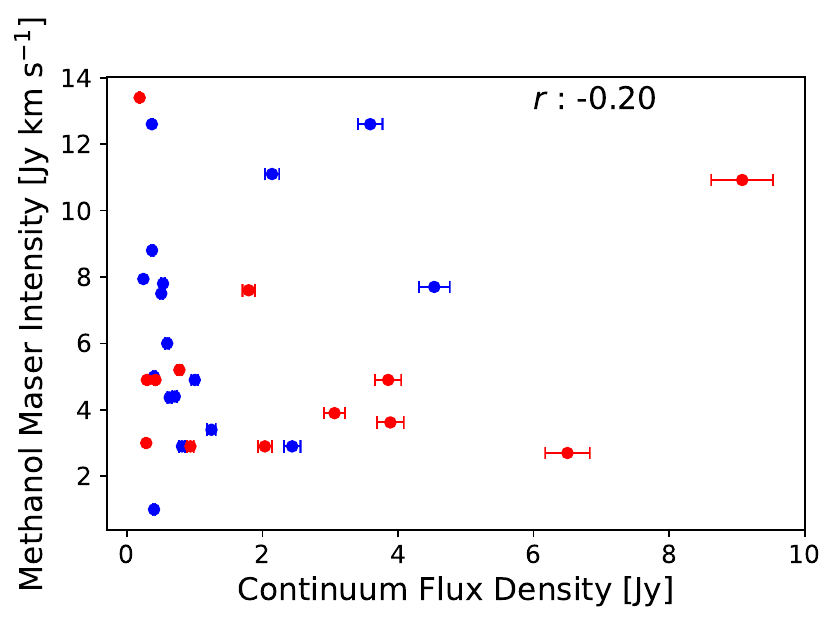}
        \caption{\textit{Left}: RRL width plotted as a function of the methanol maser line width. RRL integrated intensity (\textit{middle}) and continuum flux density (\textit{right}) plotted as a function of maser integrated intensity. Blue and red circle corresponds to a separation of less and more than 6$\arcsec$ respectively. Sources with a positional offset between the \ion{H}{ii} region and methanol maser of less than 6$\arcsec$ are depicted using filled black circles, while sources with a positional offset greater than 6$\arcsec$ but less than 12$\arcsec$ are represented by black circles. There appears to be no evident correlation between the strengths of the continuum or RRL and the maser. Additionally, no correlation is found between the line width of the RRL and the maser signals.}
        \label{fig:masserVSrrl}
    \end{figure*}

    The mean values of continuum flux densities and RRL integrated intensities for \ion{H}{ii} regions associated with methanol masers are slightly lower (1.8~Jy and 6.7~Jy~km~s$^{-1}$, respectively) compared to those without maser emissions (2.4~Jy and 9.1~Jy~km~s$^{-1}$, respectively). This finding supports the notion that methanol masers are linked to younger \ion{H}{ii} regions~\citep{2019MNRAS.482.2681Y,2021A&A...645A.110Y,1998MNRAS.301..640W}. However, it is essential to note significant differences in the sample size, which may influence the interpretation of these results. In Fig.~\ref{fig:masserVSrrl}, we present the maser integrated flux density plotted against both the continuum flux density and the RRL integrated intensity. The Spearman's rank coefficient for the correlation between maser integrated flux density and continuum flux density is $-$0.196, with a \textit{p}-value of 0.64, indicating no significant correlation between these properties. This is consistent with the conclusions drawn by \citet{2016ApJ...833...18H} and \citet{2022A&A...666A..59N}. However, \citet{2019ApJS..245...12O}, reported a strong positive correlation of luminosity between 6.7~GHz methanol maser and RRLs. In this regard, it is worth noting that their angular resolution ($\sim$3$\rm \arcmin$) was significantly poorer compared to that of GLOSTAR. Similarly, the correlation between maser integrated flux density and RRL integrated intensity yields a Spearman's rank coefficient of $-$0.203, with a \textit{p}-value of 0.43, also suggesting no significant correlation between these two variables. The methanol maser line width as a function of the RRL width is also shown in Fig.~\ref{fig:masserVSrrl} with the Spearman's rank coefficient r = $-$0.035, and a \textit{p}-value of 0.43, suggesting no relation. These findings indicate that the mechanisms responsible for powering masers and exciting RRLs are not related to each other.
    
\begin{figure*}[h!]
        \centering
        \includegraphics[scale=0.6]{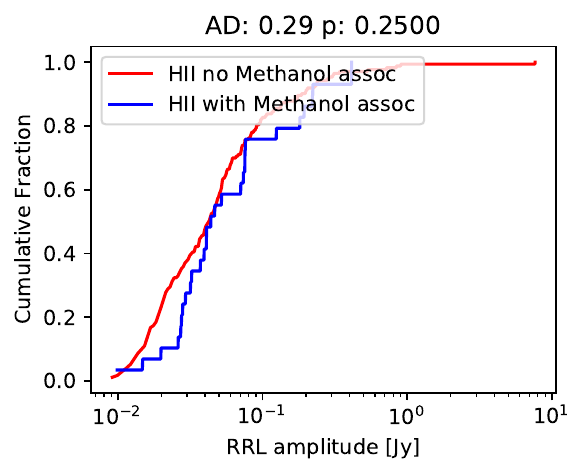}
        \includegraphics[scale=0.6]{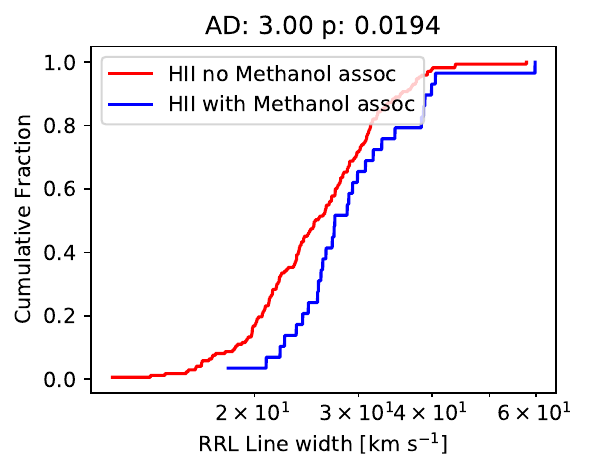}        
        \includegraphics[scale=0.6]{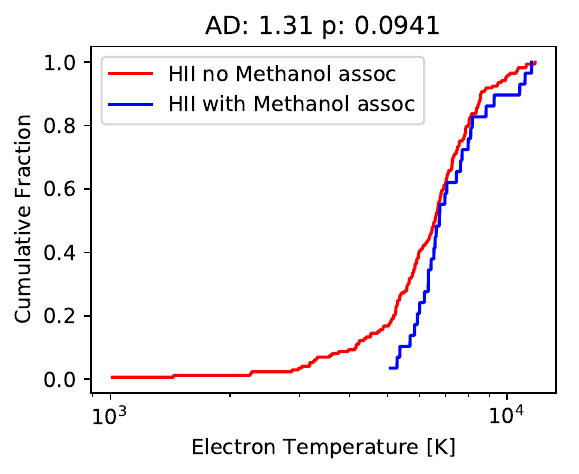}
        \includegraphics[scale=0.6]{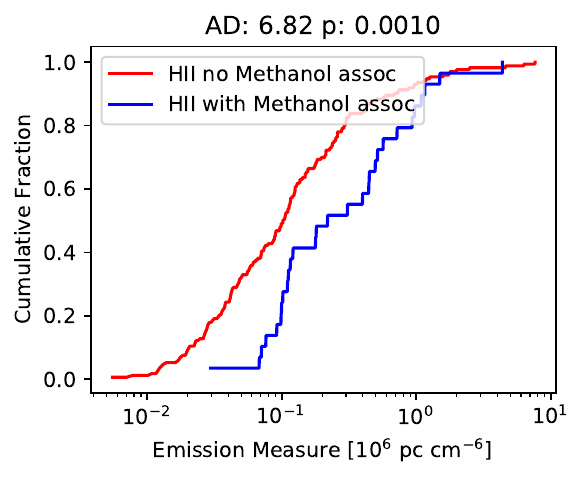}
        \includegraphics[scale=0.6]{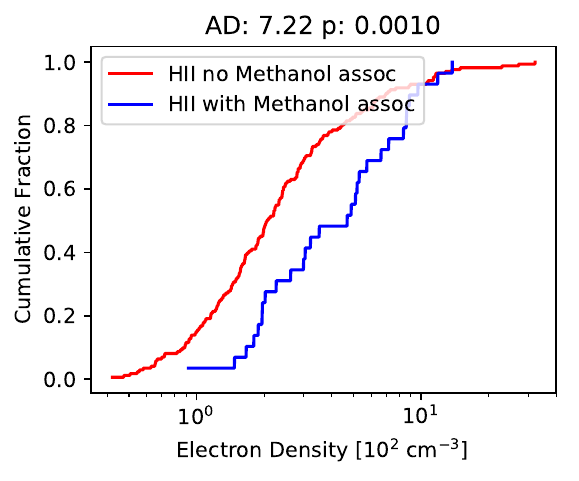}
        \includegraphics[scale=0.6]{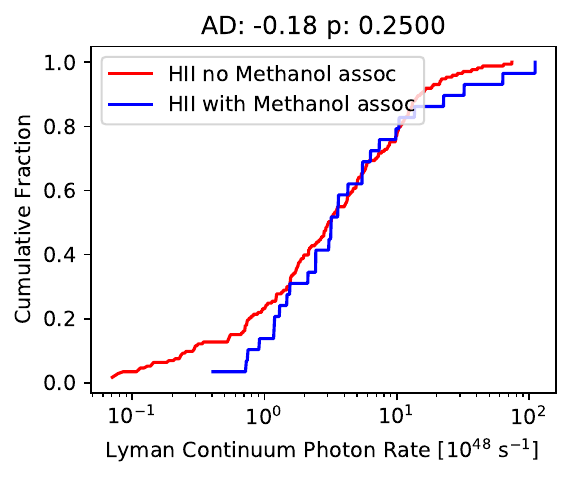}
        \includegraphics[scale=0.6]{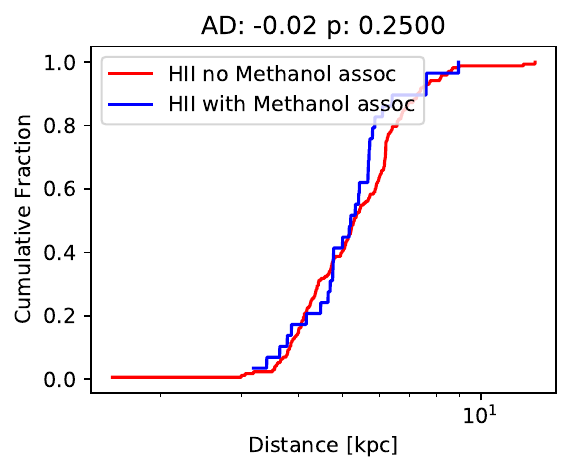}
        \caption{CDFs for various properties of GLOSTAR \ion{H}{ii} regions. We compare the sample of the GLOSTAR \ion{H}{ii} regions associated with methanol masers (in blue) to the sample of GLOSTAR \ion{H}{ii} regions not associated with methanol masers (in red). The results of the Anderson-Darling (AD) test are presented above each plot.}
        \label{fig:masser_ass}
    \end{figure*}

    In Fig.~\ref{fig:masser_ass}, we show the CDFs of \ion{H}{ii} regions categorized by the presence or absence of detected methanol masers to compare various \ion{H}{ii} region properties, such as RRL amplitude, RRL width, $\rm T_e$, EM, $\rm n_e$, $\rm N_{Lyc}$, and distance. The sample consists of 32 \ion{H}{ii} regions associated with methanol masers and 212 \ion{H}{ii} regions without methanol masers. ~\citet{2022A&A...666A..59N} compared the flux density of continuum sources that have 6.7~GHz methanol maser association and those without. They found that the radio sources with the methanol maser are significantly brighter than the general population of radio sources. From this, we can expect that the RRL amplitude of \ion{H}{ii} regions associated with methanol maser to be brighter than those without. In contrast, looking at the CDFs of RRL amplitude (top left panel of Fig.~\ref{fig:masser_ass}) and RRL width (top middle panel of Fig.~\ref{fig:masser_ass}), we did not notice a significant difference in the distribution. Similarly, for $\rm T_e$, $\rm N_{Lyc}$, and distance, we see that the distributions are indistinguishable from each other. The electron density and EM of \ion{H}{ii} regions is statistically different property at the 3$\sigma$ level (AD test results in \textit{p}-value $\ll$ 0.0013). We see that electron density and EM of the \ion{H}{ii} regions with the methanol maser are significantly higher than the \ion{H}{ii} regions without methanol masers. This suggests that methanol masers are associated with denser \ion{H}{ii} regions, which is not unexpected since methanol masers typically appear in the early phases of star formation, characterized by higher densities, as reported by \citet{2021A&A...645A.110Y}. Furthermore, we see that the mean values of \ion{H}{ii} region properties such as electron temperature, and Lyman photon rate of \ion{H}{ii} regions associated with methanol masers are slightly higher compared to those without maser emission.

\begin{figure}[h!]
        \centering
        \includegraphics[scale=0.8]{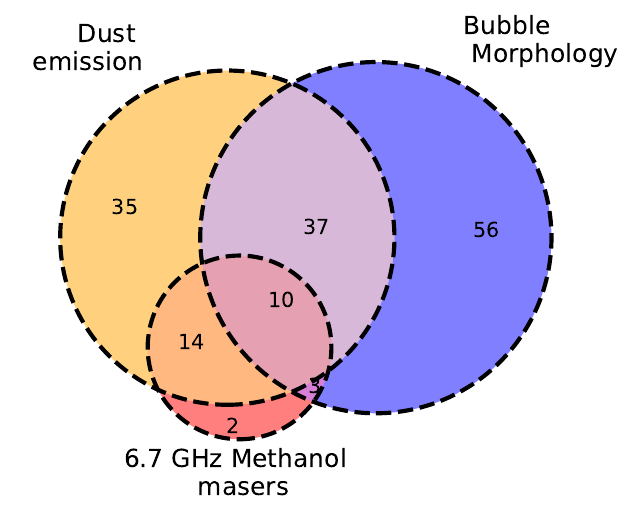}
        \caption{Venn diagram showing the correlation between the association of \ion{H}{ii} regions with the dust emission, 6.7~GHz methanol masers, and bubble morphology over a common Galactic region of 2$\degr \leq$ $\ell$ $\leq$ 60$\degr$, |\textit{b}|\,$\leq$\,1$\degr$.}
        \label{fig:venn}
    \end{figure}

    We give the average value of physical properties of \ion{H}{ii} regions associated with the dust emission, 6.7~GHz methanol maser and bubble morphology in MIR in Table~\ref{tab:summary}. The average values of $\rm n_e$, EM and $\rm N_{Lyc}$ are higher in the early stage of evolution of \ion{H}{ii} regions (methanol masers and submillimeter loud) compared to the evolved stage (submillimeter quiet), which is consistent with the standard picture. We did not notice prominent trend in the average value of $\rm T_e$ and ionized gas mass. We show a Venn diagram of the correlation between the association of \ion{H}{ii} regions with the dust emission, 6.7~GHz methanol masers, and bubble morphology over a common Galactic region of 2$\degr \leq$ $\ell$ $\leq$ 60$\degr$, |\textit{b}| $\leq$ 1$\degr$ in Fig.\ref{fig:venn}. There is a strong correlation between the \ion{H}{ii} regions associated with the dust emission and 6.7~GHz methanol masers. 
        \begin{table*}[]
            \centering
            \caption{Averaged physical properties of \ion{H}{ii} regions associated with the dust emission, 6.7~GHz methanol masers, and bubble morphology over a common Galactic region of 2$\degr \leq$ $\ell$ $\leq$ 60$\degr$, |\textit{b}| $\leq$ 1$\degr$.}
            \label{tab:summary}
            \begin{tabular}{cccccc}
            \hline \hline
            \ion{H}{ii} region association & $\rm T_e$& EM &$\rm n_e$&$\rm N_{Lyc}$&$\rm Mass_{\ion{H}{ii}}$   \\
            & [$\rm 10^3 $K]&[$\rm 10^5~pc~cm^{-6}$]&[$\rm 10^2~cm^{-3}$]&[$\rm 10^{48}~s^{-1}$]&[$\rm M_\sun$]    \\ \\
        \hline \hline 
            methanol maser &7.3$\pm$0.7 & 5.2$\pm$0.4&5.2$\pm$0.2 &10.0$\pm$0.4 &28.5$\pm$1.1 \\
            submillimeter loud & 7.3$\pm$0.8 & 5.1$\pm$0.4 &4.9$\pm$0.2 &7.9$\pm$0.3 &33.4$\pm$2.0 \\
            submillimeter quiet & 6.2$\pm$0.9& 2.8$\pm$0.2 &3.2$\pm$0.2 &5.5$\pm$0.3 &31.1$\pm$2.2 \\
            Bubble &6.5$\pm$0.9& 3.0$\pm$0.2&3.1$\pm$0.1&6.7$\pm$0.3 &38.0$\pm$2.3 \\ \hline
            Average &6.6$\pm$0.9 &3.6$\pm$0.3 &3.8$\pm$0.2 &6.3$\pm$0.3 &32.0$\pm$2.1 \\
        \hline \hline
            \end{tabular}
        \end{table*}    
        
        \subsection{Galactic electron temperature gradient}
    \cite{1975A&A....38..451C} were the first to investigate the correlation between the electron temperatures of \ion{H}{ii} regions and their distance from the center of the Galaxy (R$\rm _{Gal}$). Subsequent studies, such as~\citet{1978A&A....70..719C}, \citet{1983MNRAS.204...53S}, \citet{1983A&A...127..211W}, and \citet{2006ApJ...653.1226Q}, confirmed this gradient in the electron temperature across the Galaxy, namely that T$_e$ is lower in the Galactic center and increases with R$_{Gal}$. However, there are still uncertainties regarding the exact magnitude of this gradient and the possibility of its variations with Galactic azimuth. 
    
    In this study, we determine the electron temperatures of \ion{H}{ii} regions for a large sample of \ion{H}{ii} regions located in the first quadrant of the Galactic disk. Our \ion{H}{ii} region sample consists of 244 \ion{H}{ii} regions out of which 29 are located close to the Galactic center region ($-$2$\degr \leq$ $\ell$ $\leq$ 2$\degr$, |\textit{b}| $\leq$ 1$\degr$). Due to lack of reliable distance estimates for these \ion{H}{ii} regions (see Sect.~\ref{sec:distance}), we focused on 215 sources to investigate the gradient across the Galactic plane with R$\rm _{Gal}$ ranging from 1.5 to 9.1~kpc. Within this subset, two \ion{H}{ii} regions stand out as anomalies, deviating from the overall trend in T$_e$, exhibiting significantly lower temperatures: G025.479$-$0.174 (with a temperature of 1010~K and located at a distance of 4.95~kpc) and G030.870$-$0.099 (with a temperature of 1441~K and located at a distance of 4.85~kpc).
    
\begin{figure*}[h!]
        \centering
        \includegraphics[scale=0.5]{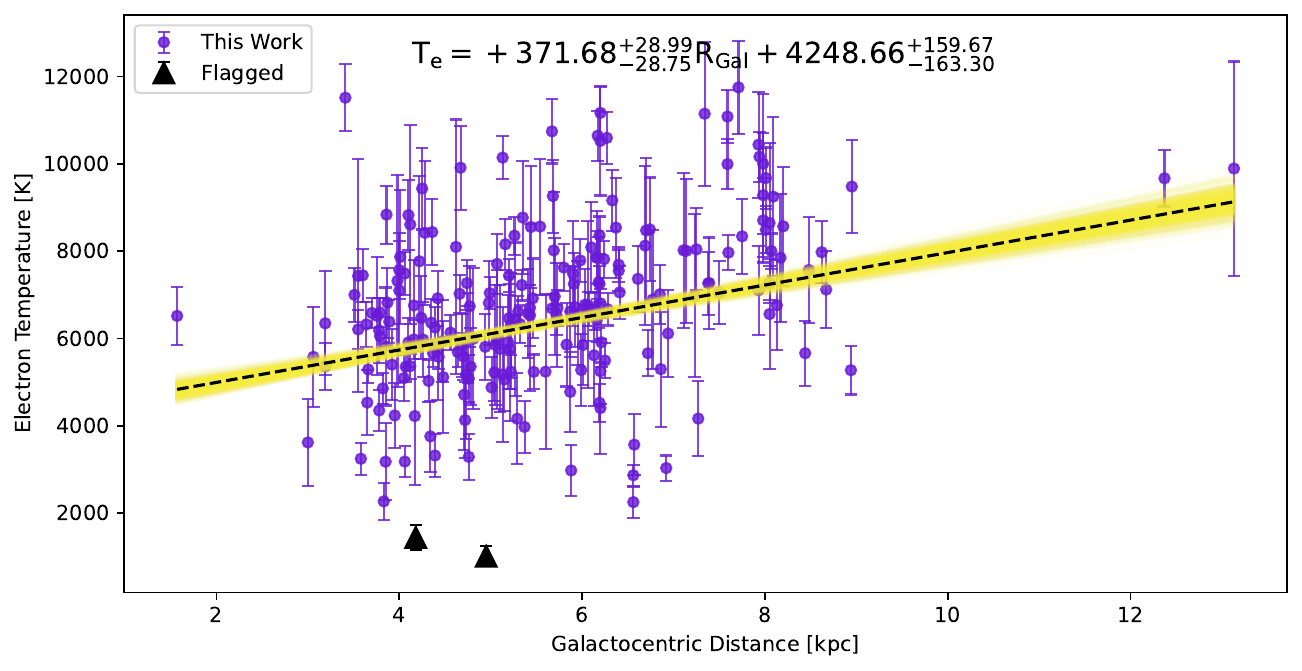}
        \includegraphics[scale=0.42]{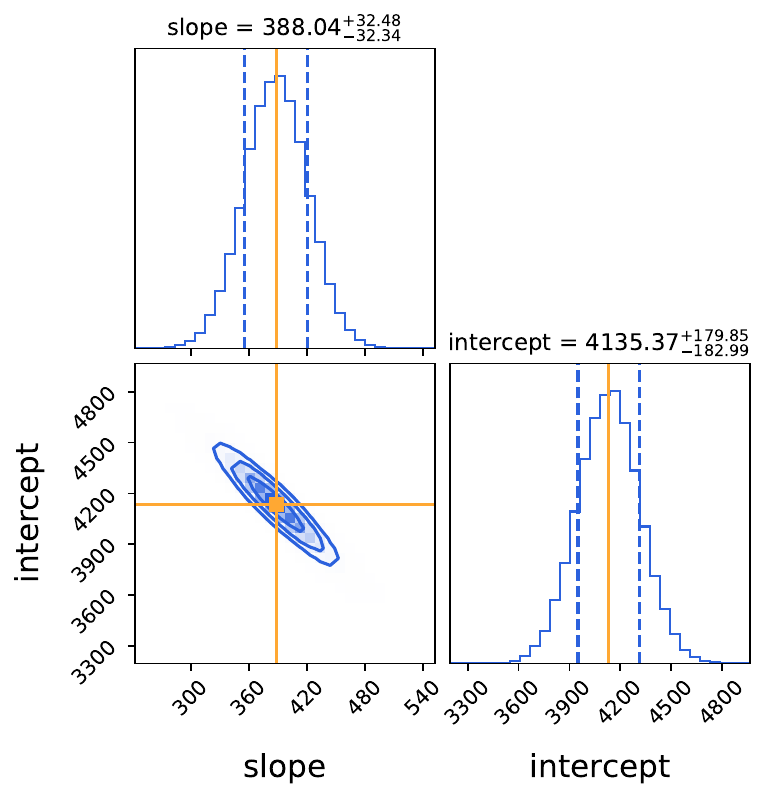}        
        \includegraphics[scale=0.5]{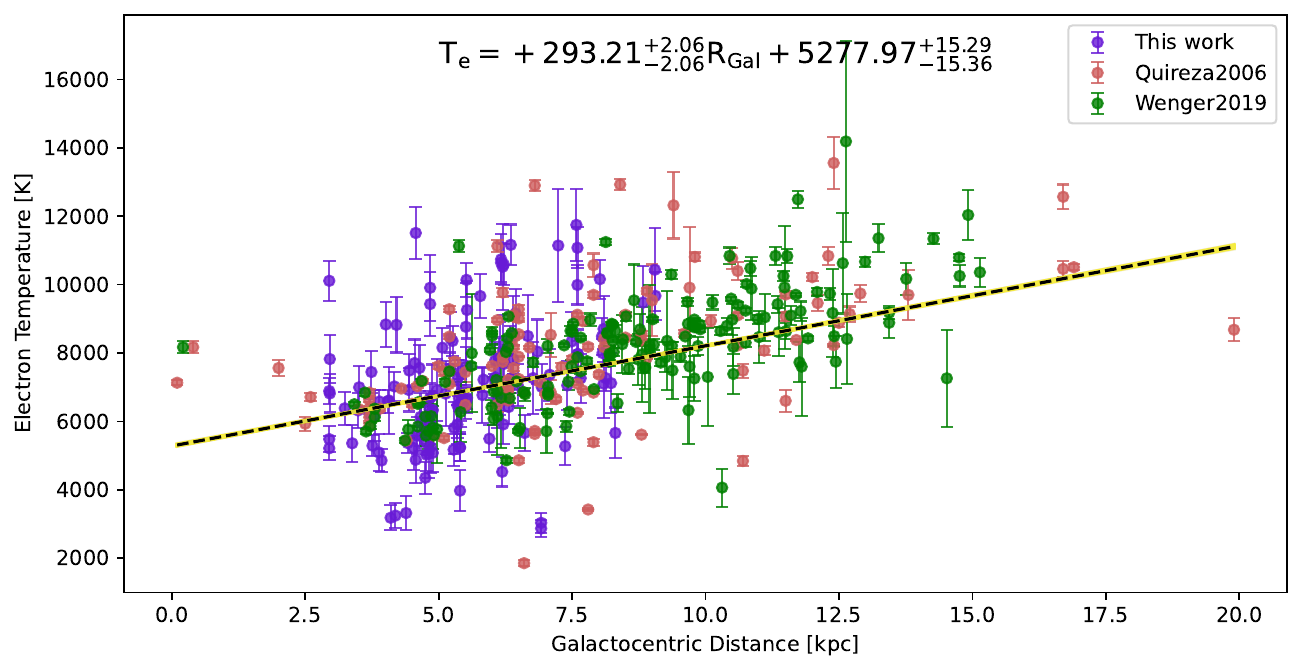}
        \includegraphics[scale=0.42]{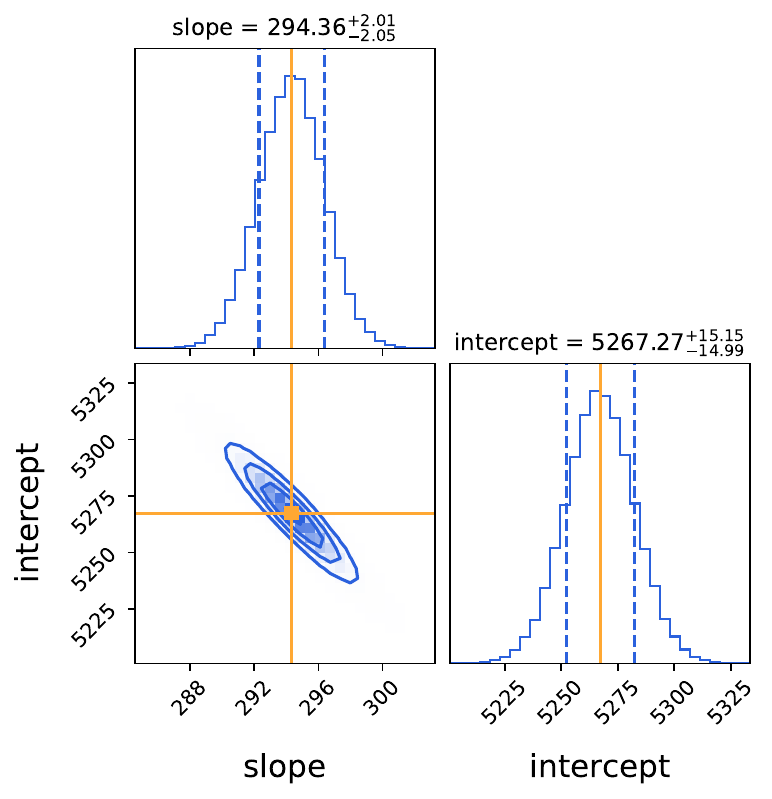}
        \caption{Most likely electron temperature gradient determined using the Bayesian linear regression with the MCMC sampling for different samples. The left panel shows the data and the most likely linear model in the dashed black line. The error bars are 1$\sigma$ uncertainty in the derived electron temperature. The shaded yellow region represents the range of ﬁts from 10000 MCMC iterations. The black triangles represent the outlier data points which are flagged before the fitting, while the filled purple circles represent the data used for fitting. The right panel shows the corresponding corner plot displaying the one-dimensional (1D) histogram of the posterior distribution for each parameter (slope and y-intercept). The histograms represent the probability density functions (PDFs) of the Monte Carlo fit parameters, while the blue curves represent the scatter plot of the joint distribution between the two parameters. The solid orange lines indicate the peaks of the PDFs, representing the most likely fit parameters, and the dotted blue lines correspond to the 1$\sigma$ confidence intervals. (\textit{Top panel}) Case 1: Consists of all 213 sources in our catalog excluding the Galactic center \ion{H}{ii} due to lack of reliable distance estimate. (\textit{Bottom panel}) Case 2: Consists of 496 sources located within R$\rm _{Gal}$ range of 0.1 to 20.0~kpc from \citet{2006ApJ...653.1226Q}, \citet{2019ApJ...887..114W}, and our catalog. The electron temperature gradient for each sample is given in the plots.}
        \label{fig:gredient_te}
    \end{figure*}

    To analyze the gradient, we derived the Galactic electron temperature gradient for two different cases. The first subset consisted of all 213 sources from this work, while the second subset included 496 sources out of these 213, 116, and 167 \ion{H}{ii} regions from this work, \citet{2006ApJ...653.1226Q} and \citet{2019ApJ...887..114W}, respectively. We performed the Bayesian linear regression with Markov chain Monte Carlo (MCMC) sampling separately for both cases. We utilized \textit{scipy.stats.linregress}\footnote{\url{https://docs.scipy.org/doc/scipy/reference/generated/scipy.stats.linregress.html}} to model the temperature gradient as a linear function, represented by T$_e$ = a$_1$ + a$_2$ R$\rm _{Gal}$~K. To implement the MCMC, we used the python package \textit{emcee}\footnote{\url{https://emcee.readthedocs.io/en/stable/}}~\citep{2013PASP..125..306F}, which allowed us to estimate the posterior probability distribution of the model parameters (slope and intercept). To visualize the results, we generated a corner plot using the python package \textit{corner.py}\footnote{\url{https://corner.readthedocs.io/en/latest/}}~\citep{2016JOSS....1...24F}. In Fig.~\ref{fig:gredient_te}, we present the fit results along with the corner plot. Each histogram in the corner plot represents the posterior distribution of a parameter, with the width of the histogram indicating the uncertainty associated with that parameter. Additionally, the scatter plot in the corner plot displays the joint distribution between two parameters.  We present the gradient fit obtained from the first sample, which comprised 213 data points with R$\rm _{Gal}$ spanning from 1.6 to 13.1~kpc. The gradient fit for this sample was determined to be T$_e$ = (372$\pm$28) R$\rm _{Gal}$ + (4248$\pm$161)~K. The gradient fit for the second sample was determined to be T$_e$ = (293$\pm$2) R$\rm _{Gal}$ + (5278$\pm$15)~K within R$\rm _{Gal}$ range of 0.1 to 20.0~kpc. The shallower slope for the latter case might be due to the exclusion of the Galactic center \ion{H}{ii} regions and a much wider range of R$\rm _{Gal}$.
    
    Upon comparing our study with previous research, we found that our derived electron temperature gradient aligns well with the reported values. \citet{2006ApJ...653.1226Q} obtained a gradient of 287 $\pm$ 46~K~kpc$\rm ^{-1}$ using 76 Galactic \ion{H}{ii} regions within the range $\rm R_{Gal}$ = 0.0$-$17.0~kpc. After excluding the Galactic center sources, they found the slope to be 340 $\pm$ 45~K~kpc$\rm ^{-1}$, using the 73 Galactic \ion{H}{ii} regions located in the range, $\rm R_{Gal}$ = 2.0$-$17.0~kpc. \citet{2019ApJ...887..114W} used Monte Carlo analysis to derive the most likely Galactic radial electron temperature gradient as 359$\rm ^{+22}_{-18}$~K~kpc$\rm ^{-1}$. \citet{2011ApJ...738...27B} reported a gradient of 299 $\pm$ 33~K~kpc$\rm ^{-1}$ using 133 sources within the range $\rm R_{Gal}$ = 0.1$-$21.9~kpc. These electron temperature gradients vary around 300~K~kpc$\rm ^{-1}$. However, a few previous studies reported very steep gradients, such as 433 $\pm$ 40~K~kpc$\rm ^{-1}$ by \citet{1983MNRAS.204...53S} with a sample of N=67 Galactic \ion{H}{ii} regions in the range R$\rm _{Gal}$ = 3.5$-$13.7~kpc, 440 $\pm$ 50~K~kpc$\rm ^{-1}$ (N=23, R$\rm _{Gal}$ = 3.6$-$12.6~kpc) by \citet{1983ApJ...266..263G}, and 372 $\pm$ 38~K~kpc$\rm ^{-1}$ (N=6, R$\rm _{Gal}$ = 6.6$-$14.8~kpc) by \citet{2000MNRAS.311..329D}.
    
    We observe a scatter in the \ion{H}{ii} region electron temperature at a given Galactocentric radius. One explanation could be the presence of systematic uncertainties in determining the electron temperature. However, in our case, we do not anticipate any systematic errors in estimating the electron temperature. This is because the VLA enables simultaneous measurement of both radio-continuum and RRL emissions, and the RRL-to-continuum flux ratio enables us to mitigate any systematic calibration or weather-related concerns. Another major factor that might contribute to the fluctuation in the $\rm T_e$ gradient is the variation in metallicity as reported by previous studies such as \citet{2006ApJ...653.1226Q}, \citet{2011ApJ...738...27B}, \citet{2019ApJ...887..114W}. The electron temperature is an outcome of heating by radiation from the massive stars and cooling by fine structure lines. With change in metallicity, the cooling function will change significantly giving a different equilibrium temperature \citep{1983MNRAS.204...53S}.

        \section{Summary} \label{sec:conclusion}
        In this study, we conducted an unbiased survey of Galactic \ion{H}{ii} regions in the  Galactic plane ($-$2$\degr \leq$ $\ell$ $\leq$ 60$\degr$ \& |\textit{b}| $\leq$ 1$\degr$ and 76$\degr \leq$ $\ell$ $\leq$ 83$\degr$ \& $-$1$\degr \leq$ \textit{b} $\leq$ 2$\degr$) that are covered by GLOSTAR. By utilizing the RRL data obtained with the VLA in D-configuration with an angular resolution of 25\arcsec and a typical 1~$\sigma$ noise level of $\sim \rm 3.0~mJy~beam^{-1}$ at a velocity resolution of $\sim \rm 5~km~s^{-1}$, we compiled a catalog of 244 Galactic \ion{H}{ii} regions, along with their physical properties. The main findings are as follows:
        \begin{itemize}
        \item The mean value of the RRL width is $\sim$ 28~km~s$^{-1}$, which is > 25~km~s$^{-1}$. Compared to previous \ion{H}{ii} region surveys, our samples exhibit comparable, if slightly broader, RRL widths. We identified 23 sources with broad line widths  $>$35~km~s$^{-1}$, of which 15 are associated with candidate HC \ion{H}{ii} regions.

        \item 48\% of the GLOSTAR \ion{H}{ii} regions show a bubble-like morphology in GLIMPSE 8.0~$\mu$m images. AD tests reveal no statistically significant differences in most of the \ion{H}{II} region properties between the bubble and non-bubble \ion{H}{ii} regions, except for a wider RRL width and a higher electron density in non-bubble HII regions that shows a significant difference at only 2$\sigma$.

        \item Comparing the \ion{H}{ii} regions with the ATLASGAL Compact Source Catalog (CSC), we find that 48\% of them are associated with the ATLASGAL dust clumps, which we term submillimeter loud \ion{H}{ii} regions. The comparison of velocities derived from RRLs and molecular lines reveals an excellent agreement between the ionized gas and molecular gas.

        \item AD tests on the CDFs of submillimeter loud \ion{H}{ii} regions and submillimeter quiet \ion{H}{ii} regions reveal that, apart from electron density, EM, and electron temperature, no other ionized gas properties change as an \ion{H}{ii} region evolves from a dust-associated \ion{H}{ii} region to a dust-quiet one.

        \item Upon crossmatching with the GLOSTAR methanol maser catalog, we found that only 15\% of \ion{H}{ii} regions in our sample were associated with methanol maser emissions. Based on the AD test, we found that \ion{H}{ii} regions with methanol masers show a significantly higher electron density and EM compared to \ion{H}{ii} regions without methanol masers.

        \item Based on a sample of 213 \ion{H}{ii} regions within R$\rm _{Gal}$ spanning from 1.6 to 13.1~kpc, our best estimate of the electron temperature gradient in the Galactic disk was determined to be 372$\pm$28~K~kpc$^\text{-1}$. This finding is consistent with previous studies of the electron temperature gradient.
        
        \end{itemize}   
        
    This study represents an initial step in investigating the relationship between ionized gas of \ion{H}{ii} regions and their associated natal dust clumps. In forthcoming research, our intention is to determine the physical characteristics and dynamics of the dense molecular environments surrounding these objects. To achieve this, we have conducted observations of multiple para-$\rm H_2CO$ transition lines with the APEX telescope. This aims to ascertain the kinetic temperatures and spatial densities of the denser \ion{H}{ii} regions through a non-LTE model and compare the pressure of the molecular and ionized gases to assess if these \ion{H}{ii} regions are approximately in pressure equilibrium. Moreover, our future investigations will involve a comparison of the nonthermal motions of molecular and ionized gases to examine whether turbulence in the ionized phase is inherited from the molecular phase.

        \begin{acknowledgements} 
        This research was partially funded by the ERC Advanced Investigator Grant GLOSTAR (247078). The National Radio Astronomy Observatory is a facility of the National Science Foundation, operated under a cooperative agreement by Associated Universities, Inc. AYY acknowledges the support from the National Key R$\&$D Program of China No. 2023YFC2206403 and National Natural Science Foundation of China (NSFC) grants No. 12303031 and No. 11988101. S.A.D. acknowledge the M2FINDERS project from the European Research Council (ERC) under the European Union's Horizon 2020 research and innovation programme (grant No 101018682).  This research made use of information from the ATLASGAL database at (\url{http://atlasgal.mpifr-bonn.mpg.de/cgi-bin/ATLASGAL_DATABASE.cgi}) supported by the MPIfR, Bonn. This work has used data from the GLIMPSE and MIPSGAL surveys of the \textit{Spitzer} Space Telescope, which was operated by the Jet Propulsion Laboratory (JPL), California Institute of Technology under the contract with NASA. This research has made use of the SIMBAD database and the VizieR catalog, operated at CDS, Strasbourg, France. This research made use of Astropy (\url{http://www.astropy.org}), a community developed core Python package for Astronomy~\citep{2013A&A...558A..33A,2018AJ....156..123A}. This document was prepared using the collaborative tool Overleaf available at:\url{https://www.overleaf.com/}. 
        \end{acknowledgements}

        \bibliographystyle{aa.bst} 
        \bibliography{mybib.bib} 
        
        \begin{appendix}
    \section{Distance determination}\label{sec:app_dist}
     To investigate the physical properties of \ion{H}{II} regions, it is crucial to know their distances. The most reliable way to calculate distances involves measuring the parallax of bright, compact objects like methanol and water masers. However, this method is very time-consuming and requires complex measurements. Additionally, because it depends on the presence of bright and compact sources, which are not always available, especially in most \ion{H}{II} regions in our sample, it cannot be applied to all of them. The Kinematic distance, though less reliable, is easier to calculate. One can determine it using the source's radial velocity relative to the local standard of rest (V$\rm _{lsr}$), obtained from molecular line observations \citep[][]{2012MNRAS.420.1656U, 2018MNRAS.473.1059U} or RRL observations \citep[][]{2003ApJ...582..756K, 2012ApJ...754...62A}, along with a model for the Galaxy's rotation curve. Numerous models for the rotation of the Milky Way Galaxy exist \citep[][]{1985ApJ...295..422C, 1993A&A...275...67B, 2014ApJ...783..130R}. We used the rotation curve determined by \citet{2014ApJ...783..130R} to estimate the kinematic distance, as it leverages all the maser parallax measurements to constrain the model. This approach has proven effective in providing kinematic distances that compare well with the maser distances. While the source velocity measurement and distance determination in the Outer Galaxy is relatively simple, the problem arises for the sources within the solar circle \citep[R$\rm _G$ < 8.34;][]{2014ApJ...783..130R}. For sources within the solar circle there are two solutions for the kinematic distance for every radial velocity and line of sight. These radial distances are equally spaced on either side of the tangent point, one on the near side and other on the far side. This effect is known as the kinematic distance ambiguity (KDA). To assign a unique distance to a particular source, the KDA must be resolved.

     \citet{2016ApJ...823...77R, 2019ApJ...885..131R} developed a Bayesian maximum likelihood method that resolves the KDA by considering the relative position of the spiral arms along the line of sight and the latitude of the sources. However, this method is biased toward placing the sources in spiral arms, whereas \ion{H}{ii} regions may well be located also in interarm regions \citep{2020A&A...633A..14R}. Therefore, we resolved the KDA using the \ion{H}{I} E/A method (discussed in the following subsection), and for sources where the KDA remained unresolved, we adopted the Bayesian maximum likelihood distance. Fig.~\ref{fig:final_dist} shows the relationship between the kinematic distances and Bayesian maximum likelihood distances, with most sources having comparable distances by both methods.

    \subsection{Resolving the kinematic distance ambiguity}
    To determine the distances to GLOSTAR \ion{H}{II} regions, we began by matching the sources to reliable distances reported in the literature. We used the following studies in preferential order based on reliability of reported distances : parallax distances~\citep{2019ApJ...885..131R}, kinematic distances of the HRDS \ion{H}{II} regions~\citep{2012ApJ...754...62A}, kinematic distances to known \ion{H}{II} regions~\citep{2009ApJ...690..706A}, and kinematic distances to the ATLASGAL clumps~\citep{2018MNRAS.473.1059U}, as these studies already went through steps removing the distance ambiguity. We compared the longitude, latitude, and velocities of our sources with the positions in these studies and adopted the distance where a correlation was found. Based on this, we adopted parallax distance~\citep{2019ApJ...885..131R} for 28 GLOSTAR \ion{H}{II} regions and kinematic distances~\citep{2009ApJ...690..706A, 2012ApJ...754...62A, 2018MNRAS.473.1059U} for 121 GLOSTAR \ion{H}{II} regions. For 13 GLOSTAR \ion{H}{II} regions located toward Cygnus~X, we adopted a distance of 1.4~kpc. We did not determine distances for 29 GLOSTAR \ion{H}{II} regions toward the Galactic center (358$\degr \leq$ $\ell$ $\leq$ 3$\degr$) because kinematic distances in this region are highly unreliable. For the rest of 53 GLOSTAR \ion{H}{II} regions we resolve the KDA.

    Several methods are discussed in the literature that can be used to resolve the KDA for Galactic \ion{H}{II} regions. In this study, we used the \ion{H}{I} E/A method, which has proven effective for resolving the KDA for \ion{H}{II} region~\citep{2003ApJ...582..756K, 2009ApJ...690..706A, 2012MNRAS.420.1656U, 2012ApJ...754...62A}. \ion{H}{II} regions emit broadband continuum free-free thermal radiation in the cm wavelength range. Neutral \ion{H}{I} gas between the \ion{H}{II} region and the observer will absorb this thermal emission if the brightness temperature of the \ion{H}{I} gas is lower than that of the \ion{H}{II} region at 21 cm. In the first quadrant, the LSR velocity increases with distance from the Sun, reaches its maximum value at the tangent point, and then decreases with further distance. Thus, \ion{H}{II} regions at the near distance will show \ion{H}{I} absorption up to their recombination line velocity. Conversely, \ion{H}{II} regions at the far distance will show \ion{H}{I} absorption from the foreground \ion{H}{I} clouds at velocities up to the tangent point velocity. The \ion{H}{I} E/A method relies on \ion{H}{I} absorption between the RRL and tangent point velocity to distinguish between the near and far distances.

    For \ion{H}{II} regions which do not show absorption features above threshold, we cannot use the \ion{H}{I} E/A method to resolve the KDA. For such sources we used \ion{H}{I} self-absorption method \citep[eg.,][]{2009ApJ...699.1153R, 2009ApJ...690..706A, 2018MNRAS.473.1059U} but these distances are much less certain and reliable~\citet{2009ApJ...690..706A}. If even the \ion{H}{I} self-absorption method is not conclusive we adopted the Bayesian maximum likelihood distance.

    We adopted the methodology explained by \citet{2012MNRAS.420.1656U, 2003ApJ...582..756K} for resolving the KDA of \ion{H}{II} regions and are briefly explained below:
    \begin{enumerate}
        \item We have extracted continuum included \ion{H}{i} spectra from the Southern Galactic Plane Survey \citep[SGPS;][]{2005ApJS..158..178M} and the HI/OH/Recombination line survey of the inner Milky Way \citep[THOR; ][]{ 2016A&A...588A..97B, 2018A&A...619A.124W}.
        \item We check the 21~cm continuum association of the GLOSTAR \ion{H}{ii} regions, by extracting the 21~cm continuum emission map from the SGPS and the THOR survey and search for positional coincidence. Subsequently, we extracted the continuum-included \ion{H}{i} spectra and spatially integrated over the GLOSTAR radio emission region to obtain a source-averaged \ion{H}{i} spectrum toward 53 GLOSTAR \ion{H}{ii} regions.
        \item As was described by \citet{2009ApJ...690..706A}, the background contribution was estimated by averaging the emission within four regions as near as possible to the 21~cm source and avoiding any other nearby continuum emission. These regions around the target are selected in such a way that it compensates for any gradient that may be present in the background emission. The final spectra were produced by subtracting the background emission from the source-averaged continuum-included \ion{H}{i} spectra. In Fig.~\ref{fig:H_spectra}, we present a sample of the background-subtracted \ion{H}{i} spectra obtained toward a GLOSTAR \ion{H}{ii} region.
        \item Before determining the reliability of any absorption feature, we first estimated two important sources of noise in the extracted continuum-included \ion{H}{i} spectra; these are the receiver noise ($\sigma_{rms}$) and \ion{H}{i} emission fluctuations. We have estimated the receiver noise by calculating the standard deviation using the absorption-free channels. The second source of noise is due to \ion{H}{i} emission fluctuation. These are present over all angular scales and can result in positive and negative wiggles in observed spectra which creates confusion in the identification of the absorption feature. The \ion{H}{i} emission fluctuation was estimated by calculating the standard deviation of the off-source spectra as a function of velocity. The receiver noise and \ion{H}{i} emission fluctuation level for each spectrum were presented by gray and purple horizontal lines, respectively in Fig.~\ref{fig:H_spectra}. We consider a reliable absorption feature if the depth of the absorption is more than four times the receiver noise and larger than the background emission fluctuation present in the spectrum. 
        \item \citet{2003ApJ...582..756K} assigned the near, far, or tangent-point kinematic distance designations for each source by calculating the velocity differences $\rm V_T - V_{lsr}$ and $\rm V_T - V_A$. They found out that there are two distinct groups of sources for which $\rm V_T - V_{lsr} > 0$. One has $\rm V_T - V_A$ close to 0 and is located at far distance. The other group has $\rm V_T -V_A$ significantly greater than 0 and are located at near distance. If any source has $\rm V_T - V_{lsr} < 0$, it must be at the tangent point. Using \citet{2003ApJ...582..756K} method, we need only three velocity measurements to resolve the KDA. We consider RRL peak velocity as source velocity. To determine the velocity of tangent point, we used the empirical relation between \ion{H}{i} termination velocities and Galactic longitude derived by \citet{2005ApJS..158..178M, 2016ApJ...831..124M}. Finally, we measure the maximum velocity of absorption for Northern Galactic Plane, which is obtained by measuring the maximum velocity where the absorption dip is larger than four times the receiver noise (standard deviation of the absorption free channels of the \ion{H}{i} spectrum) and larger than the value of the \ion{H}{i} emission fluctuations. In Table~\ref{tab:dist} we presented these velocities for each \ion{H}{ii} region. Fig.~\ref{fig:H_spectra} shows the plots of the background subtracted continuum-included \ion{H}{i} spectrum toward sources where significant \ion{H}{i} absorption is seen. On these plots the velocity of tangent point, source velocity and maximum velocity of absorption are indicated by dashed vertical red, blue, and green lines, respectively. The dashed horizontal gray and purple line shows the 4$\sigma_{rms}$ threshold level determined from the receiver noise and \ion{H}{i} emission fluctuation, respectively. The gray vertical shaded region around tangent point velocity shows the possible deviation due to streaming motions ($\pm$10~km~s$\rm ^{-1}$) and the vertical yellow shaded regions around the source velocity represent the FWHM of RRL.
        \begin{figure}[h!]
            \centering
            \includegraphics[width=0.5\textwidth]{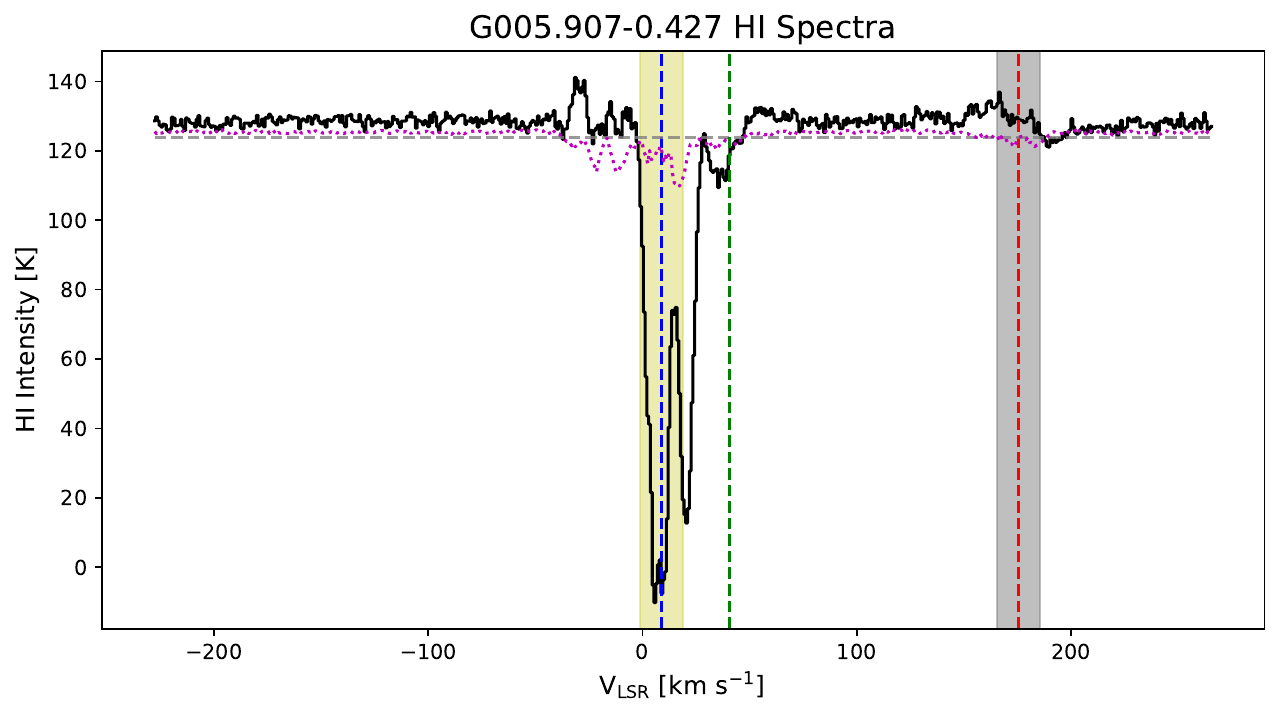}
            \includegraphics[width=0.5\textwidth]{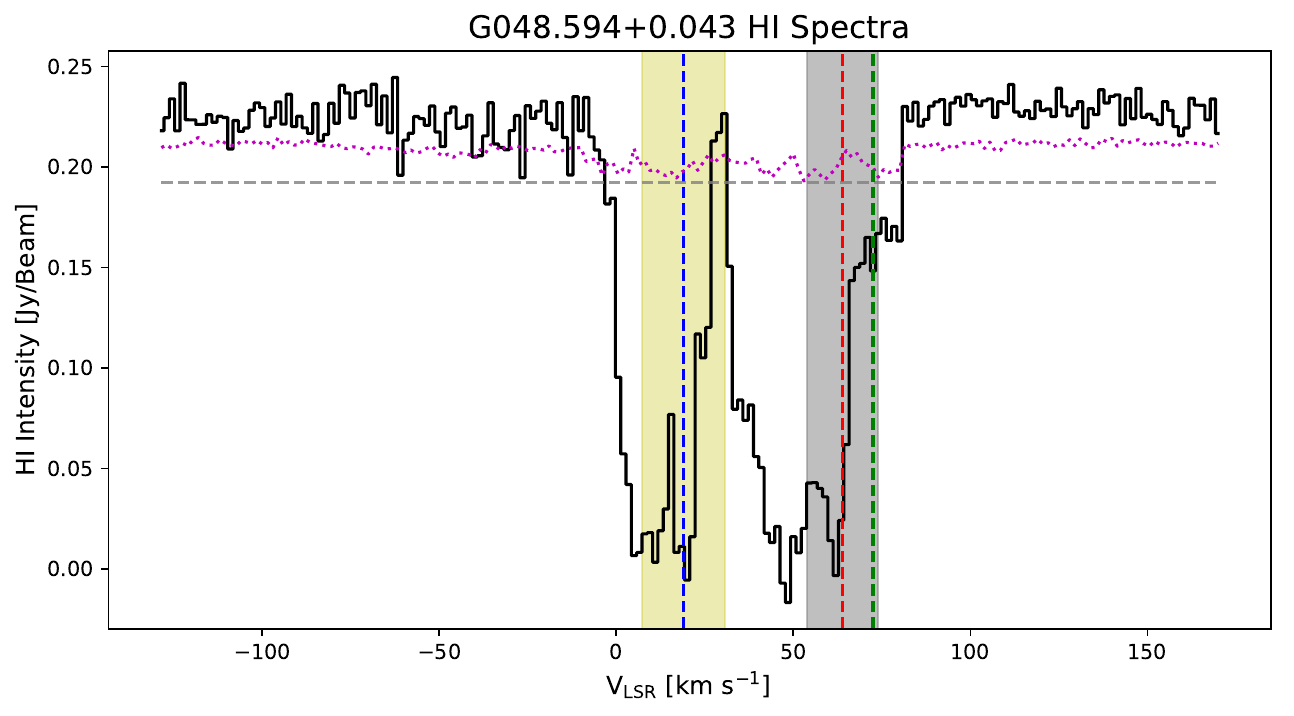}      
            \caption{Source-averaged continuum-included \ion{H}{i} spectra toward the \ion{H}{ii} regions extracted from the \ion{H}{i} SGPS (top panel) and THOR survey (bottom panel). The source velocity ($\rm V_{lsr}$), the velocity of tangent point ($\rm V_T$), and the velocity of the first absorption minimum ($\rm V_A$) are shown by the vertical dashed blue, red, and green lines, respectively. The yellow shaded region around the source velocity represents the line width of the RRL and gray shaded region around velocity of the tangent point covers 10$\rm km~s^{-1}$ either side and is provided to give an indication of the uncertainty associated with the streaming motion. The horizontal dashed gray and dotted purple lines show the $\rm 4\sigma_{rms}$ receiver noise level and the \ion{H}{I} emission fluctuations, respectively. We provide example of sources placed at the near and far distance in the top and bottom panels, respectively.}
            \label{fig:H_spectra}
        \end{figure}  
        \item To identify the positions of near, far, and tangent point sources in velocity difference space ($\rm V_T - V_{lsr}$ vs $\rm V_T - V_A$), \citet{2003ApJ...582..756K} conducted a simulation with 10000 randomly placed \ion{H}{ii} regions, with Galactocentric radii between 3 and 8~kpc and Galactic longitudes between 18$\degr$ and 50$\degr$. This simulation allowed them to approximate the boundaries in the $\rm V_T - V_{lsr}$ vs $\rm V_T - V_A$ plot that correspond to high-quality KDA resolutions, as depicted in Fig.~\ref{fig:final_dist}. In Fig.~\ref{fig:final_dist}, the diagonal and horizontal shaded regions show the expected location in the velocity parameter space of the source at the near and far distance, respectively. The dark gray region represents the location of 90 percent of the simulated sources were located, and lighter shaded region shows a 10~km~s$\rm ^{-1}$ extension to the simulated data. The darker triangular region in the lower left quadrant of the plot indicates the overlapping parameter space where distance assignments are inherently more uncertain. The dashed diagonal line divides this triangle into two regions: sources above the line are more likely to be at a near distance, while those below are more likely to be at a far distance.   
        \item In Fig.~\ref{fig:final_dist}, sources which are located in the dark shaded diagonal region were assigned near distances with high degree of confidence (``n''; Quality = A), where source located in lighter shaded diagonal region assigned a near distance with low degree of confidence (``n?''; Quality = B). With this we found out of 53 GLOSTAR \ion{H}{ii} region 22 are located at near distance. 
        \item In Fig.~\ref{fig:final_dist}, sources which are located in the dark shaded horizontal region assigned a far distance with the high degree of confidence (``f''; Quality = A), where source located in lighter shaded horizontal region assigned a far distance with low degree of confidence (``f?''; Quality = B). With this we found out of 53 GLOSTAR \ion{H}{ii} region 15 are located at far distance.
        \item For the sources which are located close to the tangent point velocity ($\rm |V_{lsr} - V_T| < 10~km~s^{-1}$), we assigned a tangent point distance to the source with high degree of confidence (``T''). If the source is located in the lower triangular region of Fig.~\ref{fig:final_dist}, we assigned a tangent point distance to the source with low degree of confidence (``T?''). With this we found out of 53 GLOSTAR \ion{H}{ii} region 11 are located at tangent point distance.
        \item For the sources where the absorption dip is not larger than the threshold and we were not able to estimate the $\rm V_A$, we used \ion{H}{i} self-absorption method to resolve the KDA. However for these sources we assigned low degree of confidence (``n?'' or ``f?''). For 5 sources we were not able to resolve the KDA, for these sources we adopted the Bayesian maximum likelihood distance and assigned the lower degree of confidence (``?'', Quality = C).
        \begin{figure}[h!]
            \centering
            \includegraphics[width=0.51\textwidth]{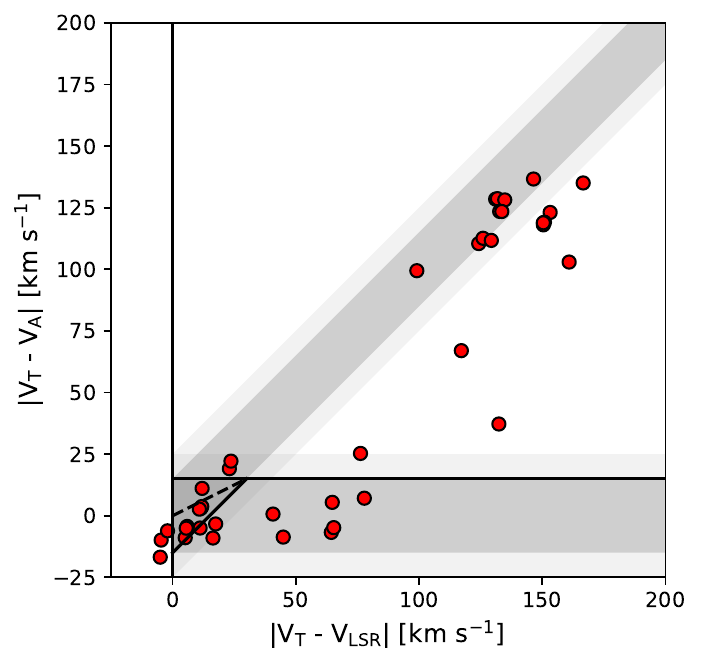}
            \includegraphics[width=0.5\textwidth]{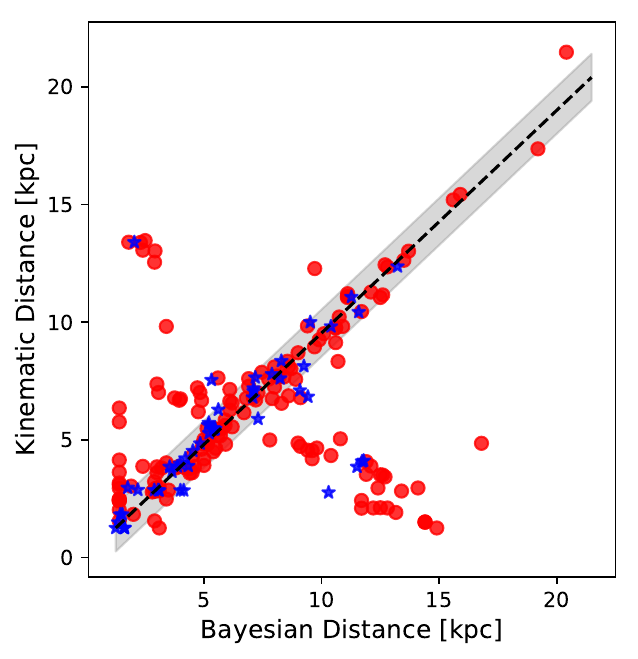}        
            \caption{Top panel: Plot of the difference in source velocity ($\rm V_{lsr}$) and the velocity of tangent point ($\rm V_T$), versus the difference in the velocity of the first absorption minimum ($\rm V_A$) and the $\rm V_T$. The diagonal and horizontal shaded regions represent the expected location of the near and far distance sources, respectively, while the solid line shows the approximate boundaries \citep{2003ApJ...582..756K, 2012MNRAS.420.1656U}. The dashed diagonal line divides the darker triangular region in the lower left quadrant of the plot into two regions: sources above the line are more likely to be at a near distance, while those below are more likely to be at a far distance. Bottom panel: Plot shows the relation between the distance determined using the Bayesian method presented by \citet{2016ApJ...823...77R, 2019ApJ...885..131R} and the near/far distance determined using the \citet{2014ApJ...783..130R} rotation curve. The red circles and blue stars represent the distances taken from literature and obtained distance for 53 \ion{H}{ii} regions, respectively. The solid black line shows the line of equality, and the gray shaded region represents $\pm$1~kpc. }
            \label{fig:final_dist}
        \end{figure} 
    \end{enumerate}
    We present the source names, radial velocities (V$\rm _{lsr}$), velocity of the tangent point (V$\rm _{T}$), velocity of the highest velocity of absorption (V$\rm _{A}$), the Bayesian distance \citep{2019ApJ...885..131R},  near and far distances determined from the rotation curve \citep{2014ApJ...783..130R}, the kinematic solution, the assigned distance to the sources and the respective uncertainty, and quality of the solution in Table~\ref{tab:dist}.

    \section{Uncertainties in the derived parameters}
    The median uncertainty on the integrated continuum flux and $\langle$Hn$\alpha \rangle$ amplitude estimation is  $\sim$ 5\% and 8\%, respectively, whereas the median uncertainty on the FWHM of the RRL is $\sim$ 9\%. The uncertainty in the electron temperature is determined by combining the uncertainties in the continuum flux, $\langle$Hn$\alpha \rangle$ amplitude, and RRL width, in quadrature. This results in a similar level of uncertainty in the derived electron temperature; that is, $\rm \sigma T_e/T_e = 12 \%$. 
    
    The average uncertainty in estimating distances to 53 \ion{H}{ii} regions using the \ion{H}{i} E/A method with the \citet{2014ApJ...783..130R} Galactic rotation curve is 17\%. This aligns with the findings of \citet{2018ApJ...856...52W}, who reported a median uncertainty of 28\% for kinematic distance estimates using the same rotation curve. The median uncertainty in parallax distances \citep{2019ApJ...885..131R} is 10\%, while the uncertainty for distances to HRDS \ion{H}{ii} regions \citep{2012ApJ...754...62A} is 5\%. In \citet{2018MNRAS.473.1059U} and \citet{2009ApJ...690..706A}, the typical distance uncertainty is about 10\%. Overall, the median distance uncertainty for GLOSTAR \ion{H}{ii} regions is $\sim$ 10\%. The resultant errors in the EM, electron density, Lyman continuum photon rate and ionized gas mass estimations obtained using \texttt{SOERP} python package are 8\%, 6\%, 21\%, and 26\%, respectively. For some sources, the uncertainty in the derived physical properties exceeds 50\% (e.g., G010.197-0.424), primarily due to very high uncertainty in the distance estimation.

    \section{Tables}
        \begin{sidewaystable*}  
        \centering
        \caption{GLOSTAR \ion{H}{ii} region catalog with continuum and stacked $\langle$Hn$\alpha \rangle$ hydrogen RRL parameters.}
        \label{tab:catalog}
        \resizebox{\linewidth}{!}{\begin{tabular}{ccccccccccccccccccc}
        \hline \hline \\
            Name\tablefootmark{a} & GLOSTAR Name\tablefootmark{b} &$\rm \ell_{peak}$ & \textit{b}$\rm _{peak}$ & S$\rm _{peak}$ & S$\rm _{C}$ &  $\sigma$S$\rm _{C}$ & S$\rm _{L}$ & $\sigma$S$\rm _{L}$ & $\Delta$V & $\sigma \Delta$V & V$\rm_{LSR}$ &  $\sigma$V$\rm_{LSR}$ & D$\rm _{sun}\tablefootmark{c} $ & $\sigma$D$\rm _{sun}$& R$\rm _{Gal}$ & Morphology\tablefootmark{d} & ATLASGAL Clump\tablefootmark{e}&$\rm CH_3OH$\tablefootmark{f} \\
            & & [$\degr$] &[$\degr$]& [$\rm Jy~Beam^{-1}$] & [mJy] & [mJy] & [mJy] & [mJy] & [$\rm km s^{-1}$] & [$\rm km s^{-1}$] & [$\rm km s^{-1}$] & [$\rm km s^{-1}$] & [kpc] & [kpc] &[kpc]& & & \\ \\
        \hline \hline
        .&.&.&.&.&.&.&.&.&.&.&.&.&.&.&.&.&. \\
        .&.&.&.&.&.&.&.&.&.&.&.&.&.&.&.&.&. \\
        .&.&.&.&.&.&.&.&.&.&.&.&.&.&.&.&.&. \\
        .&.&.&.&.&.&.&.&.&.&.&.&.&.&.&.&.&. \\
        G031.474-0.344 & G031.476-00.344  & 31.474 & -0.344 & 0.07 & 261.75 & 14.24 & 26.27 & 3.94 & 28.84 & 5.0 & 89.62 & 2.12 & 9.0 & 0.7 & 4.72 & IB & $\cdots$ & $\cdots$ \\
G032.059+0.080 & G032.063+00.078  & 32.059 & 0.08 & 0.03 & 157.28 & 11.34 & 21.23 & 3.23 & 23.9 & 4.19 & 98.86 & 1.78 & 7.2 & 0.0 & 4.34 & B & $\cdots$ & $\cdots$ \\
G032.150+0.133 & G032.151+00.133  & 32.15 & 0.133 & 0.51 & 681.96 & 34.22 & 46.24 & 1.99 & 26.49 & 1.32 & 97.46 & 0.56 & 6.2 & 0.62 & 4.39 & IB & AGAL032.149+00.134 & $\cdots$ \\
G032.272-0.225 & G032.272-00.226  & 32.272 & -0.225 & 0.29 & 320.89 & 16.15 & 12.57 & 1.31 & 23.59 & 2.85 & 21.34 & 1.21 & 12.8 & 0.5 & 7.34 & C & AGAL032.272-00.226 & $\cdots$ \\
G032.589+0.008 & G032.590+00.010  & 32.589 & 0.008 & 0.02 & 173.41 & 9.78 & 20.59 & 3.87 & 15.95 & 3.46 & 80.37 & 1.47 & 9.4 & 0.6 & 5.07 & PB & $\cdots$ & $\cdots$ \\
G032.797+0.191 & G032.798+00.191  & 32.797 & 0.191 & 2.71 & 3129.42 & 156.55 & 121.92 & 2.15 & 33.04 & 0.67 & 16.18 & 0.29 & 9.71 & 2.92 & 5.26 & C & AGAL032.797+00.191 & G032.8027+0.1926 \\
G032.927+0.606 & G032.928+00.606  & 32.927 & 0.606 & 0.29 & 328.51 & 16.51 & 12.7 & 2.24 & 27.42 & 5.58 & -34.95 & 2.37 & 19.2 & 1.5 & 13.13 & C & AGAL032.926+00.606 & $\cdots$ \\
G032.990+0.040 & G032.991+00.039  & 32.99 & 0.04 & 0.14 & 188.07 & 10.79 & 17.69 & 2.05 & 22.82 & 3.06 & 91.56 & 1.3 & 8.6 & 0.86 & 4.78 & B & $\cdots$ & G032.9918+0.0343 \\
G033.204-0.009 & IRAS 18494+0010  & 33.204 & -0.009 & 0.09 & 374.08 & 21.28 & 31.7 & 3.51 & 22.51 & 2.87 & 107.49 & 1.22 & 7.1 & 0.71 & 4.47 & IB & AGAL033.206-00.009 & G033.2037-0.0101 \\
G033.419-0.004 & G033.418-00.004  & 33.419 & -0.004 & 0.13 & 196.04 & 10.33 & 10.93 & 1.4 & 34.3 & 5.07 & 63.49 & 2.15 & 9.4 & 0.94 & 5.19 & C & $\cdots$ & $\cdots$ \\
G033.810-0.189 & G033.810-00.188  & 33.81 & -0.189 & 0.22 & 309.83 & 15.68 & 18.57 & 2.59 & 21.13 & 3.4 & 39.6 & 1.45 & 11.7 & 1.17 & 6.7 & PB & AGAL033.811-00.187 & $\cdots$ \\
G033.915+0.110 & G033.914+00.110  & 33.915 & 0.11 & 0.82 & 991.23 & 49.67 & 52.58 & 1.86 & 34.62 & 1.42 & 100.02 & 0.6 & 7.1 & 0.71 & 4.56 & IB & AGAL033.914+00.109 & $\cdots$ \\
G034.089+0.436 & G034.089+00.436  & 34.089 & 0.436 & 0.11 & 241.65 & 12.99 & 19.24 & 3.04 & 16.95 & 3.09 & 34.26 & 1.31 & 12.2 & 1.22 & 7.11 & IB & AGAL034.089+00.436 & $\cdots$ \\
G034.132+0.471 & G034.133+00.471  & 34.132 & 0.471 & 0.41 & 518.68 & 26.22 & 30.94 & 3.03 & 21.16 & 2.39 & 34.4 & 1.02 & 11.7 & 0.5 & 6.74 & B & AGAL034.133+00.471 & $\cdots$ \\
G034.255+0.146 & CORNISH G034.2544+00.1460  & 34.255 & 0.146 & 2.25 & 8087.97 & 405.24 & 532.43 & 6.7 & 25.25 & 0.37 & 54.04 & 0.16 & 3.4 & 0.34 & 5.67 & IB & $\cdots$ & $\cdots$ \\
G034.257+0.153\tablefootmark{*} & CORNISH G034.2573+00.1523  & 34.257 & 0.153 & 2.29 & 2440.25 & 122.65 & 39.35 & 1.48 & 59.79 & 2.6 & 44.09 & 1.1 & 3.4 & 0.34 & 5.67 & C & AGAL034.258+00.154 & G34.2582+0.1534 \\
G034.322+0.160 & $\cdots$  & 34.322 & 0.16 & 0.03 & 102.1 & 17.61 & 13.67 & 1.22 & 31.49 & 3.23 & 53.8 & 1.37 & 10.4 & 0.35 & 5.88 & ND & $\cdots$ & $\cdots$ \\
        .&.&.&.&.&.&.&.&.&.&.&.&.&.&.&.&.&. \\
        .&.&.&.&.&.&.&.&.&.&.&.&.&.&.&.&.&.\\
        .&.&.&.&.&.&.&.&.&.&.&.&.&.&.&.&.&. \\
        .&.&.&.&.&.&.&.&.&.&.&.&.&.&.&.&.&.\\        
        \hline \hline 
        \end{tabular}}
        \tablefoot{The entire table is accessible in a machine-readable format at the CDS. A section of the table is provided here as a reference for its format and content.\\
        \tablefoottext{a}{Glll.lll+b.bbb where l and b is the position of peak pixel.}
        \tablefoottext{b}{Name of the corresponding GLOSTAR continuum source located within angular separations of 36$\arcsec$ in the GLOSTAR VLA D-configuration continuum source catalog~\citep[][in prep]{2019A&A...627A.175M}. For sources where no crossmatch was found due to contamination from extended large-scale structures, we provided the name of the nearest \ion{H}{ii} region in the \texttt{SIMBAD} database.}
        \tablefoottext{c}{References for distances: (1)~\citet{2019ApJ...885..131R}; (2) \citet{2012ApJ...754...62A}; (3) \citet{2009ApJ...690..706A}; (4) \citet{2018MNRAS.473.1059U}.}
        \tablefoottext{d}{Classification of source based on \textit{Spitzer} GLIMPSE 8~$\mu$m emissions, as is explained in Section.~\ref{sec:mir_morphology}.}
        \tablefoottext{e}{Name of the associated ATLASGAL dust clump~\citep{2014A&A...568A..41U} (see Section.~\ref{sec:dust_asso}).}
        \tablefoottext{f}{Name of the associated 6.7~GHz methanol ($\rm CH_3OH$}) maser~\citep{2022A&A...666A..59N} (see Section.~\ref{sec:maser_asso}).
        \tablefoottext{*}{Candidate HC \ion{H}{ii} regions (see Section.~\ref{sec:line_width})}.
}

     \end{sidewaystable*}

     \begin{sidewaystable*}  
        \centering
        \caption{Physical properties of the GLOSTAR \ion{H}{ii} regions.}
        \label{tab:phy_prop}
        \resizebox{\linewidth}{!}{\begin{tabular}{ccccccccccccc}
        \hline \hline \\
        Name&$\rm \theta_{source}$&$\rm d_{eff}$&$\rm T_e$&$\rm \sigma T_e$& EM & $\sigma$EM &$\rm n_e$&$\rm \sigma n_e$&$\rm N_{Lyc}$&$\rm \sigma N_{Lyc}$&$\rm Mass_{\ion{H}{ii}}$&$\rm \sigma Mass_{\ion{H}{ii}}$    \\
        &[$\arcsec$]&[pc]&[$\rm 10^3$~K]&[$\rm 10^3$~K]&[$\rm 10^5~pc~cm^{-6}$]&[$\rm 10^5~pc~cm^{-6}$]&[$\rm 10^2~cm^{-3}$]&[$\rm 10^2~cm^{-3}$]&[$\rm 10^{48}~s^{-1}$]&[$\rm 10^{48}~s^{-1}$]&[$\rm M_\sun$]&[$\rm M_\sun$]    \\ \\
        \hline \hline 
        .&.&.&.&.&.&.&.&.&.&.&.&. \\
        .&.&.&.&.&.&.&.&.&.&.&.&. \\
        .&.&.&.&.&.&.&.&.&.&.&.&. \\        
G031.474-0.344 & 81.24 & 3.54 & 4.1 & 0.9 & 0.28 & 0.03 & 0.89 & 0.06 & 2.82 & 0.52 & 50.99 & 11.89 \\
G032.059+0.080 & 99.73 & 3.48 & 3.8 & 0.8 & 0.11 & 0.01 & 0.55 & 0.03 & 1.13 & 0.11 & 32.64 & 4.15 \\
G032.150+0.133 & 56.24 & 1.69 & 6.3 & 0.5 & 1.75 & 0.12 & 3.21 & 0.2 & 2.89 & 0.59 & 14.4 & 3.66 \\
G032.272-0.225 & 50.66 & 3.14 & 11.1 & 1.7 & 1.24 & 0.1 & 1.99 & 0.09 & 4.47 & 0.47 & 36.04 & 4.74 \\
G032.589+0.008 & 100.87 & 4.6 & 6.0 & 1.6 & 0.14 & 0.02 & 0.54 & 0.04 & 1.73 & 0.3 & 50.82 & 11.47 \\
G032.797+0.191 & 49.75 & 2.34 & 8.4 & 0.4 & 11.33 & 0.75 & 6.96 & 1.12 & 28.58 & 17.61 & 65.74 & 51.9 \\
G032.927+0.606 & 52.44 & 4.88 & 9.9 & 2.5 & 1.14 & 0.12 & 1.53 & 0.1 & 10.87 & 2.09 & 114.04 & 28.16 \\
G032.990+0.040 & 55.42 & 2.31 & 5.4 & 0.9 & 0.47 & 0.04 & 1.43 & 0.1 & 1.65 & 0.36 & 18.47 & 5.0 \\
G033.204-0.009 & 100.87 & 3.47 & 5.9 & 0.9 & 0.29 & 0.03 & 0.92 & 0.06 & 2.13 & 0.46 & 37.16 & 9.99 \\
G033.419-0.004 & 50.21 & 2.29 & 5.9 & 1.1 & 0.62 & 0.06 & 1.64 & 0.11 & 1.96 & 0.43 & 19.08 & 5.24 \\
G033.810-0.189 & 56.64 & 3.21 & 8.5 & 1.7 & 0.87 & 0.08 & 1.65 & 0.11 & 4.08 & 0.9 & 39.7 & 11.06 \\
G033.915+0.110 & 52.44 & 1.81 & 6.1 & 0.4 & 2.9 & 0.2 & 4.01 & 0.24 & 5.56 & 1.14 & 22.21 & 5.63 \\
G034.089+0.436 & 75.09 & 4.44 & 8.0 & 1.8 & 0.38 & 0.04 & 0.92 & 0.07 & 3.55 & 0.8 & 61.5 & 17.58 \\
G034.132+0.471 & 54.58 & 3.1 & 8.5 & 1.2 & 1.57 & 0.13 & 2.25 & 0.1 & 6.83 & 0.74 & 48.5 & 6.55 \\
G034.255+0.146 & 99.04 & 1.63 & 6.7 & 0.3 & 6.84 & 0.45 & 6.47 & 0.39 & 10.01 & 2.03 & 24.75 & 6.23 \\
G034.257+0.153\tablefootmark{*} & 46.91 & 0.77 & 10.7 & 0.7 & 10.86 & 0.74 & 11.85 & 0.72 & 2.44 & 0.5 & 3.3 & 0.84 \\
G034.322+0.160 & 72.29 & 3.64 & 3.0 & 0.6 & 0.12 & 0.03 & 0.58 & 0.07 & 1.7 & 0.23 & 47.07 & 5.32 \\
        .&.&.&.&.&.&.&.&.&.&.&.&. \\
        .&.&.&.&.&.&.&.&.&.&.&.&. \\
        .&.&.&.&.&.&.&.&.&.&.&.&. \\         
        \hline \hline 
        \end{tabular}}
        \tablefoot{The entire table is accessible in a machine-readable format at the CDS. A section of the table is provided here as a reference for its format and content.\\
        \tablefoottext{*}{Candidate HC \ion{H}{ii} regions (see Section.~\ref{sec:line_width})}
        }

     \end{sidewaystable*}

    \begin{table*}[h]
        \centering
        \caption{Kinematic distance analysis of 53 GLOSTAR \ion{H}{II} regions.}
        \label{tab:dist}
        \resizebox{\linewidth}{!}{\begin{tabular}{cccccccccccccc}
        \hline \hline \\
            Name&$\rm \ell_{peak}$ & \textit{b}$\rm _{peak}$ & V$\rm _{LSR}$ & V$\rm _{T}$ & V$\rm _{A}$ & \multicolumn{3}{c}{Reid Distance} & n/f/? &D$\rm _{sun}$&$\sigma$D$\rm _{sun}$&Quality\\
            \cline{7-9}
            &&&&&&Bayesian&Near&Far&&&&& \\ 
            &[$\degr$]&[$\degr$]&[$\rm km~s^{-1}$]&[$\rm km~s^{-1}$]&[$\rm km~s^{-1}$]&[kpc]&[kpc]&[kpc]&&[kpc]&[kpc]&& \\ \\
        \hline \hline
        G005.907-0.427  &       5.907   &       -0.427  &       8.83    &       175.45  &       40.4    &       2.96    &       1.76    &       13.8    &       n?      &       1.76    &       0.74    &       B       \\
        G008.662-0.342  &       8.662   &       -0.342  &       36.12   &       167.22  &       38.75   &       2.85    &       4.14    &       12.02   &       n       &       4.14    &       0.45    &       A       \\
        G008.872-0.319  &       8.872   &       -0.319  &       34.77   &       166.6   &       37.93   &       2.85    &       4.0         &       12.14   &       n       &       4.0     &       0.48    &       A       \\
        G010.597-0.384  &       10.597  &       -0.384  &       0.56    &       161.48  &       58.54   &       4.84    &       0.73    &       15.66   &       ?       &       4.84    &       0.48    &       C       \\
        G010.197-0.424  &       10.197  &       -0.424  &       9.45    &       162.66  &       39.58   &       1.25    &       1.25    &       14.42   &       n?      &       1.25    &       1.04    &       B       \\
        G010.142-0.422  &       10.142  &       -0.422  &       11.92   &       162.82  &       43.7    &       1.25    &       1.61    &       14.13   &       n?      &       1.61    &       0.96    &       B       \\
        G010.069-0.425  &       10.069  &       -0.425  &       16.58   &       163.04  &       26.39   &       2.87    &       2.19    &       13.66   &       n       &       2.19    &       0.8         &    A       \\
        G010.229-0.297  &       10.229  &       -0.297  &       12.11   &       162.57  &       44.52   &       1.25    &       1.62    &       14.12   &       n?      &       1.62    &       0.94    &       B       \\
        G010.229-0.349  &       10.229  &       -0.349  &       12.14   &       162.57  &       43.7    &       1.25    &       1.62    &       14.12   &       n?      &       1.62    &       0.94    &       B       \\
        G012.805-0.224  &       12.805  &       -0.224  &       30.75   &       154.97  &       44.52   &       2.86    &       3.05    &       12.86   &       n       &       3.05    &       0.5         &    A       \\
        G012.192-0.104  &       12.192  &       -0.104  &       24.36   &       156.77  &       119.55  &       12.37   &       2.64    &       13.23   &       f?      &       13.23   &       0.51    &       B       \\
        G012.724-0.225  &       12.724  &       -0.225  &       38.04   &       155.21  &       88.22   &       3.85    &       3.54    &       12.43   &       n       &       3.54    &       0.51    &       A       \\
        G012.912-0.282  &       12.912  &       -0.282  &       28.67   &       154.65  &       42.05   &       2.86    &       2.89    &       13.0    &       n       &       2.89    &       0.52    &       A       \\
        G013.184+0.050  &       13.184  &       0.05    &       54.75   &       153.86  &       54.42   &       3.89    &       4.33    &       11.69   &       n       &       4.33    &       0.31    &       A       \\
        G014.597+0.020  &       14.597  &       0.02    &       20.39   &       149.73  &       38      &       13.4    &       2.04    &       13.66   &       n       &       2.04    &       0.61    &       A               \\
        G014.619+0.118  &       14.619  &       0.118   &       34.85   &       149.66  &               &       2.84    &       3.1     &       12.72   &       n?      &       3.1     &       0.44    &       B               \\
        G015.137-0.529  &       15.137  &       -0.529  &       13.37   &       148.16  &       20      &       1.49    &       1.35    &       14.24   &       n       &       1.35    &       0.7     &       A               \\
        G015.189-0.599  &       15.189  &       -0.599  &       15.27   &       148     &       24.5    &       1.83    &       1.53    &       14.08   &       n       &       1.53    &       0.67    &       A               \\
        G015.205-0.627  &       15.205  &       -0.627  &       14.36   &       147.96  &       24.5    &       1.82    &       1.44    &       14.16   &       n       &       1.44    &       0.69    &       A       \\
        G019.117-0.337  &       19.117  &       -0.337  &       65.99   &       136.71  &               &       4.21    &       4.21    &       11.37   &       n       &       4.21    &       0.28    &       A               \\
        G020.740-0.085  &       20.74   &       -0.085  &       54.31   &       132.11  &       125     &       4.11    &       3.6     &       11.79   &       f       &       11.79   &       0.3     &       A               \\
        G020.682-0.157  &       20.682  &       -0.157  &       56.04   &       132.28  &       107     &       4.08    &       3.68    &       11.72   &       f?      &       11.72   &       0.3     &       B               \\
        G023.437-0.209  &       23.437  &       -0.209  &       101.56  &       124.59  &       105.5   &       7.54    &       5.32    &       9.86    &       n       &       5.32    &       0.26    &       A       \\
        G023.384-0.142  &       23.384  &       -0.142  &       98.61   &       124.74  &               &       5.28    &       5.22    &       9.97    &       n       &       5.22    &       0.26    &       A               \\
        G024.507-0.222  &       24.507  &       -0.222  &       97.98   &       121.65  &       99.5    &       5.62    &       5.18    &       9.87    &       n       &       5.18    &       0.27    &       A       \\
        G024.402+0.071  &       24.402  &       0.071   &       110.49  &       121.93  &       119     &       6.83    &       5.66    &       9.42    &       f?      &       9.42    &       0.27    &       B               \\
        G024.712-0.125  &       24.712  &       -0.125  &       109.12  &       121.09  &       110     &       6.28    &       5.61    &       9.43    &       n?      &       5.61    &       0.28    &       B               \\
        G025.692+0.031  &       25.692  &       0.031   &       53.63   &       118.42  &       113     &       3.85    &       3.32    &       11.51   &       f       &       11.51   &       0.31    &       A               \\
        G025.479-0.174  &       25.479  &       -0.174  &       55.8    &       118.99  &               &       3.87    &       3.43    &       11.43   &       ?       &       3.87    &       0.44    &       C               \\
        G025.275-0.317  &       25.275  &       -0.317  &       67.45   &       119.55  &               &       3.64    &       3.95    &       10.96   &       ?       &       3.64    &       0.35    &       C               \\
        G027.994-0.022  &       27.994  &       -0.022  &       100.64  &       112.23  &               &       5.89    &       5.34    &       9.27    &       T?      &       7.3     &       0.32    &       B               \\
        G028.569+0.020  &       28.569  &       0.02    &       98.88   &       110.71  &       107     &       8.13    &       5.28    &       9.25    &       f?      &       9.25    &       0.32    &       B               \\
        G028.977-0.607  &       28.977  &       -0.607  &       51.67   &       109.63  &               &       3.98    &       3.13    &       11.26   &       ?       &       3.98    &       0.8     &       C               \\
        G029.859-0.059  &       29.859  &       -0.059  &       97.99   &       107.32  &               &       7.65    &       5.3     &       9.05    &       T       &       7.18    &       0.36    &       A               \\
        G030.252+0.053  &       30.252  &       0.053   &       70.29   &       106.29  &               &       4.53    &       4       &       10.26   &       ?       &       4.53    &       0.83    &       C               \\
        G030.687-0.072  &       30.687  &       -0.072  &       94.32   &       105.17  &       102.5   &       7.06    &       5.15    &       9.07    &       T?      &       7.11    &       0.36    &       B       \\
        G030.750+0.013  &       30.75   &       0.013   &       93.85   &       105     &       110     &       7.1     &       5.13    &       9.08    &       f       &       9.08    &       0.36    &       A               \\
        G030.789-0.100  &       30.789  &       -0.1    &       108.74  &       104.9   &               &       7.18    &       6.05    &       8.17    &       T       &       7.11    &       0.47    &       A               \\
        G030.854+0.151  &       30.854  &       0.151   &       40.34   &       104.73  &       111.5   &       10.43   &       2.52    &       11.58   &       f       &       11.58   &       0.36    &       A       \\
        G031.147+0.275  &       31.147  &       0.275   &       108.97  &       103.98  &               &       6.8     &       6.13    &       8.03    &       T       &       7.08    &       0.5     &       A               \\
        G034.322+0.160  &       34.322  &       0.16    &       53.8    &       95.94   &               &       9.82    &       3.19    &       10.4    &       f       &       10.4    &       0.35    &       A               \\
        G035.035-0.504  &       35.035  &       -0.504  &       53.44   &       94.17   &       93.5    &       2.76    &       3.18    &       10.3    &       f       &       10.3    &       0.36    &       A       \\
        G043.182-0.027  &       43.182  &       -0.027  &       9.78    &       75.16   &       80      &       11.07   &       0.63    &       11.27   &       f       &       11.27   &       0.44    &       A               \\
        G043.185-0.525  &       43.185  &       -0.525  &       57.69   &       75.15   &       78.5    &       7.61    &       3.77    &       8.24    &       f       &       8.24    &       0.53    &       A       \\
        G045.132+0.143  &       45.132  &       0.143   &       54.57   &       70.94   &       80      &       7.79    &       3.71    &       7.89    &       f       &       7.89    &       0.53    &       A               \\
        G048.594+0.043  &       48.594  &       0.043   &       18.93   &       63.82   &       72.5    &       10      &       1.3     &       9.52    &       f       &       9.52    &       0.47    &       A               \\
        G048.914-0.285  &       48.914  &       -0.285  &       68.21   &       63.18   &       80      &       5.46    &       5.48    &       5.48    &       T       &       5.48    &       0.82    &       A               \\
        G049.205-0.342  &       49.205  &       -0.342  &       67.27   &       62.61   &       72.5    &       5.44    &       5.45    &       5.45    &       T       &       5.45    &       0.82    &       A       \\
        G049.477-0.329  &       49.477  &       -0.329  &       56.96   &       62.07   &       71      &       5.48    &       5.42    &       5.42    &       T       &       5.42    &       0.82    &       A               \\
        G049.585-0.385  &       49.585  &       -0.385  &       63.94   &       61.86   &       68      &       5.43    &       5.41    &       5.41    &       T       &       5.41    &       0.82    &       A               \\
        G049.440-0.459  &       49.44   &       -0.459  &       56.26   &       62.15   &       66.5    &       5.48    &       5.42    &       5.42    &       T       &       5.42    &       0.82    &       A       \\
        G051.365-0.012  &       51.365  &       -0.012  &       52.93   &       58.45   &       63.5    &       5.72    &       5.21    &       5.21    &       T       &       5.21    &       0.84    &       A       \\
        G059.800+0.231  &       59.8    &       0.231   &       -2.07   &       44.11   &               &       8.35    &       0.1     &       8.29    &       f?      &       8.29    &       0.55    &       B               \\

        \hline \hline 
        \end{tabular}}
   \end{table*}

        \end{appendix}
        
\end{document}